\documentclass[%
                aip,
                amsmath,amssymb,
                reprint,%
                nofootinbib, 
              ]{revtex4-1}

\usepackage[utf8]{inputenc}

\usepackage{etoolbox}

\draft 

\usepackage{import}

\usepackage[english]{babel}
\addto{\captionsenglish}{%
  
}

\usepackage{hyperref}
\hypersetup{
	colorlinks=true,
	linkcolor=black,
	filecolor=magenta,
	urlcolor=cyan,
	citecolor=black,
}
\usepackage[capitalise]{cleveref}

\usepackage{nicefrac}

\usepackage{cancel}

\usepackage{subcaption}

\usepackage{afterpage}

\usepackage{placeins}

\usepackage{makecell}

\usepackage{textcomp}

\usepackage[textwidth=2.1cm, colorinlistoftodos=true, color=green]{todonotes}

\definecolor{acsblue}{RGB}{17,76,139}
\definecolor{acswhite}{RGB}{255,255,255}
\definecolor{acsyellow}{RGB}{255,241,204}

\usepackage{mdframed}
\definecolor{shadecolor}{RGB}{110,160,179}

\mdfdefinestyle{mdfpreamble}{%
	linecolor=acswhite, linewidth=0.8pt,
	backgroundcolor=shadecolor,
	leftline=false, rightline=false,
	innertopmargin=0.2cm, innerbottommargin=0.2cm,
	innerleftmargin=1.25cm, innerrightmargin=0.25cm,
}

\usepackage{textpos}

\newcommand{\figwidth}{3.2in}				
\newcommand{\smallfigwidth}{2.0in}			
\newcommand{\largerfigwidth}{3.225in}		
\newcommand{\midlargerfigwidth}{3.75in}		

\definecolor{vpm}{RGB}{79,147,206}
\definecolor{fclr}{RGB}{79,147,206}
\definecolor{gclr}{RGB}{160,140,110}

\newcommand{\f}{{\color{fclr}{f}}}
\newcommand{\g}{{\color{gclr}{g}}}

\newcommand{\filter}[1]{ \overline{#1} }

\newcommand{\filters}[2]{ \filter{#1} \, \filter{#2}}

\newcommand{\ave}[1]{ \left< #1 \right> }

\graphicspath{{figures/}}

\newtoggle{brief}

\toggletrue{brief}

\begin{document}

    \title[Reviving the VPM: A Stable Formulation for Meshless LES]{Reviving the Vortex Particle Method:\\A Stable Formulation for Meshless Large Eddy Simulation}

    \author{Eduardo J. Alvarez}
        \email{edoalvarezr@gmail.com}
    \author{Andrew Ning}%
        \email{aning@byu.edu}
    \affiliation{Ira A. Fulton College of Engineering, Brigham Young University, Utah, USA}%


    \begin{abstract}
        \vspace{0mm}
        The vortex particle method (VPM) is a mesh-free approach to computational fluid dynamics (CFD) solving the Navier-Stokes equations in their velocity-vorticity form.
        The VPM uses a Lagrangian scheme, which not only avoids the hurdles of mesh generation, but it also conserves vortical structures over long distances with minimal numerical dissipation while being orders of magnitude faster than conventional mesh-based CFD.
        However, VPM is known to be numerically unstable when vortical structures break down close to the turbulent regime.
        In this study, we reformulate the VPM as a large eddy simulation (LES) in a scheme that is numerically stable, without increasing its computational cost.
        A new set of VPM governing equations are derived from the LES-filtered Navier-Stokes equations.
        The new equations reinforce conservation of mass and angular momentum by reshaping the vortex elements subject to vortex stretching.
        In addition to the VPM reformulation, a new anisotropic dynamic model of subfilter-scale (SFS) vortex stretching is developed.
        This SFS model is well suited for turbulent flows with coherent vortical structures where the predominant cascade mechanism is vortex stretching.
        Extensive validation is presented, asserting the scheme comprised of the reformulated VPM and SFS model as a meshless LES that accurately resolves large-scale features of turbulent flow.
        Advection, viscous diffusion, and vortex stretching are validated through simulation of isolated and leapfrogging vortex rings.
        Mean and fluctuating components of turbulent flow are validated through simulation of a turbulent round jet, where Reynolds stresses are resolved directly and compared to experimental measurements.
        Finally, the computational efficiency of the scheme is showcased in the simulation of an aircraft rotor in hover, showing our meshless LES to be 100x faster than a mesh-based LES with similar fidelity, while being 10x faster than a low-fidelity unsteady Reynolds-average Navier-Stokes simulation and 1000x faster than a high-fidelity detached-eddy simulation.
        The implementation of our meshless LES scheme is hereby released as an open-source software, called \href{https://github.com/byuflowlab/FLOWVPM.jl}{FLOWVPM}.
    \end{abstract}


    \maketitle

    \begin{textblock*}{1.125\textwidth}(-2.25cm,-13.9cm)
    	\begin{mdframed}[style=mdfpreamble]
    		\begin{minipage}{\textwidth}

        \begin{minipage}{0.1\textwidth}
    			\href{http://flow.byu.edu/}{
    				\hspace{-5mm}
    				\includegraphics[width=\textwidth]{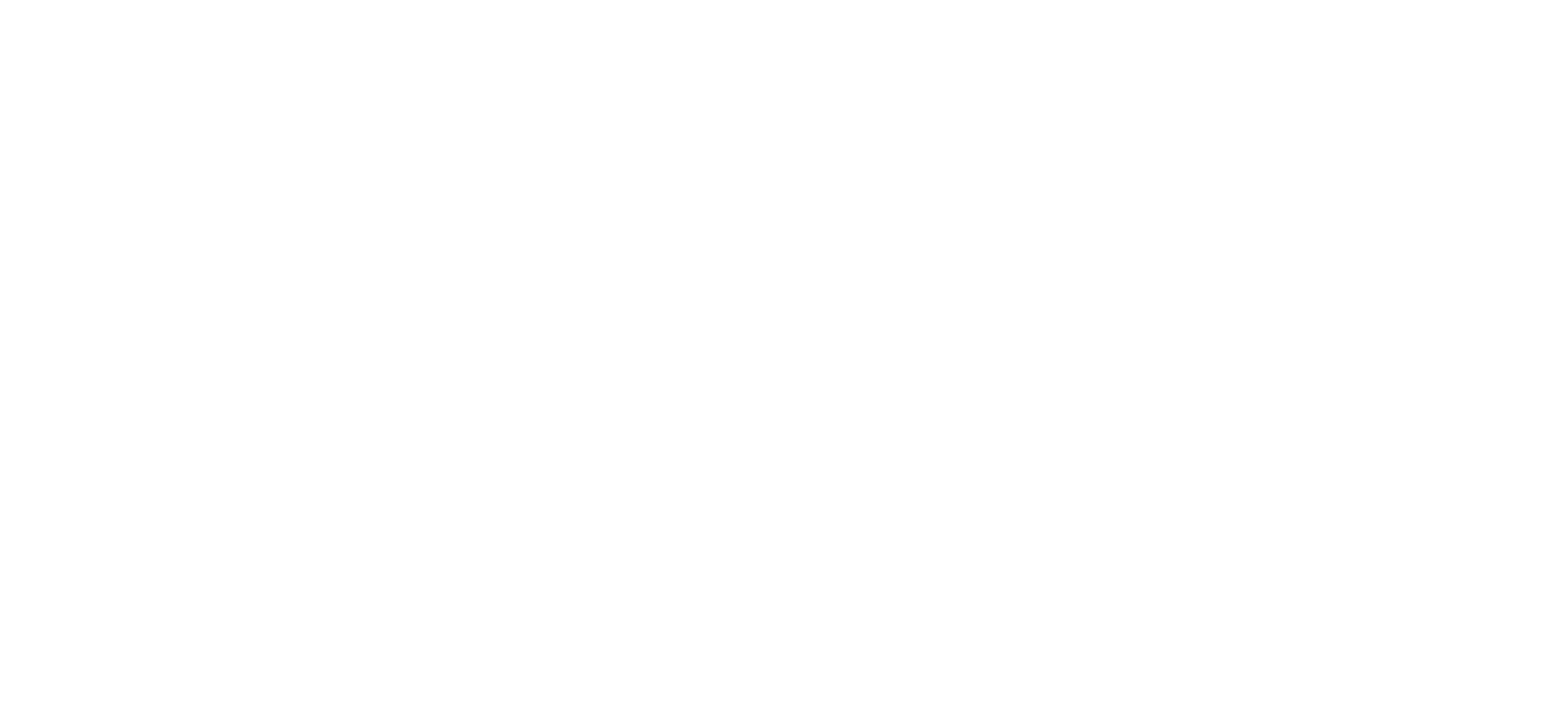}
          }%
        \end{minipage}%
    			\begin{minipage}{0.9\textwidth}
    				\begin{center}
              {\footnotesize
      					\textcolor{white}{
      						\noindent Preprint of E. J. Alvarez and A. Ning, ``Stable Vortex Particle Method Formulation for Meshless Large-Eddy Simulation,'' \textit{AIAA Journal}, Nov 2023. doi: 10.2514/1.J063045\\
                            The content of this paper may differ from the final publisher version.
      					}%
    					}%
    				\end{center}%
    			\end{minipage}%

    		\end{minipage}%
    	\end{mdframed}%
    \end{textblock*}

    \vspace{-5mm}
    \section{Introduction}

    Large eddy simulation (LES) is a class of computational fluid dynamics (CFD) that filters the Navier-Stokes equations to decompose small-scale fluctuations of the flow from large ones.
    The large scales are then resolved directly, while the effects of the smaller scales are modeled.
    The most common LES methods use a mesh or grid to discretize the space and calculate fluxes and derivatives, classified as finite volume/element/difference methods.
    Significant user effort is spent in the generation and manipulation of the mesh, with studies showing that about 67\% of engineering time in mesh-based CFD is spent in these efforts.\cite{Hardwick2005}

    Vortex methods \cite{Leonard1980,Cottet2000a,Winckelmans2004,Mimeau2021} are a class of meshless CFD solving the Navier-Stokes equations in their velocity-vorticity form.
    This form is solved in a Lagrangian scheme, which not only avoids the hurdles of mesh generation, but also conserves vortical structures over long distances with minimal numerical dissipation.

    The vortex particle method \cite{Winckelmans1993} (VPM) is a vortex method that uses particles to discretize the Navier-Stokes equations, with the particles representing radial basis functions that construct a continuous vorticity field \cite{Barba2004}.
    This meshless CFD has several advantages over conventional mesh-based CFD.
    In the absence of a mesh, the VPM (1) does not suffer from the conventional Courant--Friedrichs--Lewy (CFL) condition, (2) does not suffer from the numerical dissipation introduced by a mesh, (3) derivatives are calculated exactly rather than approximated through a stencil, and (4) it has been shown to be 100x to 1000x faster than mesh-based CFD with comparable fidelity \cite{Alvarez2020a}.
    Furthermore, the VPM is spatially second-order accurate \cite{Yokota2013}, it is highly efficient since elements are placed only where vorticity exists, the spatial discretization can be automatically adapted \cite{Barba2005,Lakkis2009,Stock2021} in the fashion of an adaptive mesh refinement, and simulations are highly-parallelizable in heterogeneous CPU and GPU architectures \cite{Yokota2013a,Hu2014}.

    \begin{figure*}[t]
        \begin{minipage}{0.49\textwidth}
            \includegraphics[width=\linewidth]{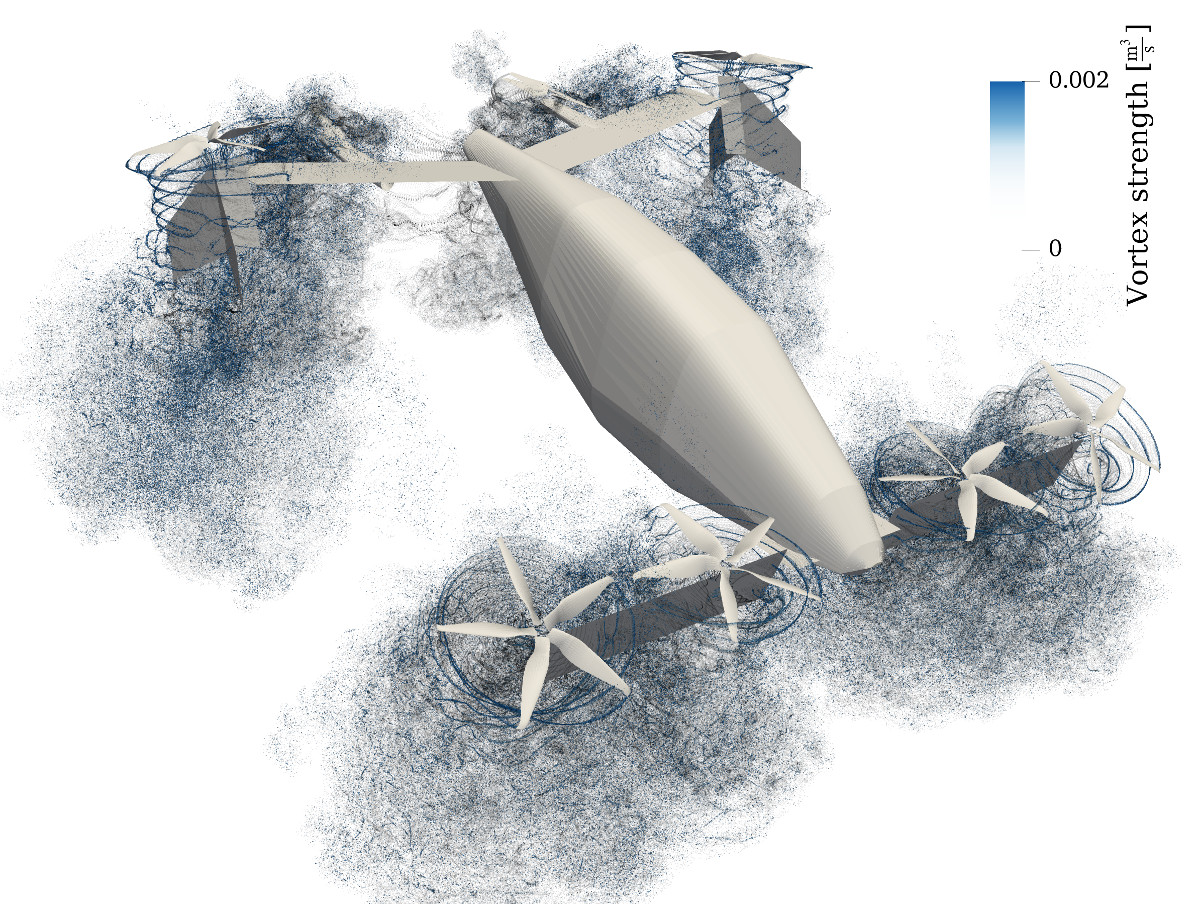}
        \end{minipage}
        \begin{minipage}{0.49\textwidth}
            \includegraphics[width=\linewidth]{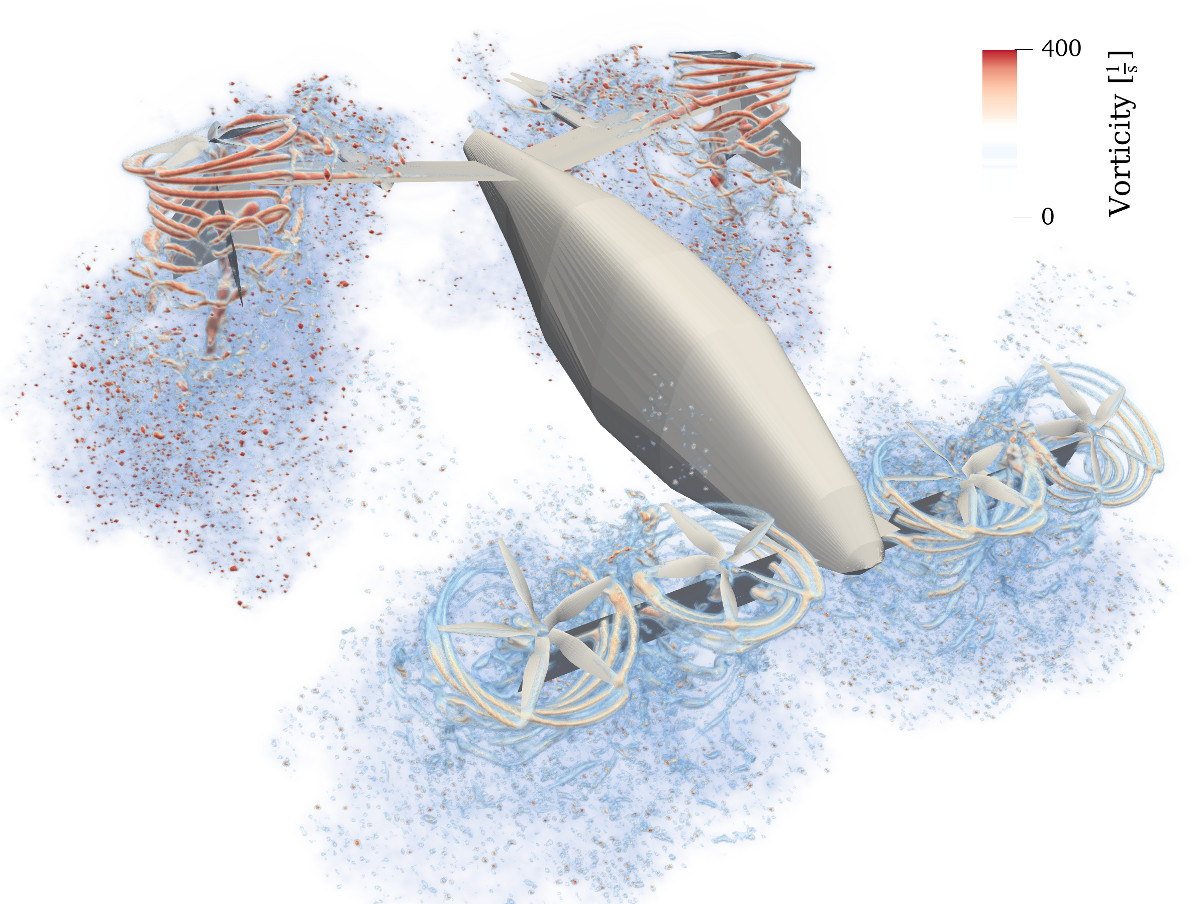}
        \end{minipage}

        \caption{Meshless LES of eVTOL aircraft using reformulated VPM: (left) computational elements (vortex particles and strength), and (right) volume rendering of vorticity field.}
        \label{fig:evtol}

    \end{figure*}

    VPM has gained popularity in recent years due to a growing need to predict complex aerodynamic interactions in modern electric aircraft \cite{Kasliwal2019,nasem2020,Schwab2021,McDonald2021} at a computational cost fit for conceptual design \cite{Silva2018,Brelje2018}.
    For example, VPM has been used for high-fidelity simulation of rotorcraft forward flight \cite{Stock2010} and multirotor interactions \cite{Alvarez2020,Alvarez2020a}, and as a mid and low-fidelity tool for ground effect \cite{FengTan2020,Tan2021}, electrical vertical takeoff and landing (eVTOL) \cite{Yucekayal2020}, distributed electric propulsion \cite{Alvarez2018a,Teixeira2019}, aeromechanics of unconventional rotorcraft \cite{Singh2021,Jacobellis2021a,Corle2021}, rotor-rotor interactions \cite{Alvarez2019,Lee2020}, and wind energy \cite{Lee2019a,Lee2019b,Mehr2020}.
    Furthermore, a couple of opensource VPM codes have been recently released \cite{Kuzmina2019,Stock2020,Tugnoli2021}.
    \cref{fig:evtol} shows the simulation of an eVTOL aircraft using VPM.

    In spite of its growing popularity, VPM is known to be numerically unstable when vortical structures break down close to the turbulent regime.
    This has limited its range of applications in the aforementioned studies to mostly benign cases with well-behaved numerics (\textit{e.g.,} coarse simulations of rotors with axial inflow).
    The instability is triggered when vortex stretching gives raise to a rapid increase of local vorticity, which the low numerical dissipation of the method fails to damp out.
    Such numerical instability is believed to be caused by a combination of Lagrangian distortion, a vorticity field that is not divergence-free, and the absence of subfilter-scale (SFS) turbulent diffusion, among other reasons.

    Multiple meshless schemes have been developed over the years to address Lagrangian distortion \cite{Barba2005a,Lakkis2009,Kirchhart2021}, divergence of the vorticity field \cite{Pedrizzetti1992,Gharakhani1997}, and SFS effects \cite{Gharakhani2002,Gharakhani2005a}.
    While some of these schemes are effective in two-dimensional cases, they have not succeeded at making the three-dimensional VPM numerically stable.
    The need to circumvent this challenge motivated further development of the vortex-in-cell (VIC) method \cite{Christiansen1997,Cottet2004,Koumoutsakos2005}, also known as vortex particle-mesh method.
    In this method, the particles are projected onto a background mesh at every time step, and vortex stretching, viscous diffusion, and the Biot-Savart law are computed in a mesh-based scheme.
    This approach has shown to be numerically stable, recently enabling the study of wake dynamics with unprecedented fidelity.\cite{Chatelain2008,Chatelain2016,Caprace2020a,Nguyen2019,Nguyen2021,Colognesi2021,Ramos-Garcia2021}
    However, the introduction of this mesh also forfeits a few of the aforementioned benefits of a purely Lagrangian (meshless) scheme.

    In this study, we propose an LES formulation of the VPM that is both numerically stable and meshless.
    A new set of VPM governing equations are derived from the LES-filtered Navier-Stokes equations in~\cref{sec:fundamentals}.
    In~\cref{sec:classicvpm}, the classic VPM is shown to be only one out of multiple possible formulations of these governing equations.
    Such formulation is shown to locally violate conservation of mass and angular momentum under vortex stretching, explaining why the classic VPM tends to be numerically unstable.
    In~\cref{sec:reformulatedvpm}, we consider multiple possible formulations using different element shapes and conservation laws.
    One formulation is selected, which uses the particle size to reinforce local conservation of both mass and angular momentum.
    This formulation, referred to as the \textit{reformulated VPM}, is implemented and hereby released as the open-source software \href{https://github.com/byuflowlab/FLOWVPM.jl}{FLOWVPM}.\footnote{The code is available at \href{https://github.com/byuflowlab/FLOWVPM.jl}{https://github.com/byuflowlab/FLOWVPM.jl}}

    In addition to the reformulated VPM, a novel anisotropic structural model of subfilter-scale (SFS) vortex stretching is developed in~\cref{sec:sfs}.
    The means for backscatter control are also provided, along with a dynamic procedure for the automatic computation of the model coefficient.
    This SFS model is apt for both meshless and mesh-based CFD, and is well suited for turbulent flows with coherent vortical structures where the predominant cascade mechanism is vortex stretching.

    The scheme comprised of the reformulated VPM and SFS model is extensively tested in~\cref{sec:res}, validating the scheme as an LES method.
    Advection, viscous diffusion, and vortex stretching are validated through simulation of isolated and leapfrogging vortex rings.
    Large-scale turbulent dynamics are validated through simulation of a turbulent round jet, where predicted fluctuations and Reynolds stress are compared to experimental measurements.
    Finally, the computational efficiency of our meshless LES is showcased in an engineering application through the simulation of an aircraft rotor in hover.

    \clearpage
    \section{Fundamentals of \\the Vortex Particle Method} \label{sec:fundamentals}

    A new set of VPM governing equations are hereby derived from the LES-filtered Navier-Stokes equations in their vorticity form.
    Along the way, the derivation here presented introduces all the fundamental concepts necessary to bring the reader up to speed with LES and vortex methods.

    \subsection{Vorticity Form of the Navier-Stokes Equations} \label{sec:fundamentals:ns}

        In a Newtonian, incompressible fluid with constant viscosity, the linear momentum of a differential fluid element is governed by the Navier-Stokes equation
        \begin{align} \label{eq:NS0}
                \frac{\partial {\textbf u}}{\partial t} +
                    ({\textbf u}\cdot\nabla){\textbf u} = -\frac{1}{\rho}\nabla p + \nu\nabla^2{\textbf u}
        ,\end{align}
        where ${\textbf u}({\textbf x}, t)$ is the velocity field, $p({\textbf x}, t)$ is the pressure field, and $\rho$ and $\nu$ are the density and kinematic viscosity of the fluid, respectively.

        Taking the curl over \cref{eq:NS0}, the pressure dependence disappears obtaining
        \begin{align} \label{eq:NS3}
            \frac{\text{D} }{\text{D} t} \boldsymbol{\omega} = (\boldsymbol{\omega} \cdot \nabla ){\textbf u} +
                \nu\nabla^2\boldsymbol{\omega}
        ,\end{align}
        where $\frac{\text{D} }{\text{D} t}$ denotes the material derivative operator, $\frac{\text{D} }{\text{D} t}() \equiv \frac{\partial }{\partial t}() + {({\textbf u}~\cdot~\nabla)()}$, and $\boldsymbol{\omega}({\textbf x}, t) = \nabla\times{\textbf u}({\textbf x}, t)$ is the vorticity field.

        \cref{eq:NS3} is the vorticity form of the Navier-Stokes linear-momentum equation.
        This equation depends on $\boldsymbol{\omega}$ alone since $\textbf{u}$ can be calculated from $\boldsymbol{\omega}$ through the Biot-Savart law.
        While~\cref{eq:NS0} stems from conservation of linear momentum, ~\cref{eq:NS3} can be interpreted as some form of conservation of angular momentum\cite{Chatwin1973}, which is further discussed in~\cref{sec:reformulatedvpm:physicalimplications}.
        From the right-hand side, we see that the evolution of the vorticity field is governed by vortex stretching (first term) and viscous diffusion (second term).

    \subsection{Large Eddy Simulation Equation} \label{sec:fundamentals:les}

        We now derive the filtered vorticity Navier-Stokes equation---here referred to as \textit{large-eddy simulation (LES) equation}---and define the stress tensor that transfers energy between resolved scales and subfilter scales.

        Let $\phi$ be a field and $\zeta_\sigma$ a filter kernel, the filtered field is denoted by a bar and defined as
        \begin{align*}
                \filter{\phi} \left( \mathbf{x} \right)
            & \equiv
                \int\limits_{-\infty}^\infty
                    \phi(\mathbf{y})\zeta_\sigma(\mathbf{x}-\mathbf{y})
                \,\mathrm{d}\mathbf{y}
        ,\end{align*}
        where the filter $\zeta_\sigma$ is associated to a certain cutoff length scale $\sigma$.
        In this study, $\zeta_\sigma$ is defined as $\zeta_\sigma(\mathbf{x}) \equiv \frac{1}{\sigma^3} \zeta (\frac{\Vert \mathbf{x} \Vert}{\sigma})$ where $\zeta$ is a radial basis function, and is required to have a volume integral of unity, $\int\limits_{-\infty}^\infty \zeta_\sigma(\mathbf{y})\,\mathrm{d}\mathbf{y} = 1$.

        In order to derive the LES version of the vorticity equation,~\cref{eq:NS3} is written in tensor notation and filtered as
        \begin{align} \label{eq:filtered:angular:original}
                \filter{ \frac{\partial \omega_i}{\partial t} }
                + \filter{ u_j \frac{\partial \omega_i}{\partial x_j} }
                = \filter{ \omega_j \frac{\partial u_i}{\partial x_j} } +
                    \nu \filter{ \nabla^2 \omega_i }
        .\end{align}
        In this equation, $\filter{ u_j\frac{\partial \omega_i}{\partial x_j} }$ and $\filter{ \omega_j\frac{\partial u_i}{\partial x_j} }$ are non-linear terms that cannot be calculated from resolved quantities, but are rather approximated through a tensor $T_{ij}$ that encapsulates the error between $\filter{ u_i \omega_j }$ and $\filters{ u_i }{ \omega_j }$ as
        \begin{align*} 
            \filter{ u_i \omega_j } = \filters{ u_i }{ \omega_j } + T_{ij}
        .\end{align*}
        The gradient of $T_{ij}$ and its transpose result in
        \begin{align*}
            &
                \frac{\partial T'_{ij}}{\partial x_j} =
                \filter{ u_j \frac{\partial \omega_i}{\partial x_j}} -
                \filter{ u_j } \frac{\partial \filter{ \omega_i }}{\partial x_j}
            \\
            &
                \frac{\partial T_{ij}}{\partial x_j} =
                \filter{ \omega_j \frac{\partial u_i}{\partial x_j} } -
                \filter{ \omega_j } \frac{\partial \filter{ u_i }}{\partial x_j}
        \end{align*}
        Replacing this into \cref{eq:filtered:angular:original} and using the derivative-filter commutation property, we obtain the LES vorticity equation:
        \begin{subequations} \label{eq:filtered:angular:tensor}
        \begin{align*}
            \tag{\ref{eq:filtered:angular:tensor}}
                \frac{\partial \filter{ \omega_i }}{\partial t}
                + \filter{ u_j } \frac{\partial \filter{ \omega_i }}{\partial x_j}
            & =
            \\
                \filter{ \omega_j } \frac{\partial \filter{ u_i }}{\partial x_j}
            & +
                \nu \nabla^2 \filter{ \omega_i } -
                \frac{\partial T'_{ij}}{\partial x_j} +
                \frac{\partial T_{ij}}{\partial x_j}
        .\end{align*}
        \end{subequations}
        The term $\frac{\partial T'_{ij}}{\partial x_j}$ represents the subfilter-scale (SFS) contributions arising from the advective term (vorticity advection), while $\frac{\partial T_{ij}}{\partial x_j}$ represents the contributions arising from vortex stretching.
        $T_{ij}$ is associated to the SFS vorticity stress that encapsulates the interactions between large scale dynamics and SFS dynamics and has to be modeled in terms of resolved quantities.
        The accuracy of LES hinges on the modeling of this tensor.
        Its divergence represents the rate at which enstrophy---a measure of rotational kinetic energy---is transferred from resolved scales to subfilter scales (diffusion) and from subfilter scales to resolved scales (backscatter).
        In~\cref{sec:sfs} we will develop a model that approximates the vortex-stretching component of this tensor, and provide the means for backscatter control.

        For convenience, we write~\cref{eq:filtered:angular:tensor} in vector notation as
        \begin{align}  \label{eq:filtered:angular:vector}
            \frac{\text{d} }{\text{d} t} \filter{ \boldsymbol\omega }
            = \left( \filter{ \boldsymbol\omega } \cdot \nabla \right) \filter{ \mathbf{u} } +
            \nu \nabla^2 \filter{ \boldsymbol\omega }
            - \mathbf{E}_\mathrm{adv} - \mathbf{E}_\mathrm{str}
        ,\end{align}
        where $\left( \mathbf{E}_\mathrm{adv} \right)_i \equiv \frac{\partial T'_{ij}}{\partial x_j}$ is the SFS vorticity advection, $\left( \mathbf{E}_\mathrm{str} \right)_i \equiv - \frac{\partial T_{ij}}{\partial x_j}$ is the SFS vortex stretching, and the $\frac{\text{d} }{\text{d} t}$ operator is a linearized version of the filtered material derivative, $\frac{\text{d} }{\text{d} t} () \equiv \frac{\partial }{\partial t}() + {(\filter{ {\mathbf u} } \cdot \nabla)()}$.

    \subsection{Viscous Diffusion} \label{sec:fundamentals:splitting}
        One common practice for solving non-linear partial differential equations (PDE), like~\cref{eq:filtered:angular:vector}, is that of splitting the PDE operator into a linear sum of its non-linear pieces.
        This permits discretizing and solving each non-linear piece in a separate numerical scheme.
        In this study we will split the PDE in~\cref{eq:filtered:angular:vector} into inviscid and viscous pieces as
        \begin{align*}  
                \frac{\text{d} }{\text{d} t} \filter{ \boldsymbol\omega }
            & =
                \left(
                    \frac{\text{d} }{\text{d} t} \filter{ \boldsymbol\omega }
                \right)_\mathrm{inviscid}
                +
                \left(
                    \frac{\text{d} }{\text{d} t} \filter{ \boldsymbol\omega }
                \right)_\mathrm{viscous}
        ,\end{align*}
        where
        \begin{align*}
                \left(
                    \frac{\text{d} }{\text{d} t} \filter{ \boldsymbol\omega }
                \right)_\mathrm{inviscid}
            & =
                \left( \filter{ \boldsymbol\omega } \cdot \nabla \right) \filter{ \mathbf{u} }
                -
                \mathbf{E}_\mathrm{adv} - \mathbf{E}_\mathrm{str}
            \\
                \left(
                    \frac{\text{d} }{\text{d} t} \filter{ \boldsymbol\omega }
                \right)_\mathrm{viscous}
            & =
                \nu \nabla^2 \filter{ \boldsymbol\omega }
        .\end{align*}

        Over the years, multiple Lagrangian schemes have been developed that accurately resolve the viscous component, like the vortex redistribution method \cite{Shankar1996,Gharakhani2001}, particle strength exchange \cite{Degond1989}, and core spreading \cite{Rossi1996}, to name a few.
        In this study, viscous diffusion will be solved through the core spreading method coupled with the radial basis function interpolation approach for spatial adaptation developed by Barba \cite{Barba2005a, Barba2004, Torres2009}.
        This viscous scheme diffuses the vorticity by thickening each particle's core size $\sigma$ over time, while using an RBF interpolation to reset core sizes when they have overgrown.
        As shown by Rossi \cite{Rossi1996}, the core spreading method has second-order spatial convergence, while showing linear convergence when coupled with spatial adaptation.

        In the following sections we will focus on the inviscid part of the PDE, developing a scheme for its numerical solution.

    \subsection{Lagrangian Discretization: The Vortex Particle} \label{sec:fundamentals:discretization}

        \begin{figure*}[t]
            \begin{subfigure}{0.30\textwidth} \centering
                \includegraphics[width=\linewidth]{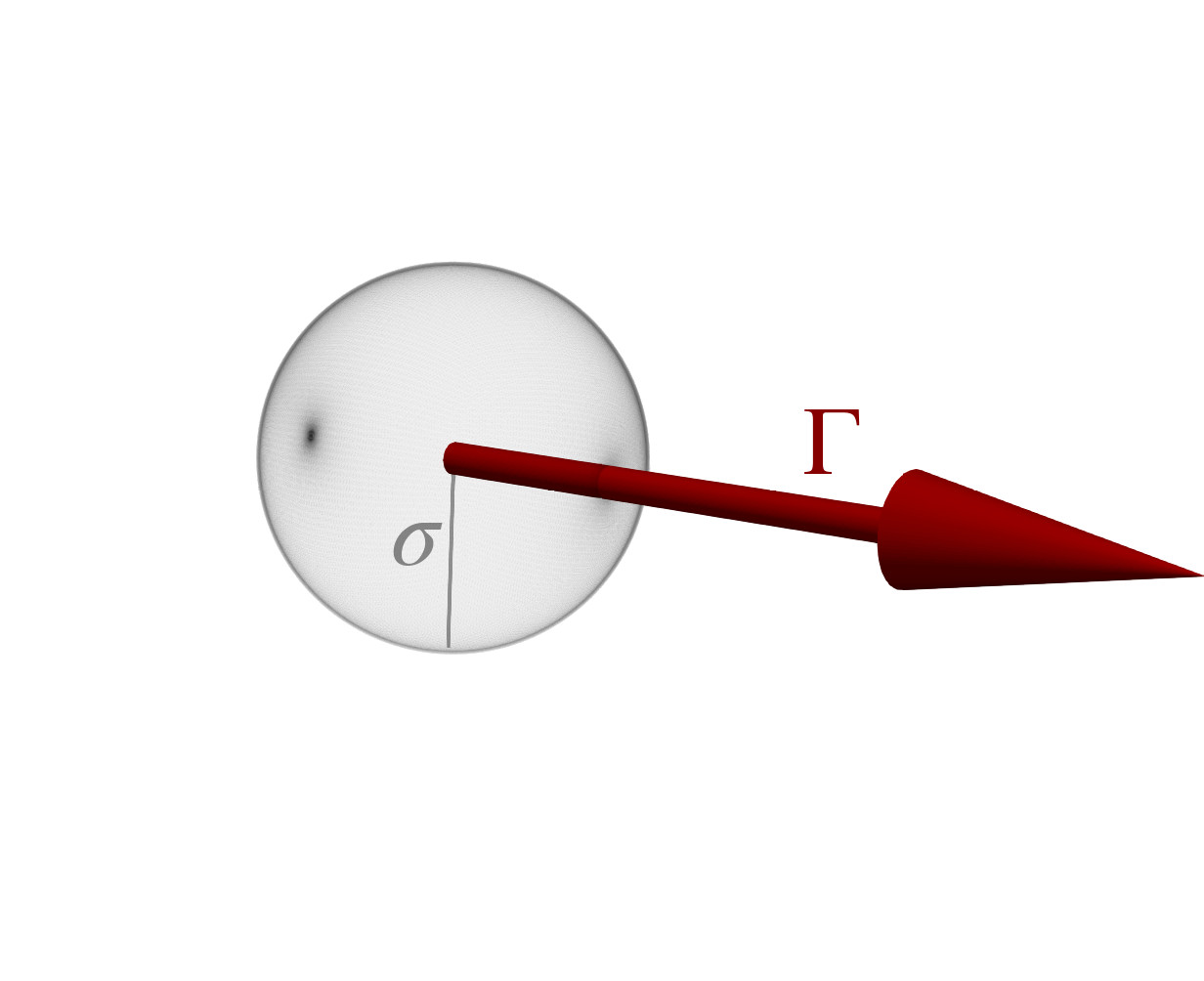}
            \end{subfigure}
            \begin{subfigure}{0.30\textwidth} \centering
                \includegraphics[width=\linewidth]{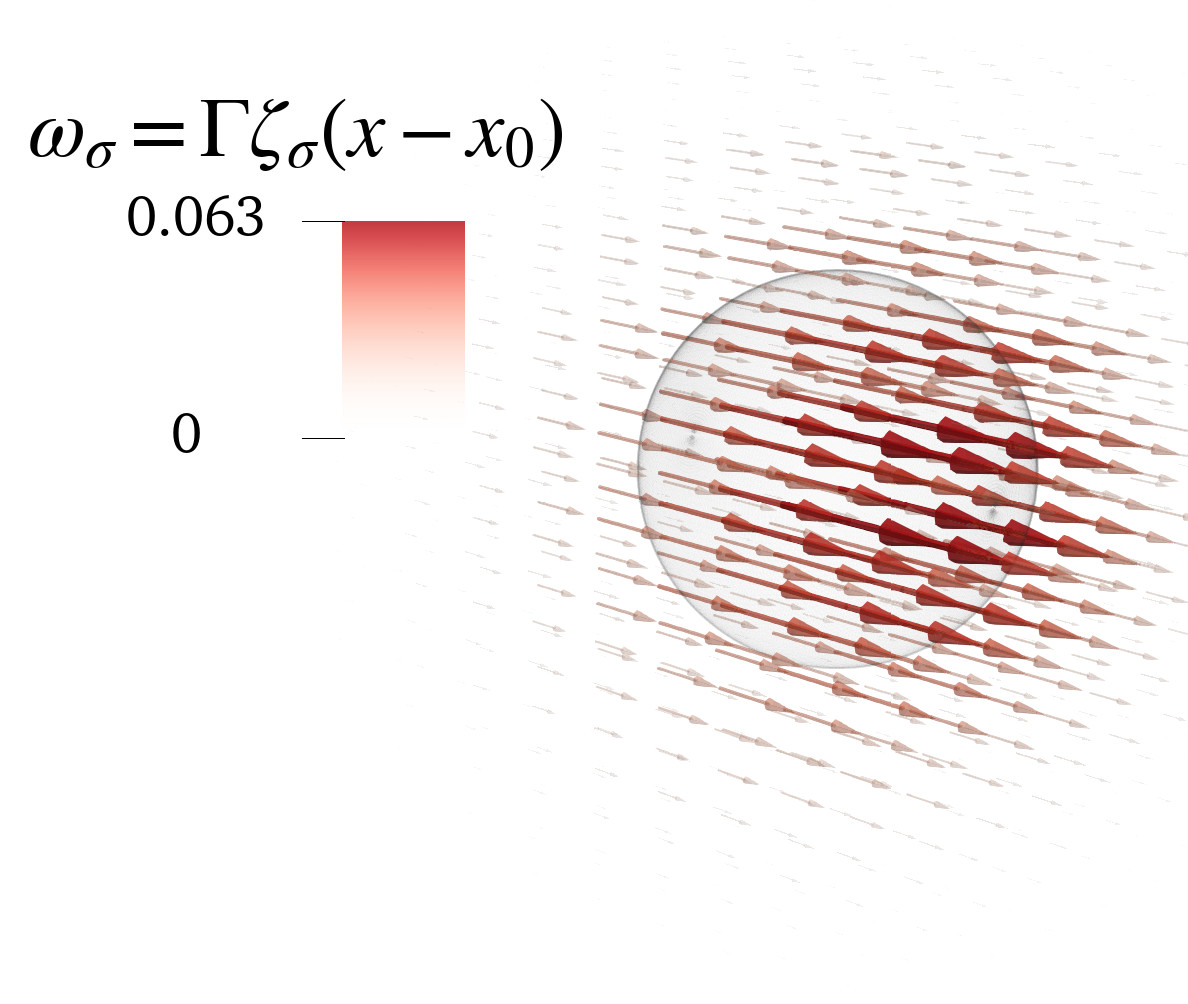}
            \end{subfigure}
            \hspace{2mm}
            \begin{subfigure}{0.35\textwidth} \centering
                \includegraphics[width=\linewidth]{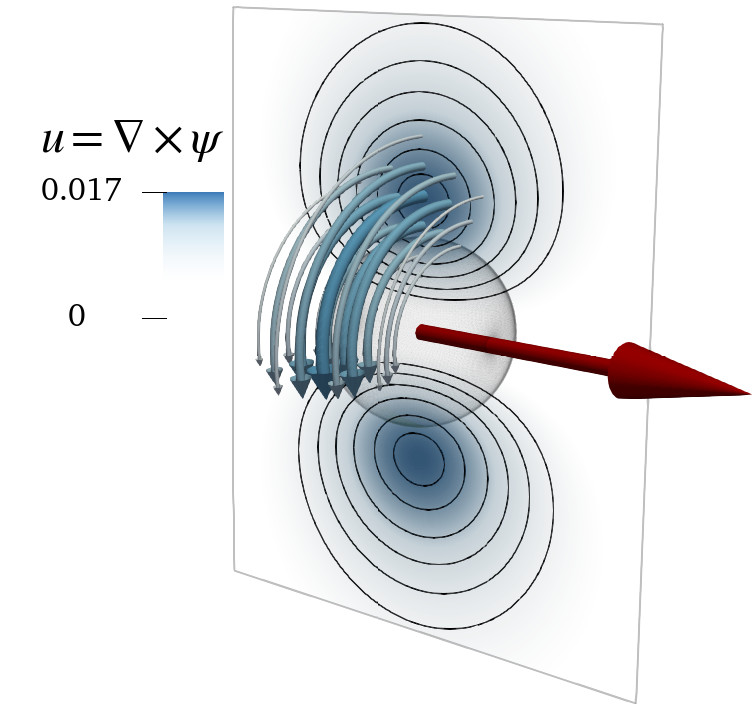}
            \end{subfigure}
            \caption{Vortex particle with unit size and strength: (left) core size $\sigma$ and vortex strength $\Gamma$, (middle) vorticity field, and (right) contours of velocity field and streamlines (represented as curved arrows).}
            \label{fig:vortexparticle}
        \end{figure*}

        The material derivative in~\cref{eq:NS3} and the material-conservative nature of the vorticity makes the $\boldsymbol{\omega}$ field especially well fit for a Lagrangian description.
        We now discretize the vorticity equation with Lagrangian elements, termed \textit{vortex particles}.
        Each particle represents a volume of fluid that is convected by the velocity field carrying an integral quantity of vorticity.

        The unfiltered $\boldsymbol{\omega}$ field is discretized with singular vortex particles of positions ${\textbf x}_p$ and coefficients $\boldsymbol{\Gamma}_p$, approximating $\boldsymbol{\omega}$ as
        \begin{align} \label{eq:particle:dirac}
            \boldsymbol{\omega}({\textbf x},t) \approx \sum
                \limits_p \boldsymbol{\Gamma}_p (t)
                    \delta  ({\textbf x} - {\textbf x}_p(t))
        ,\end{align}
        where $\delta$ is the Dirac delta.
        Each particle travels with the local velocity as
        \begin{align*}
            \frac{\text{d}}{\text{d}t}{\textbf x}_p = {\textbf u}({\textbf x}_p)
        ,\end{align*}
        where ${\textbf x}_p$ is the position of the $p$-th particle.
        Thus, each coefficient $\boldsymbol{\Gamma}_p$ (termed \textit{vortex strength}) represents the average vorticity that is carried in the volume of each particle since
        \begin{align*}
            \int\limits_{-\infty}^\infty
                \boldsymbol{\omega}({\textbf x},t)
            \,\mathrm{d}\mathbf{x}
            \approx \sum
                \limits_p \boldsymbol{\Gamma}_p(t)
        .\end{align*}

        Using the singular particle approximation in the filtered vorticity $\filter{ \boldsymbol{\omega} }$,
        \begin{align*}
                \filter{ \boldsymbol{\omega} } \left( \mathbf{x} \right)
            & =
                \int\limits_{-\infty}^\infty
                    \boldsymbol{\omega}\left( \mathbf{y} \right)
                    \zeta_\sigma(\mathbf{x}-\mathbf{y})
                \,\mathrm{d}\mathbf{y}
            \\ & \approx
                \int\limits_{-\infty}^\infty
                    \left(
                        \sum\limits_p
                            \boldsymbol{\Gamma}_p
                            \delta ({\textbf y} - {\textbf x}_p)
                    \right)
                    \zeta_\sigma(\mathbf{x}-\mathbf{y})
                \,\mathrm{d}\mathbf{y}
        ,\end{align*}
        we obtain an approximation of the filtered vorticity field,
        \begin{align} \label{eq:particle:blob}
                \filter{ \boldsymbol{\omega} } \left( \mathbf{x} \right)
            \approx
                \sum\limits_p
                    \boldsymbol{\Gamma}_p
                    \zeta_\sigma(\mathbf{x}-\mathbf{x}_p)
        .\end{align}

        As seen in Eq. (\ref{eq:particle:blob}), the filter operator has the effect of spreading the vortex strength $\boldsymbol{\Gamma}_p$ in space, regularizing the singularity originally introduced by the Dirac delta in Eq. (\ref{eq:particle:dirac}).
        Thus, the filter kernel takes the role of a basis function that is used to discretize and approximate the filtered vorticity field through particles.
        We let the filter width $\sigma$ change in time and space according to the evolution of each individual particle.
        Here on, the filter width is referred to as \textit{smoothing radius} or \textit{core size}, denoted $\sigma_p$, and defined as $\sigma_p(t) \equiv \sigma(\mathbf{x}_p, t)$.
        We approximate the filtered vorticity as $\filter{ \boldsymbol{\omega} } \approx \boldsymbol\omega_\sigma$, with
        \begin{align*}
                \boldsymbol\omega_\sigma\left( \mathbf{x},\,t \right)
            & \equiv
                \sum\limits_p
                    \boldsymbol{\Gamma}_p (t)
                    \zeta_{\sigma_p}(\mathbf{x}-\mathbf{x}_p(t))
        \end{align*}
        and
        \begin{align*}
                \zeta_{\sigma_p}(\mathbf{x}-\mathbf{x}_p(t))
            & =
                \frac{1}{\sigma_p^3 (t)} \zeta \left(\frac{\Vert \mathbf{x}-\mathbf{x}_p(t) \Vert}{\sigma_p(t)} \right)
        .\end{align*}

        Replacing the original filter $\zeta_\sigma$ with the variable-width filter $\zeta_{\sigma_p}$ introduces an error from commuting filter and differentiation operators in the LES equations.
        However, this error is second order in the filter width \cite{Ghosal1995}, hence it is assumed to be negligible or part of the discretization error of the VPM.
        Also, \cref{eq:particle:blob} introduces numerical issues by approximating $\filter{ \boldsymbol{\omega} }$ through a field that is in general not divergence-free.
        Cottet \cite{Cottet2000a} suggested that the divergence of $\filter{ \boldsymbol{\omega} }$ arises from unphysical small scales produced by the computation, which could be properly addressed with subfilter-scale diffusion.
        The divergence of $\filter{ \boldsymbol{\omega} }$ will be treated using the relaxation proposed by Pedrizzetti \cite{Pedrizzetti1992}, which, as shown in~\cref{sec:res}, will be sufficient to attain both numerical stability and physical accuracy with the LES formulation and subfilter-scale model developed in this study.

        The rotational part of the velocity field is calculated from the particles using the Helmholtz decomposition $\filter{\mathbf{u}} = \nabla \times \boldsymbol \psi$ where $\boldsymbol\psi$ is some vector potential.
        This is done by analytically solving the Poisson equation $\nabla^2 \boldsymbol\psi = -\filter{\boldsymbol\omega}$, which solution is a regularized Biot-Savart law.
        \cref{fig:vortexparticle} shows a vortex particle with unit size and strength, the spreading of the vortex strength by the filter kernel, and the resulting velocity field.

    \subsection{General VPM Governing Equations} \label{sec:fundamentals:GE}
        We will now use vortex particles to discretize the LES vorticity equation,~\cref{eq:filtered:angular:vector}, and derive the equations governing the evolution of the Lagrangian elements.
        For ease of notation, here on we denote the filtered velocity field $\filter{ \mathbf{u} }$ simply $\mathbf{u}$ and use $\frac{\mathrm{D}}{\mathrm{D} t}$ and $\frac{\mathrm{d}}{\mathrm{d} t}$ interchangeably.
        Also, time dependence is no longer explicitly indicated, but $\mathbf{x}_p$, $\boldsymbol\Gamma_p$, and $\sigma_p$ are time-dependent variables.

        Starting from the inviscid part of the LES-filtered vorticity equation,
        \begin{align*}
                \frac{\text{d} }{\text{d} t} \filter{ \boldsymbol\omega }
            & =
                \left( \filter{ \boldsymbol\omega } \cdot \nabla \right) \mathbf{u}
                - \mathbf{E}_\mathrm{adv} - \mathbf{E}_\mathrm{str}
        ,\end{align*}
        we write the filter operator explicitly, 
        \begin{align*}
            &
                \frac{\text{d} }{\text{d} t}
                \left(
                    \int\limits_{-\infty}^\infty
                        \boldsymbol{\omega}\left( \mathbf{y} \right)
                        \zeta_\sigma(\mathbf{x}-\mathbf{y})
                    \,\mathrm{d}\mathbf{y}
                \right)
            =
            \\
            & \qquad \qquad \qquad
                \left[
                    \left(
                        \int\limits_{-\infty}^\infty
                            \boldsymbol{\omega}\left( \mathbf{y} \right)
                            \zeta_\sigma(\mathbf{x}-\mathbf{y})
                        \,\mathrm{d}\mathbf{y}
                    \right)
                    \cdot \nabla 
                \right] \mathbf{u} \left( \mathbf{x} \right)
            \\
            & \qquad \qquad \qquad \qquad \qquad \qquad  \qquad
                - \mathbf{E}_\mathrm{adv} \left( \mathbf{x} \right) 
                - \mathbf{E}_\mathrm{str} \left( \mathbf{x} \right)
        .\end{align*}
        Using the singular particle approximation, 
        \begin{align*}
            \boldsymbol{\omega}({\textbf y}) \approx \sum
                \limits_q \boldsymbol{\Gamma}_q
                    \delta  ({\textbf y} - {\textbf x}_q)
        ,\end{align*}
        both integrals collapse resulting in
        \begin{subequations} \label{eq:filtered:angular:vector:particle1}
        \begin{align*}
            &
            \tag{\ref{eq:filtered:angular:vector:particle1}}
                \frac{\mathrm{d}}{\mathrm{d} t}
                \left(
                    \sum \limits_q
                        \boldsymbol{\Gamma}_q
                        \zeta_{\sigma_q} ({\mathbf x} - {\mathbf x}_q)
                \right)
            =
            \\
            &
                \left[
                    \left(
                        \sum
                        \limits_q \boldsymbol{\Gamma}_q
                        \zeta_{\sigma_q} ({\mathbf x} - {\mathbf x}_q)
                    \right)
                \cdot \nabla \right]{\mathbf u}({\mathbf x})
                - \mathbf{E}_\mathrm{adv}({\mathbf x}) - \mathbf{E}_\mathrm{str}({\mathbf x})
        .\end{align*}
        \end{subequations}
        With the help of~\cref{sec:app:der:dzetadt}, the left-hand side unfolds as
        \begin{align*}
            &
                \frac{\mathrm{d}}{\mathrm{d} t}
                \left(
                    \sum \limits_q
                        \boldsymbol{\Gamma}_q
                        \zeta_{\sigma_q} ({\mathbf x} - {\mathbf x}_q)
                \right)
            \\
            & =
                \sum\limits_q
                \left[
                    \frac{\mathrm{d} \boldsymbol{\Gamma}_q}{\mathrm{d} t}
                    \zeta_{\sigma_q} ({\mathbf x} - {\mathbf x}_q) +
                    \boldsymbol{\Gamma}_q \frac{\mathrm{d}}{\mathrm{d} t} \left(
                        \zeta_{\sigma_q} ({\mathbf x} - {\mathbf x}_q)
                    \right)
                \right]
            \\
            & =
                \sum\limits_q
                \left\{
                    \frac{\mathrm{d} \boldsymbol{\Gamma}_q}{\mathrm{d} t}
                    \zeta_{\sigma_q} ({\mathbf x} - {\mathbf x}_q) +
                    \boldsymbol{\Gamma}_q \frac{\partial \zeta_{\sigma_q} }{\partial t} ({\mathbf x} - {\mathbf x}_q)
                \right.
            \\
            & \qquad \qquad \qquad
                \left.
                +   \boldsymbol{\Gamma}_q \left[
                        \left(\mathbf{u} ({\mathbf x}) -\mathbf{u} ({\mathbf x}_q)\right) \cdot \nabla
                    \right] \zeta_{\sigma_q} ({\mathbf x} - {\mathbf x}_q)
                \right\}
        \end{align*}
        while vortex stretching is rearranged as
        \begin{align*}
            &
              \left[
                      \left(
                          \sum
                          \limits_q \boldsymbol{\Gamma}_q
                          \zeta_{\sigma_q} ({\mathbf x} - {\mathbf x}_q)
                      \right)
                  \cdot \nabla
              \right]{\mathbf u}({\mathbf x})
            =
            \\
            & \qquad \qquad \qquad \qquad \qquad
                \sum \limits_q
                    \zeta_{\sigma_q} ({\mathbf x} - {\mathbf x}_q)
                    \left(
                        \boldsymbol{\Gamma}_q
                        \cdot \nabla
                    \right)
                    {\mathbf u}({\mathbf x})
        .\end{align*}
        Evaluating at the position of the $p$-th particle, $\mathbf{x} = \mathbf{x}_p$, and pulling the $p$-th term out of each sum,
        \begin{align*}
            &
                \frac{\mathrm{d}}{\mathrm{d} t}
                \left(
                    \sum \limits_q
                        \boldsymbol{\Gamma}_q
                        \zeta_{\sigma_q} ({\mathbf x}_p - {\mathbf x}_q)
                \right)
            \\
            & =
                \frac{\mathrm{d} \boldsymbol{\Gamma}_p}{\mathrm{d} t}
                \zeta_{\sigma_p} ({\mathbf 0}) +
                \boldsymbol{\Gamma}_p \frac{\partial \zeta_{\sigma_p}}{\partial t} ({\mathbf 0})
            \\
            & \quad
                +
                \sum\limits_{q \neq p}
                \left\{
                    \frac{\mathrm{d} \boldsymbol{\Gamma}_q}{\mathrm{d} t}
                    \zeta_{\sigma_q} ({\mathbf x}_p - {\mathbf x}_q)
                    +
                    \boldsymbol{\Gamma}_q \frac{\partial \zeta_{\sigma_q} }{\partial t} ({\mathbf x}_p - {\mathbf x}_q)
            \right.
            \\
            & \qquad \qquad \qquad
            \left.
                    + \,
                    \boldsymbol{\Gamma}_q \left[
                        \left(\mathbf{u} ({\mathbf x}_p) -\mathbf{u} ({\mathbf x}_q)\right) \cdot \nabla
                    \right] \zeta_{\sigma_q} ({\mathbf x}_p - {\mathbf x}_q)
                \right\}
        \end{align*}
        and
        \begin{align*}
            &
                \left[
                        \left(
                            \sum
                            \limits_q \boldsymbol{\Gamma}_q
                            \zeta_{\sigma_q} ({\mathbf x}_p - {\mathbf x}_q)
                        \right)
                    \cdot \nabla
                \right]{\mathbf u}({\mathbf x}_p)
            \\
            & =
                \zeta_{\sigma_p} ({\mathbf 0})
                \left(
                    \boldsymbol{\Gamma}_p
                    \cdot \nabla
                \right)
                {\mathbf u}({\mathbf x}_p)
                +
                \sum \limits_{q \neq p}
                    \zeta_{\sigma_q} ({\mathbf x}_p - {\mathbf x}_q)
                    \left(
                        \boldsymbol{\Gamma}_q
                        \cdot \nabla
                    \right)
                    {\mathbf u}({\mathbf x}_p)
        .\end{align*}
        Thus,~\cref{eq:filtered:angular:vector:particle1} evaluated at $\mathbf{x} = \mathbf{x}_p$ becomes
        \begin{subequations} \label{eq:filtered:angular:vector:particle2}
        \begin{align*}
            \tag{\ref{eq:filtered:angular:vector:particle2}}
                \frac{\mathrm{d} \boldsymbol{\Gamma}_p}{\mathrm{d} t}
                \zeta_{\sigma_p} ({\mathbf 0})
            &
                +
                \boldsymbol{\Gamma}_p \frac{\partial \zeta_{\sigma_p}}{\partial t} ({\mathbf 0})
                +
                \mathbf{M}_p^0
            \\
            & =
                \zeta_{\sigma_p} ({\mathbf 0})
                \left(
                    \boldsymbol{\Gamma}_p \cdot \nabla
                \right)
                \mathbf{u} ({\mathbf x}_p)
            \\
            & \quad
            + (\mathbf{M}_p^1 + \mathbf{M}_p^2)
            - \left( \mathbf{E}_\mathrm{adv} ({\mathbf x}_p) + \mathbf{E}_\mathrm{str} ({\mathbf x}_p) \right)
        ,\end{align*}
        \end{subequations}
        where
        \begin{align}
            \label{eq:M0}
            \bullet \quad &
                    \mathbf{M}_p^0
                \equiv
                    \sum\limits_{q \neq p}
                    \left(
                        \frac{\mathrm{d} \boldsymbol{\Gamma}_q}{\mathrm{d} t}
                        \zeta_{\sigma_q} ({\mathbf x}_p - {\mathbf x}_q)
                        +
                        \boldsymbol{\Gamma}_q \frac{\partial \zeta_{\sigma_q} }{\partial t} ({\mathbf x}_p - {\mathbf x}_q)
                    \right)
            \\
            \label{eq:M1}
            \bullet \quad &
                    \mathbf{M}_p^1
                \equiv
                    - \sum\limits_{q \neq p}
                        \boldsymbol{\Gamma}_q
                        \left(\mathbf{u} ({\mathbf x}_p) -\mathbf{u} ({\mathbf x}_q)\right)
                        \cdot
                        \nabla \zeta_{\sigma_q} ({\mathbf x}_p - {\mathbf x}_q)
            \\
            \label{eq:M2}
            \bullet \quad &
                    \mathbf{M}_p^2
                \equiv
                    \sum \limits_{q \neq p}
                        \zeta_{\sigma_q} ({\mathbf x}_p - {\mathbf x}_q)
                        \left(
                            \boldsymbol{\Gamma}_q
                            \cdot \nabla
                        \right)
                        {\mathbf u}({\mathbf x}_p)
        \end{align}
        These $\mathbf{M}$-terms pose an interdependence between the $p$-th particle and neighboring particles, which arises from filtering the vorticity equation.
        In this study we will neglect this interdependence as explained in~\cref{sec:classicvpm}.
        However, we will see in~\cref{sec:sfs} that $\mathbf{M}_p^1$ and $\mathbf{M}_p^2$ are closely related to advection and vortex stretching at the subfilter scale.

        Recalling that $\zeta_\sigma$ is defined as $\zeta_\sigma (\mathbf{x}) = \frac{1}{\sigma^3} \zeta (\frac{\Vert \mathbf{x} \Vert}{\sigma})$, its time derivative is calculated as
        \begin{align*}
                \frac{\partial \zeta_{\sigma}}{\partial t} ({\mathbf x})
            & =
                - 3 \frac{1}{\sigma^4}\frac{\partial \sigma}{\partial t}
                \zeta \left( \frac{\Vert \mathbf{x} \Vert}{\sigma} \right)
                +
                \frac{1}{\sigma^3} \frac{\partial }{\partial t}
                \left(
                    \zeta \left( \frac{\Vert \mathbf{x} \Vert}{\sigma} \right)
                \right)
            \\ & =
                - 3 \frac{1}{\sigma}
                \frac{\partial \sigma}{\partial t} \zeta_\sigma (\mathbf{x})
                - \frac{1}{\sigma^3}
                \frac{\partial \zeta}{\partial r} \left( \frac{\Vert \mathbf{x} \Vert}{\sigma} \right)
                \frac{\Vert \mathbf{x} \Vert}{\sigma^2}
                \frac{\partial \sigma}{\partial t}
        .\end{align*}
        Assuming that $\zeta(r)$ reaches a maximum at $r=0$ (\textit{i.e.}, $\frac{\partial \zeta}{\partial r}(0) = 0$) and evaluating at $\mathbf{x} = \mathbf{0}$, we get
        \begin{align} \label{eq:dzetadt0}
                \frac{\partial \zeta_{\sigma}}{\partial t} ({\mathbf 0})
            & =
                - 3 \frac{1}{\sigma}
                \frac{\partial \sigma}{\partial t} \zeta_\sigma (\mathbf{0})
        .\end{align}
        Since viscous effects have been set aside through operator splitting, $\frac{\partial \sigma}{\partial t}$ in this derivation only accounts for inviscid effects.
        For clarity, this means that core spreading due to viscous diffusion must not be included in~\cref{eq:dzetadt0}.

        Finally, substituting~\cref{eq:dzetadt0} into~\cref{eq:filtered:angular:vector:particle2} and assuming $\zeta_{\sigma_p} ({\mathbf 0}) \neq 0$, we arrive to the equation governing the evolution of vortex strength,
        \begin{subequations} \label{eq:GE:Gamma:general}
        \begin{align*}
            \tag{\ref{eq:GE:Gamma:general}}
                \frac{\mathrm{d} \boldsymbol{\Gamma}_p}{\mathrm{d} t}
            =
                \left(
                    \boldsymbol{\Gamma}_p \cdot \nabla
                \right)
                \mathbf{u} ({\mathbf x}_p)
            & +
                3 \boldsymbol{\Gamma}_p \frac{1}{\sigma_p}
                \frac{\partial \sigma_p}{\partial t}
            \\
            & +
                \frac{1}{\zeta_{\sigma_p} ({\mathbf 0})}
                (-\mathbf{M}_p^0 + \mathbf{M}_p^1 + \mathbf{M}_p^2)
            \\
            & -
                \frac{1}{\zeta_{\sigma_p} ({\mathbf 0})}
                \left( \mathbf{E}_\mathrm{adv} ({\mathbf x}_p) + \mathbf{E}_\mathrm{str} ({\mathbf x}_p) \right)
        .\end{align*}
        \end{subequations}
        Thus, particle convection as in
        \begin{align} \label{eq:GE:x}
            \frac{\text{d}}{\text{d}t}{\mathbf x}_p = {\mathbf u}({\mathbf x}_p)
        ,\end{align}
        strength evolution as in \cref{eq:GE:Gamma:general}, some expression for $\frac{\partial \sigma_p}{\partial t}$, and the viscous diffusion equation make up the general governing equations of the vortex particle method that solve the LES-filtered Navier-Stokes vorticity equation.

        We see in~\cref{eq:GE:Gamma:general} that the evolution of the vortex strength is dictated by vortex stretching (first term), inviscid expansion/contraction of the particle size $\sigma_p$ (second term), a dependence on other particles through the $\mathbf{M}$ terms, and SFS contributions through the $\mathbf{E}$ terms.
        One obtains different formulations of the VPM depending on how $\frac{\partial \sigma}{\partial t}$, the $\mathbf{M}$ terms, and the SFS $\mathbf{E}$ terms are handled.
        In particular, we will see in~\cref{sec:classicvpm} that the classic VPM is equivalent to assuming\footnote{The only instance that the classic VPM uses $\frac{\partial \sigma}{\partial t} \neq 0$ is in solving the viscous diffusion equation through the core spreading scheme, but even in this case the effects of $\frac{\partial \sigma}{\partial t}$ on vortex strength are neglected.} $\frac{\partial \sigma}{\partial t} = 0$ while neglecting all $\mathbf{M}$ terms.
        In~\cref{sec:reformulatedvpm} we propose a new formulation that uses $\frac{\partial \sigma}{\partial t} \neq 0$ to reinforce conservations laws in spherical elements.
        In~\cref{sec:sfs} we discuss existing SFS models for meshless schemes and develop a new anisotropic dynamic model of SFS vortex stretching, $\mathbf{E}_\mathrm{str}$.

    \subsection{Numerical Schemes}

        In \href{https://github.com/byuflowlab/FLOWVPM.jl}{FLOWVPM}---the solver used for this study---vortex stretching is solved in the transposed scheme \cite{Winckelmans1989,Winckelmans1993} and the divergence of the vorticity field is treated through the relaxation scheme developed by Pedrizzeti \cite{Pedrizzetti1992}.
        The time integration of the governing equations is done through a low-storage third-order Runge-Kutta scheme \cite{Williamson1980}.
        A Gaussian kernel is used as the LES filter $\zeta_\sigma$ (or VPM radial basis function).
        The fast multipole method \cite{Greengard1987,Cheng1999} (FMM) is used for the computation of the regularized Biot-Savart law, approximating the velocity field and vortex stretching through spherical harmonics with computational complexity $\mathcal{O}(N)$, where $N$ is the number of particles.
        The FMM computation of vortex stretching is performed through an efficient complex-step derivative approximation \cite{Alvarez2020}, implemented in a modified version of the open-source, parallelized code ExaFMM\cite{Yokota2011,Wang2021}.
        \href{https://github.com/byuflowlab/FLOWVPM.jl}{FLOWVPM} is implemented in the Julia language \cite{Bezanson2017}, which is a modern, high-level, dynamic programming language for high-performance computing.

\section{Classic Vortex Particle Method} \label{sec:classicvpm}
    Before further exploring the general equation governing vortex stretching,~\cref{eq:GE:Gamma:general}, we pause to analyze the equation that has been used extensively in the literature throughout the years:
    \begin{align} \label{eq:GE:Gamma:classic}
            \frac{\mathrm{d} \boldsymbol{\Gamma}_p}{\mathrm{d} t}
        =
            \left(
                \boldsymbol{\Gamma}_p \cdot \nabla
            \right)
            \mathbf{u} (\mathbf{x}_p)
    .\end{align}
    We denote this equation as the \textit{Classic VPM}.

    Multiple variations of~\cref{eq:GE:Gamma:classic} have been used over the years.
    For instance, Gharakhani \cite{Gharakhani1997} introduced a new term accounting for the divergence of the approximated vorticity field, Winckelmans and Leonard \cite{Winckelmans1993} replaced the differential operator in vortex stretching with its transpose, and Mansfield et al. \cite{Mansfield1998,Mansfield1999} and Cottet \cite{Cottet1996} reintroduced the SFS contributions.
    However, all these variations and the classic equation have in common that they all neglect $\frac{\partial \sigma_p}{\partial t}$ and are free of the interdependence posed by the $\mathbf{M}$-terms in~\cref{eq:GE:Gamma:general}.

    We now delineate the assumption underlying the classic VPM that will justify neglecting the $\mathbf{M}$-terms, termed \textit{localized-vorticity assumption}, as follows.

    Given a vorticity field $\boldsymbol\omega$ that is compact in a small volume $\mathrm{Vol}$, the average vorticity $\ave{\boldsymbol\omega}$ is calculated as
    \begin{align*}
        \ave{\boldsymbol\omega} =
            \frac{\int\limits_\mathrm{Vol}\boldsymbol\omega(\mathbf{y}) \,\mathrm{d}\mathbf{y}}{\mathrm{Vol}}
    .\end{align*}
    The field can be approximated through a radially-symmetric field $\tilde{\boldsymbol\omega}$ defined as
    \begin{align*}
            \tilde{\boldsymbol\omega}(\mathbf{x})
        \equiv
            \ave{\boldsymbol{\omega}}\mathrm{Vol}
            \zeta_\sigma(\mathbf{x}-\mathbf{x}_0)
    ,\end{align*}
    where $\zeta_\sigma$ is a radial basis function of spread $\sigma$ and center $\mathbf{x}_0$ approximating the vorticity distribution of the original field $\boldsymbol\omega$.
    If $\zeta_\sigma$ is normalized such as to have a volume integral of unity, $\tilde{\boldsymbol\omega}$
    approximates $\boldsymbol\omega$ in an average sense since
    \begin{align*}
            \int\limits_{-\infty}^\infty\tilde{\boldsymbol\omega}(\mathbf{y}) \,\mathrm{d}\mathbf{y}
        & =
            \int\limits_\mathrm{Vol}\boldsymbol\omega(\mathbf{y}) \,\mathrm{d}\mathbf{y}
    .\end{align*}

    Defining $\boldsymbol\Gamma = \ave{\boldsymbol{\omega}}\mathrm{Vol}$, the localized-vorticity field $\boldsymbol\omega$ can be approximated with a single particle as
    \begin{align*}
            \boldsymbol\omega \left( \mathbf{x} \right)
        \approx
            \boldsymbol\Gamma
            \zeta_{\sigma}(\mathbf{x}-\mathbf{x}_0)
    .\end{align*}
    Replacing this in the inviscid part of the vorticity equation,
    \begin{align*}
        \frac{\text{d} }{\text{d} t} \boldsymbol{\omega} = (\boldsymbol{\omega} \cdot \nabla ){\mathbf u}
    ,\end{align*}
    evaluating at $\mathbf{x}=\mathbf{x}_0$, and following similar steps as in~\cref{sec:fundamentals:GE}, we get
    \begin{align} \label{eq:GE:localized}
            \frac{\mathrm{d} \boldsymbol{\Gamma}}{\mathrm{d} t}
        =
            \left(
                \boldsymbol{\Gamma}  \cdot \nabla
            \right)
            {\mathbf u} (\mathbf{x}_0)
            +
            3 \boldsymbol{\Gamma} \frac{1}{\sigma} \frac{\partial \sigma}{\partial t}
    .\end{align}
    This is the evolution equation of the vortex strength approximating the field of localized vorticity, discretized with only one particle.
    In contrast, in~\cref{sec:fundamentals:GE} we showed that a more general vorticity field that is discretized using multiple particles leads to a governing equation that is slightly different:
    \begin{subequations} \label{eq:nonlocalizednoSFS}
    \begin{align*}
        \tag{\ref{eq:nonlocalizednoSFS}}
            \frac{\mathrm{d} \boldsymbol{\Gamma}_p}{\mathrm{d} t}
        =
            \left(
                \boldsymbol{\Gamma}_p \cdot \nabla
            \right)
            \mathbf{u} (\mathbf{x}_p)
        & +
            3 \boldsymbol{\Gamma}_p \frac{1}{\sigma_p}
            \frac{\partial \sigma_p}{\partial t}
        \\
        & +
            \frac{1}{\zeta_{\sigma_p} ({\mathbf 0})}
            (-\mathbf{M}_p^0 + \mathbf{M}_p^1 + \mathbf{M}_p^2)
    ,\end{align*}
    \end{subequations}
    after ignoring the SFS contributions.
    Here, the $\mathbf{M}$-terms, as defined in~\cref{eq:M0,eq:M1,eq:M2}, pose a dependence on neighboring particles that arises from having filtered the vorticity equation.

    Discarding these $\mathbf{M}$-terms is equivalent to assuming that the vorticity field can be approximated by the superposition of blobs of fluid with compact vorticity that evolve somewhat independently from each other.\footnote{The interactions neglected here are only the ones pertaining to the evolution of vortex strength, while other interactions are still accounted for in the convection and viscous diffusion of the particles.}
    This assumption, which we call \textit{localized-vorticity assumption}, reduces~\cref{eq:nonlocalizednoSFS} into~\cref{eq:GE:localized} and is valid as long as there is no significant particle overlap.
    In both derivations of the classic VPM by Winckelmans and Leonard \cite{Winckelmans1993} and Cottet and Koumutsakos \cite{Koumoutsakos2001}, these $\mathbf{M}$-terms are not present due to the construction of the method:
    In the classic derivation, the unfiltered vorticity equation (\cref{eq:NS3}) is discretized with singular particles and only the velocity field is filtered to obtain a regularized field, while in our derivation we discretized the LES-filtered vorticity equation (\cref{eq:filtered:angular:vector}) which leads to~\cref{eq:nonlocalizednoSFS}.
    In order to bring both approaches into agreement, in~\cref{sec:reformulatedvpm} we will use the localized-vorticity assumption in the derivation of our reformulated VPM to neglect these $\mathbf{M}$-terms.

    Interestingly, discarding the $\mathbf{M}$-terms resembles the LES decomposition approach of truncated basis functions \cite{Meneveau2000}.
    In such an approach, the flow field is expanded using orthonormal basis functions.
    The summation of bases is then truncated to define the large-scale field, and the discarded modes represent the range of subfilter scales.
    In the localized-vorticity assumption, the sum over all the particles is truncated after the leading term, $p$, while neglecting the contributions of neighboring particles.
    Hence, the localized-vorticity assumption can be regarded as a secondary LES filter with the neglected $\mathbf{M}$-terms becoming part of the subfilter-scale contributions.

    While it is justifiable to neglect the $\mathbf{M}$-terms through the localized-vorticity assumption, it is unclear to us what the basis is for the classic VPM to assume $\frac{\partial \sigma_p}{\partial t}=0$.
    In fact, we hypothesize that this last assumption is the cause of the numerical instabilities that pervade the classic VPM.

    The classic VPM simply regards $\sigma$ as a numerical parameter with no physical significance.
    However, Leonard \cite{Leonard1980} suggested that $\sigma$ should change according to conservation of mass, and Nakanishi, Ojima, and Maremoto \cite{NAKANISHI1993,OJIMA2000} suggested that $\sigma$ should change according to Kelvin's theorem, which was more recently implemented by Kornev \textit{et al.} \cite{Kornev2020}
    Even though these authors let $\sigma$ evolve in time, they did not include these effects back in the equation that governs $\boldsymbol\Gamma$, effectively assuming $\frac{\partial \sigma_p}{\partial t} = 0$ in the evolution of $\boldsymbol\Gamma$.
    In the following chapter we propose a formulation that uses $\frac{\partial \sigma_p}{\partial t} \neq 0$ in the governing equation of $\boldsymbol\Gamma$, while letting $\sigma$ change as to reinforce conservation of both mass and angular momentum, which will be shown to lead to remarkable numerical stability.

    \section{Reformulated Vortex Particle Method} \label{sec:reformulatedvpm}

    The general governing equations derived in~\cref{sec:fundamentals:GE} call for an expression for $\frac{\partial \sigma_p}{\partial t}$.
    While the classic VPM simply assumes $\frac{\partial \sigma_p}{\partial t} = 0$, we will now go back to first principles to find some plausible expressions for $\frac{\partial \sigma_p}{\partial t}$.
    As a preamble, in~\cref{sec:reformulatedvpm:physicalimplications} we put forth two physical implications that distill from the vorticity equation, which will then guide our search for possible candidate formulations of $\frac{\partial \sigma_p}{\partial t}$ in~\cref{sec:reformulatedvpm:candidateformulations}.
    The process for constructing candidate formulation is then generalized in~\cref{sec:reformulatedvpm:generalized}, and the entire space of possible formulation is analyzed in~\cref{sec:reformulatedvpm:analysis}.
    This will then lead us to one formulation, termed \textit{reformulated VPM} and summarized in~\cref{sec:reformulatedvpm:GE}, which reinforces conservation of both mass and angular momentum by reshaping the vortex elements as they are subject to vortex stretching.

	\subsection{Physical Implications of the Vorticity Equation} \label{sec:reformulatedvpm:physicalimplications}

    In~\cref{sec:fundamentals:ns} we showed that the linear-momentum Navier-Stokes equation can be transformed into an expression that only depends on vorticity, namely~\cref{eq:NS3}, here repeated:
    \begin{align*} \tag{\ref{eq:NS3}}
            \frac{\text{d} }{\text{d} t} \boldsymbol{\omega}
        =
            (\boldsymbol{\omega} \cdot \nabla )\mathbf{u}
            +
            \nu\nabla^2\boldsymbol{\omega}
    .\end{align*}
    We now point our attention to two laws that follow from~\cref{eq:NS3}: conservation of mass and angular momentum.
    The discussion that follows is inspired by the writing of P. A. Davidson \cite{Davidson2001,Davidson2019}.

    \subsubsection{Conservation of Angular Momentum} \label{sec:reformulatedvpm:physicalimplications:momentum}

        Consider a spherical differential fluid element carrying a mean vorticity $\boldsymbol\omega$ with moment of inertia $I$.
        Due to $\boldsymbol\omega$, the element is then rotating at an angular velocity of $\boldsymbol\omega/2$ and its angular momentum $\mathbf{L}$ is calculated as \cite[p.~82]{Batchelor1967}
        \begin{align} \label{eq:L}
            \mathbf{L} = \frac{I \boldsymbol\omega}{2}
        .\end{align}
        Given that the element is spherical (and before strain distorts the element), the pressure field exerts no torque on the element and the only torque-producing forces are due to viscous effects, namely $\boldsymbol\tau_\mathrm{viscous}$.
        The change of angular momentum is then calculated as
        \begin{align} \label{eq:dLdt}
            \frac{\mathrm{d} }{\mathrm{d} t} \mathbf{L} = \boldsymbol\tau_\mathrm{viscous}
        .\end{align}
        Replacing~\cref{eq:L} in~\cref{eq:dLdt}, we arrive to the following expression of the total derivative of vorticity\footnote{More generally, this equation should be expressed with an inertia tensor, but for a spherical body this tensor reduces to a scalar due to the symmetry of the body.},
        \begin{align} \label{eq:domegadt}
                \frac{\mathrm{d} }{\mathrm{d}t} \boldsymbol\omega
            =
                - \frac{\boldsymbol\omega}{I}  \frac{\mathrm{d} }{\mathrm{d} t} I
                +
                \boldsymbol\tau_\mathrm{viscous}^*
        ,\end{align}
        where $\boldsymbol\tau_\mathrm{viscous}^*= 2\boldsymbol\tau_\mathrm{viscous}/I$.
        When viscous effects are ignored,~\cref{eq:domegadt} implies that, in order to conserve angular momentum, the angular velocity (or vorticity) must decrease whenever the moment of inertia increases, and vice versa.

\begin{figure}[t]
            \centering
            \includegraphics[width=\smallfigwidth]{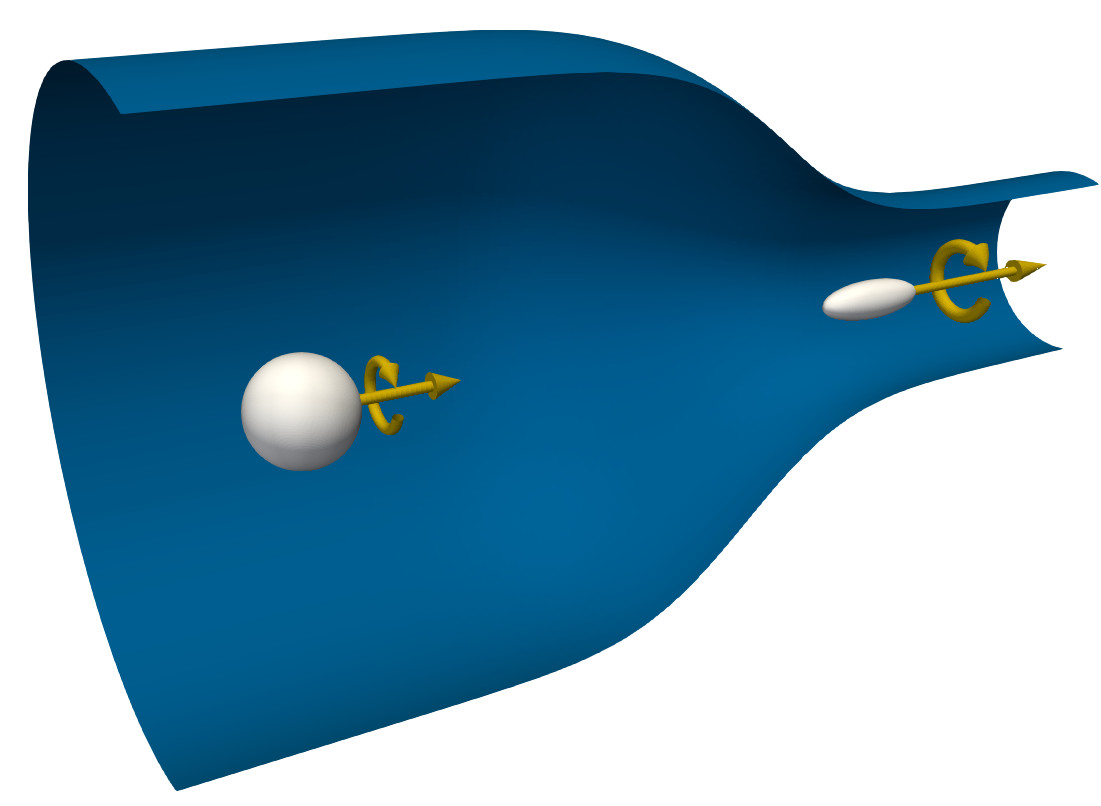}
            \caption{Stretching of a spherical fluid element and increase in vorticity (or angular velocity).}
            \label{fig:tubevortexstretching}
        \end{figure}

        \cref{eq:NS3,eq:domegadt} suggest that the vorticity Navier-Stokes equation is simply an expression of the conservation of angular momentum in a spherical fluid element\cite{Chatwin1973}.
        Furthermore, the first term in the right hand side of~\cref{eq:NS3} accounts for the change of moment of inertia as
        \begin{align} \label{eq:inertia}
                (\boldsymbol\omega \cdot \nabla )\mathbf{u}
            =
                -\frac{\boldsymbol\omega}{I}\frac{\mathrm{d} }{\mathrm{d} t} I
        ,\end{align}
        which leads to an increase/decrease of vorticity to conserve momentum as the moment of inertia decreases/increases.
        Hence, $(\boldsymbol{\omega} \cdot \nabla )\mathbf{u}$ is referred to as vortex stretching as it accounts for the deformation exerted by the velocity field on the fluid element, intensifying the vorticity in the direction that the element is stretched, as illustrated in~\cref{fig:tubevortexstretching}.

    \subsubsection{Conservation of Mass}

\begin{figure}[t]
            \centering
            \includegraphics[width=\figwidth]{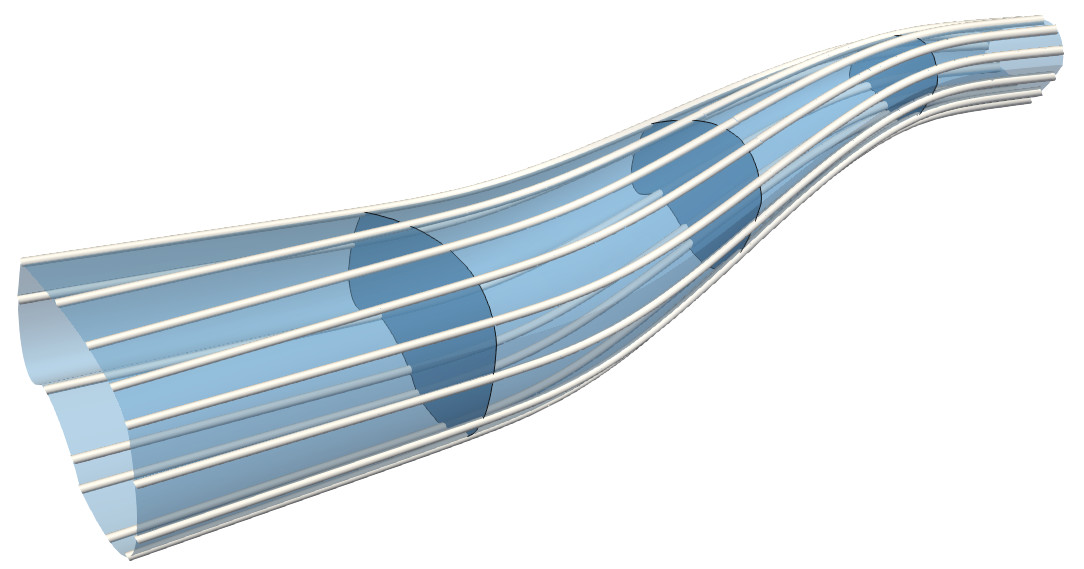}
            \caption{Vortex tube formed by the vortex lines passing through a closed curve.}
            \label{fig:tubevortexlines}
        \end{figure}

        Another physical implication of~\cref{eq:NS3} is the existence of vortex lines and vortex tubes, described as follows.
        We define a short material line $\boldsymbol{\ell}$ that moves with the fluid, with start and end points $\mathbf{x}$ and $\mathbf{x} + \boldsymbol\ell$, respectively.
        If $\boldsymbol{\ell}$ is infinitesimally small, the evolution of $\boldsymbol{\ell}$ is given by
        \begin{align} \label{eq:vortexline}
                \frac{\mathrm{d} }{\mathrm{d} t} \boldsymbol\ell
            =
                \mathbf{u}\left( \mathbf{x} + \boldsymbol\ell \right)
                -
                \mathbf{u}\left( \mathbf{x} \right)
            =
                \left( \boldsymbol\ell \cdot \nabla \right) \mathbf{u}\left( \mathbf{x} \right)
        .\end{align}
        Defining $\boldsymbol{\ell}$ as being tangent to the local vorticity, and comparing~\cref{eq:vortexline} to the inviscid part of~\cref{eq:NS3}, we see that vorticity evolves as the material line, being identically stretched and reoriented by the velocity field.
        This led Helmholtz to conclude the existence of lines of vorticity, termed \textit{vortex lines}, that move with the fluid.
        We then define a \textit{vortex tube} as the surface formed by all the vortex lines passing through a closed curve, depicted in~\cref{fig:tubevortexlines}.

        The fact that vortex lines move with the fluid implies that the volume of a vortex tube segment must be conserved in incompressible flow.
        For instance, let us discretize and approximate a vortex tube through Lagrangian cylindrical elements of length $\ell$ and cross section $r$.
        As the length of the element is stretched, the cross section $r$ must shrink to conserve the same volume $\mathrm{Vol} = \pi r^2 \ell$. This is,
        \begin{align*}
            \frac{\mathrm{d} }{\mathrm{d} t} \mathrm{Vol} = 0
        ,\end{align*}
        leading to
        \begin{align} \label{eq:tubestretching}
            \frac{\mathrm{d} }{\mathrm{d} t} r   = -\frac{r}{2 \ell} \frac{\mathrm{d} }{\mathrm{d} t} \ell
        .\end{align}

    \subsubsection{Implications for the VPM} \label{sec:reformulatedvpm:physicalimplications:implications}
        The aforementioned laws of conservation point our attention to two major pitfalls of the classic VPM, if particles are considered as material elements.
        First, \cref{eq:inertia} establishes that vortex stretching must lead to a change in the shape of the fluid element.
        Otherwise, if the vorticity (angular velocity) has changed due to vortex stretching while the shape of the element (moment of inertia) is kept constant, the conservation of angular momentum has been locally violated.
        This is exactly the case when the classic VPM assumes a constant core size, $\frac{\mathrm{d}}{\mathrm{d} t}\sigma=0$.

        Second, in an incompressible tube of vorticity,~\cref{eq:vortexline,eq:NS3} establish that an increase in vorticity due to vortex stretching is associated to a lengthening of the tube, which must cause the cross section of the tube to shrink according to~\cref{eq:tubestretching}.
        Otherwise, the tube has kept a constant cross section while being stretched, thus locally violating conservation of mass.
        Again, this is exactly the case when the classic VPM assumes $\frac{\mathrm{d}}{\mathrm{d} t}\sigma=0$.

        In summary, we believe that the classic VPM violates conservation of both angular momentum and mass when it assumes $\frac{\mathrm{d}}{\mathrm{d} t}\sigma=0$.
        We also hypothesize that this is the reason for its poor numerical stability.
        In the following sections we will consider some possible formulations that use $\frac{\mathrm{d}}{\mathrm{d} t}\sigma$ as a degree of freedom to locally reinforce these conservation laws.

\subsection{Candidate Formulations} \label{sec:reformulatedvpm:candidateformulations}

    Starting from the general equation governing vortex strength (\cref{eq:GE:Gamma:general}), neglecting SFS terms, and using the localized-vorticity assumption, we obtain
    \begin{align} \label{eq:GE:Gamma:localized}
            \frac{\mathrm{d} }{\mathrm{d} t} \boldsymbol{\Gamma}_p
        =
            \left(
                \boldsymbol{\Gamma}_p \cdot \nabla
            \right)
            \mathbf{u} (\mathbf{x}_p)
            +
            3 \boldsymbol{\Gamma}_p \frac{1}{\sigma_p}
            \frac{\partial \sigma_p}{\partial t}
    .\end{align}
    We will now explore multiple combinations of element shapes and conservation laws to derive possible expressions for $\frac{\partial \sigma_p}{\partial t}$.

    \subsubsection{Sphere with Conservation of Momentum} \label{sec:reformulatedvpm:sphereformulation}

        Given that the moment of inertia of a solid sphere is $I=\frac{2}{5}mr^2$, where $m=\frac{4}{3}\rho\pi r^3$ is the mass of the sphere and $r$ its radius, the inertial term in~\cref{eq:inertia} becomes
        \begin{align*}
                \frac{1}{I}\frac{\mathrm{d} }{\mathrm{d} t} I
                =
                    \frac{1}{r^5}
                    \frac{\mathrm{d} }{\mathrm{d} t} \left( r^5 \right)
                =
                    \frac{5}{r}
                    \frac{\mathrm{d} }{\mathrm{d} t} r
        .\end{align*}
        Replacing this into~\cref{eq:inertia} and taking the dot product with $\boldsymbol\omega$,
        \begin{align*}
                - \left( \boldsymbol\omega \cdot \boldsymbol\omega \right)
                \frac{5}{r} \frac{\mathrm{d} }{\mathrm{d} t} r
            =
                \left[
                    (\boldsymbol\omega \cdot \nabla )\mathbf{u}
                \right] \cdot \boldsymbol\omega
        ,\end{align*}
        results in
        \begin{align*}
                \frac{1}{r} \frac{\mathrm{d} }{\mathrm{d} t} r
            & =
                - \frac{1}{5} \frac{1}{ \Vert \boldsymbol\omega \Vert^2}
                \left[
                    (\boldsymbol\omega \cdot \nabla )\mathbf{u}
                \right] \cdot \boldsymbol\omega
            \\
            & =
                - \frac{1}{5}
                \left[
                    (\hat{\boldsymbol\omega} \cdot \nabla )\mathbf{u}
                \right] \cdot \hat{\boldsymbol\omega}
        .\end{align*}

        When $\boldsymbol\omega$ is filtered and discretized with vortex particles, and the localized-vorticity assumption is applied, the spherical element becomes the vortex particle itself.
        Vortex strengths are roughly aligned with the vorticity field\footnote{This is especially true when using Pedrizzetti's relaxation scheme.\cite{Pedrizzetti1992}}, meaning $\hat{\boldsymbol\Gamma}_p \approx \hat{\boldsymbol\omega} \left( \mathbf{x}_p \right)$, and the size of the particle can be expressed as $\sigma_p=\alpha r$ with $\alpha$ some scaling factor, obtaining
        \begin{align*}
                \frac{1}{\sigma_p} \frac{\mathrm{d} }{\mathrm{d} t} \sigma_p
            & =
                - \frac{1}{5}
                \left[
                    (\hat{\boldsymbol\Gamma}_p \cdot \nabla )\mathbf{u} \left( \mathbf{x}_p \right)
                \right] \cdot \hat{\boldsymbol\Gamma}_p
        ,\end{align*}
        or, equivalently
        \begin{align} \label{eq:sphere:dsigmadt}
                \frac{\mathrm{d} }{\mathrm{d} t} \sigma_p
            & =
                - \frac{1}{5} \frac{\sigma_p}{\Vert \boldsymbol\Gamma_p \Vert}
                \left[
                    (\boldsymbol\Gamma_p \cdot \nabla )\mathbf{u} \left( \mathbf{x}_p \right)
                \right] \cdot \hat{\boldsymbol\Gamma}_p
        .\end{align}

        Noticing that, in a Lagrangian scheme, $\sigma_p$ is only a function of time (\textit{i.e.}, $\frac{\mathrm{d} }{\mathrm{d} t} \sigma_p = \frac{\partial \sigma_p}{\partial t}$), we replace~\cref{eq:sphere:dsigmadt} into~\cref{eq:GE:Gamma:localized} to obtain the equation governing $\boldsymbol\Gamma_p$:
        \begin{align} \label{eq:sphere:dGammadt}
                \frac{\mathrm{d} }{\mathrm{d} t} \boldsymbol{\Gamma}_p
            =
                \left(
                    \boldsymbol{\Gamma}_p \cdot \nabla
                \right)
                \mathbf{u} (\mathbf{x}_p)
                -
                \frac{3}{5}
                \left\{
                    \left[
                        (\boldsymbol\Gamma_p \cdot \nabla )\mathbf{u} \left( \mathbf{x}_p \right)
                    \right] \cdot \hat{\boldsymbol\Gamma}_p
                \right\} \hat{\boldsymbol\Gamma}_p
        .\end{align}
        Thus, ~\cref{eq:sphere:dsigmadt,eq:sphere:dGammadt} are the governing equations required for the particles to preserve angular momentum.

    \subsubsection{Tube with Conservation of Mass} \label{sec:reformulatedvpm:tubeformulation}

        Assuming that the field $\boldsymbol\omega$ encompasses a tube of vorticity of radius $r$, we now use conservation of mass to derive an expression for $\frac{\mathrm{d}}{\mathrm{d} t}\sigma$.
        When $\boldsymbol\omega$ is discretized through cylindrical vortex elements of length $\ell$ and radius $\sigma_t=\alpha_t r$, with $\alpha_t$ some scaling factor, \cref{eq:tubestretching} becomes
        \begin{align} \label{eq:tube:dsgmdt:VTM}
            \frac{1}{\sigma_t}\frac{\mathrm{d} }{\mathrm{d} t} \sigma_t   = -\frac{1}{2 \ell} \frac{\mathrm{d} }{\mathrm{d} t} \ell
        .\end{align}
        The length $\ell$ of the cylindrical element relates to the evolution of its vorticity $\boldsymbol{\omega_t}$ through vortex stretching as
        \begin{align*}
                \frac{1}{\ell}\frac{\mathrm{d} }{\mathrm{d}t} \ell
            =
                \frac{1}{\Vert \boldsymbol{\omega_t} \Vert}
                    \frac{\mathrm{d} \boldsymbol{\omega_t}}{\mathrm{d}t}  \cdot \hat{\boldsymbol{\omega_t}}
        .\end{align*}
        Thus, Ojima and Maremoto \cite{OJIMA2000} proposed that this lengthening of $\ell$ and contraction of $\sigma_t$ caused by vortex stretching in a cylindrical element translates into the following relationship in order for a vortex particle to conserve mass:
        \begin{align} \label{eq:tube:dsgmdt}
                \frac{1}{\sigma_p}\frac{\mathrm{d} }{\mathrm{d} t} \sigma_p
            =
                -\frac{1}{2 \Vert \boldsymbol{\Gamma}_p \Vert}
                \frac{\mathrm{d} \boldsymbol\Gamma_p }{\mathrm{d}t} \cdot \hat{\boldsymbol{\Gamma}}_p
        .\end{align}

        Now, using~\cref{eq:tube:dsgmdt} in the vortex strength equation,~\cref{eq:GE:Gamma:localized}, we obtain
        \begin{align} \label{eq:tube:ge:aux1}
                \frac{\mathrm{d} }{\mathrm{d} t} \boldsymbol{\Gamma}_p
            =
                \left(
                    \boldsymbol{\Gamma}_p \cdot \nabla
                \right)
                \mathbf{u} (\mathbf{x}_p)
                -
                \frac{3}{2}
                \left(
                    \frac{\mathrm{d} \boldsymbol\Gamma_p }{\mathrm{d}t} \cdot \hat{\boldsymbol{\Gamma}}_p
                \right) \hat{\boldsymbol{\Gamma}}_p
        ,\end{align}
        and from its dot product with $\hat{\boldsymbol{\Gamma}}_p$ we get
        \begin{align*}
                \frac{\mathrm{d} \boldsymbol\Gamma_p }{\mathrm{d}t} \cdot \hat{\boldsymbol{\Gamma}}_p
            =
                \frac{2}{5}
                \left[
                    \left(
                        \boldsymbol{\Gamma}_p \cdot \nabla
                    \right)
                    \mathbf{u} (\mathbf{x}_p)
                \right] \cdot \hat{\boldsymbol{\Gamma}}_p
        .\end{align*}
        Finally, substituting this back into~\cref{eq:tube:dsgmdt,eq:tube:ge:aux1}, we obtain the governing equations required for the particles to preserve mass in a tube of vorticity:
        \begin{align*}
                \frac{\mathrm{d} }{\mathrm{d} t} \sigma_p
            & =
                - \frac{1}{5}
                \frac{\sigma_p}{\Vert \boldsymbol{\Gamma}_p \Vert}
                \left[
                    \left(
                        \boldsymbol{\Gamma}_p \cdot \nabla
                    \right)
                    \mathbf{u} (\mathbf{x}_p)
                \right] \cdot \hat{\boldsymbol{\Gamma}}_p
            \\
                \frac{\mathrm{d} }{\mathrm{d} t} \boldsymbol{\Gamma}_p
            & =
                \left(
                    \boldsymbol{\Gamma}_p \cdot \nabla
                \right)
                \mathbf{u} (\mathbf{x}_p)
                -
                \frac{3}{5}
                \left\{
                    \left[
                        \left(
                            \boldsymbol{\Gamma}_p \cdot \nabla
                        \right)
                        \mathbf{u} (\mathbf{x}_p)
                    \right] \cdot \hat{\boldsymbol{\Gamma}}_p
                \right\} \hat{\boldsymbol{\Gamma}}_p
        \end{align*}
        Surprisingly, we have arrived to the same governing equations of the momentum-conserving sphere formulation\footnote{With the caveat that this is so only because we have neglected the SFS term. For this reason, in~\cref{sec:reformulatedvpm:generalized} we will revert back to~\cref{eq:tube:dsgmdt}.},~\cref{eq:sphere:dsigmadt,eq:sphere:dGammadt}.
        This stems from the fact that both conservation laws are physical consequences of the same equation: the vorticity Navier-Stokes equation.
        These new governing equations,~\cref{eq:sphere:dsigmadt,eq:sphere:dGammadt}, ensure local conservation of both angular momentum and mass, thus overcoming the pitfalls of the classic VPM.

    \subsubsection{Other Formulations} \label{sec:reformulatedvpm:otherformualtions}

        Combining the two conservations laws with the two element shapes, one can devise up to six different formulations: Each permutation between sphere/tube elements, momentum/mass conservation, and the assumption that mass conservation has or has not been already ensured by the convection of the elements.
        Two of them have already been discussed and turn out to be the same.
        For completeness, we now derive and briefly discuss the four remaining formulations.

        \vspace{0.25cm}\noindent\textit{Mass-Conserving Sphere}

            The mass of a spherical fluid element is conserved in incompressible flow as long as its volume remains constant.
            This translates into vortex particles of constant radii,
            \begin{align} \label{eq:masssphere:dsgmdt}
                \frac{\mathrm{d} }{\mathrm{d} t} \sigma_p = 0
            ,\end{align}
            which reduces the vortex strength equation to
            \begin{align*}
                    \frac{\mathrm{d} }{\mathrm{d} t} \boldsymbol{\Gamma}_p
                & =
                    \left(
                        \boldsymbol{\Gamma}_p \cdot \nabla
                    \right)
                    \mathbf{u} (\mathbf{x}_p)
            .\end{align*}
            Notice that these are the equations of the classic VPM.
            Even though this formulation preserves mass of spherical chunks of the flow, it is deemed unphysical as it violates local conservation of mass when the vorticity field encompasses a tube of vorticity that is being stretched, as discussed in~\cref{sec:reformulatedvpm:physicalimplications}.

        \vspace{0.25cm}\noindent\textit{Momentum-Conserving Tube}

            Consider a cylindrical element that rotates with angular velocity ${\boldsymbol\Omega = \boldsymbol\omega/2}$.
            Let $\mathbb{I}$ denote its inertia tensor and $I_z$ the moment of inertia about the centerline axis.
            If we assume that $\boldsymbol\Omega$ is always aligned with its centerline axis, the inertia tensor product $\mathbb{I}\boldsymbol\Omega$ becomes $\mathbb{I}\boldsymbol\Omega = I_z \boldsymbol\Omega$ since $\boldsymbol\Omega$ is an eigenvector of $\mathbb{I}$, with $I_z$ the corresponding eigenvalue.
            In order to conserve angular momentum, neglecting pressure torque and following the discussion in~\cref{sec:reformulatedvpm:physicalimplications:momentum}, the moment of inertia must change according to
            \begin{align} \label{eq:inertia:2}
                    -\frac{\boldsymbol\omega}{I_z}\frac{\mathrm{d} }{\mathrm{d} t} I_z
                =
                    (\boldsymbol\omega \cdot \nabla )\mathbf{u}
            .\end{align}
            Using $I_z = \frac{1}{2} m r^2$ and without assuming that the mass $m = \rho \pi r^2 \ell$ is automatically conserved, the dot product of~\cref{eq:inertia:2} with $\boldsymbol\omega$ leads to
            \begin{align*}
                    \frac{\mathrm{d} }{\mathrm{d} t} r
                =
                    - \frac{r}{4\ell} \frac{\mathrm{d} }{\mathrm{d} t} \ell
                    -
                    \frac{r}{4 \Vert \boldsymbol\omega \Vert^2 }
                    \left[
                        (\boldsymbol\omega \cdot \nabla )\mathbf{u}
                    \right]
                    \cdot \boldsymbol\omega
            .\end{align*}
            In the vortex particle scheme this becomes
            \begin{align} \label{eq:momentumtube:dsigmadt}
                    \frac{\mathrm{d} }{\mathrm{d} t} \sigma_p
                =
                    - \frac{1}{4}
                    \frac{\sigma_p}{\Vert \boldsymbol{\Gamma}_p \Vert}
                    \frac{\mathrm{d} \boldsymbol\Gamma_p }{\mathrm{d}t}  \cdot \hat{\boldsymbol{\Gamma}}_p
                    -
                    \frac{1}{4} \frac{\sigma_p}{\Vert \boldsymbol\Gamma_p \Vert }
                    \left[
                        (\boldsymbol\Gamma_p \cdot \nabla )\mathbf{u}
                    \right]
                    \cdot \hat{\boldsymbol\Gamma}_p
            ,\end{align}
            which in conjunction to the vortex strength equation,~\cref{eq:GE:Gamma:localized}, we obtain the following governing equations:
            \begin{align*}
                    \frac{\mathrm{d} }{\mathrm{d} t} \sigma_p
                & =
                    -
                    \frac{2}{7} \frac{\sigma_p}{\Vert \boldsymbol\Gamma_p \Vert }
                    \left[
                        (\boldsymbol\Gamma_p \cdot \nabla )\mathbf{u}
                    \right]
                    \cdot \hat{\boldsymbol\Gamma}_p
                \\
                    \frac{\mathrm{d} }{\mathrm{d} t} \boldsymbol{\Gamma}_p
                & =
                    \left(
                        \boldsymbol{\Gamma}_p \cdot \nabla
                    \right)
                    \mathbf{u} (\mathbf{x}_p)
                    -
                    \frac{6}{7}
                    \left\{
                        \left[
                            (\boldsymbol\Gamma_p \cdot \nabla )\mathbf{u}
                        \right]
                        \cdot \hat{\boldsymbol\Gamma}_p
                    \right\}
                    \hat{\boldsymbol\Gamma}_p
            .\end{align*}

        \vspace{0.25cm}\noindent\textit{Assumed Mass Conservation}

            Vortex methods typically assume that the continuity equation is implicitly solved by convecting the elements with the velocity field, automatically satisfying mass conservation.
            Repeating the momentum-conserving tube derivation, but this time assuming that the convection of the particles automatically ensures $\frac{\mathrm{d} }{\mathrm{d} t} m = 0$, we get the following governing equations:
            \begin{subequations} \label{eq:assumedmass:dsgmdt}
                \begin{align*}
                    \tag{\ref{eq:assumedmass:dsgmdt}}
                        \frac{\mathrm{d} }{\mathrm{d} t} \sigma_p
                    & =
                        -
                        \frac{1}{2} \frac{\sigma_p}{\Vert \boldsymbol\Gamma_p \Vert }
                        \left[
                            (\boldsymbol\Gamma_p \cdot \nabla )\mathbf{u}
                        \right]
                        \cdot \hat{\boldsymbol\Gamma}_p
                    \\
                        \frac{\mathrm{d} }{\mathrm{d} t} \boldsymbol{\Gamma}_p
                    & =
                        \left(
                            \boldsymbol{\Gamma}_p \cdot \nabla
                        \right)
                        \mathbf{u} (\mathbf{x}_p)
                        -
                        \frac{3}{2}
                        \left\{
                            \left[
                                (\boldsymbol\Gamma_p \cdot \nabla )\mathbf{u}
                            \right]
                            \cdot \hat{\boldsymbol\Gamma}_p
                        \right\}
                        \hat{\boldsymbol\Gamma}_p
                .\end{align*}
            \end{subequations}

            Taking the dot product of $\hat{\boldsymbol\Gamma}_p$ over the vortex strength equation, we get
            \begin{align*}
                    \frac{\mathrm{d} \boldsymbol{\Gamma}_p }{\mathrm{d} t} \cdot \hat{\boldsymbol\Gamma}_p
                & =
                    -
                    \frac{1}{2}
                    \left[
                        (\boldsymbol\Gamma_p \cdot \nabla )\mathbf{u}
                    \right]
                    \cdot \hat{\boldsymbol\Gamma}_p
            .\end{align*}
            In this last equation, notice that the derivative of the strength projected on its own direction is negative when vortex stretching is positive.
            This means that the magnitude of vortex strength (and associated vorticity) decreases when the particle is stretched, which is an unphysical result.

            Along the same lines, we can repeat the momentum-conserving sphere derivation from~\cref{sec:reformulatedvpm:sphereformulation}, but this time assuming automatic mass conservation, $\frac{\mathrm{d} }{\mathrm{d} t} m = 0$.
            This turns out to lead to the same governing equations in~\cref{eq:assumedmass:dsgmdt}.
            Hence, we conclude that it is not possible to impose momentum conservation with either tube or sphere elements while assuming automatic mass conservation.

	\subsection{Generalized Formulation} \label{sec:reformulatedvpm:generalized}

        In the preceding section, the conservation laws were imposed in six different ways to derive various expressions of $\frac{\mathrm{d} }{\mathrm{d} t} \sigma_p$, namely~\cref{eq:sphere:dsigmadt,eq:tube:dsgmdt,eq:masssphere:dsgmdt,eq:momentumtube:dsigmadt,eq:assumedmass:dsgmdt}.
        In turn, each version of $\frac{\mathrm{d} }{\mathrm{d} t} \sigma_p$ was used with the vortex strength equation,~\cref{eq:GE:Gamma:localized}, to obtain a distinct VPM formulation.
        We can generalize this formulation procedure noticing that all versions of $\frac{\mathrm{d} }{\mathrm{d} t} \sigma_p$ take the functional form
        \begin{subequations} \label{eq:dsigmadt:fg}
            \begin{align*}
                \tag{\ref{eq:dsigmadt:fg}}
                    \frac{\mathrm{d} }{\mathrm{d}t} \sigma_p
                & =
                    - \f
                        \sigma_p \frac{1}{\Vert \boldsymbol{\Gamma}_p \Vert}
                        \frac{\mathrm{d} \boldsymbol\Gamma_p}{\mathrm{d}t} \cdot \hat{\boldsymbol{\Gamma}}_p
                \\
                & \qquad
                    - \g
                        \sigma_p \frac{1}{\Vert \boldsymbol{\Gamma}_p \Vert}
                            \left[
                                \left(
                                        \boldsymbol{\Gamma}_p  \cdot \nabla
                                    \right)
                                    \mathbf{u} {\small (\mathbf{x}_p)}
                            \right]
                            \cdot \hat{\boldsymbol{\Gamma}}_p
            ,\end{align*}
        \end{subequations}
        where the parameters $\f, \g \in \mathbb{R}$ are derived applying the conservation laws.

        We will now use the $\f$--$\g$ form of $\frac{\mathrm{d} }{\mathrm{d}t} \sigma_p $,~\cref{eq:dsigmadt:fg}, to derive a new set of governing equation of the VPM, also in the $\f$--$\g$ form.
        We start from the general equation governing vortex strength,~\cref{eq:GE:Gamma:general}, here repeated
        \begin{align} \tag{\ref{eq:GE:Gamma:general}}
                \frac{\mathrm{d} \boldsymbol{\Gamma}_p}{\mathrm{d} t}
            =
                \left(
                    \boldsymbol{\Gamma}_p \cdot \nabla
                \right)
                \mathbf{u} (\mathbf{x}_p)
                +
                3 \boldsymbol{\Gamma}_p \frac{1}{\sigma_p}
                \frac{\partial \sigma_p}{\partial t}
                +
                \mathbf{M}_p
        ,\end{align}
        where
        \begin{align*}
                \mathbf{M}_p
            & \equiv
                \frac{1}{\zeta_{\sigma_p} (\mathbf{0})}
                (-\mathbf{M}_p^0 + \mathbf{M}_p^1 + \mathbf{M}_p^2)
            \\
            & \qquad
                -
                \frac{1}{\zeta_{\sigma_p} (\mathbf{0})}
                \left( \mathbf{E}_\mathrm{adv} (\mathbf{x}_p) + \mathbf{E}_\mathrm{str} (\mathbf{x}_p) \right)
        .\end{align*}
        Recalling that $\frac{\mathrm{d} \sigma_p}{\mathrm{d}t} = \frac{\partial \sigma_p}{\partial t}$ since $\sigma_p$ is a Lagrangian quantity, we substitute~\cref{eq:dsigmadt:fg} into~\cref{eq:GE:Gamma:general} and take its dot product with $\hat{\boldsymbol{\Gamma}}_p$ to get
        \begin{align} \label{eq:analysis:stretching:fg}
                \frac{\mathrm{d} \boldsymbol{\Gamma}_p}{\mathrm{d} t} \cdot \hat{\boldsymbol{\Gamma}}_p
            =
                \frac{1 - 3\g}{1 + 3\f}
                \left[
                    \left(
                            \boldsymbol{\Gamma}_p  \cdot \nabla
                        \right)
                        \mathbf{u} {\small (\mathbf{x}_p)}
                \right]
                \cdot \hat{\boldsymbol{\Gamma}}_p
                +
                \frac{1}{1 + 3\f}
                \mathbf{M}_p \cdot \hat{\boldsymbol{\Gamma}}_p
        .\end{align}

        Substituting~\cref{eq:analysis:stretching:fg} into~\cref{eq:dsigmadt:fg}, we arrive to the equation governing the evolution of particle size,
        \begin{subequations} \label{eq:ge:rvpm:dsigmadt}
            \begin{align*}
                \tag{\ref{eq:ge:rvpm:dsigmadt}}
                    \frac{\mathrm{d} }{\mathrm{d}t} \sigma_p
                & =
                    - \left(
                        \frac{\g + \f}{1 + 3\f}
                    \right)
                    \frac{\sigma_p}{\Vert \boldsymbol{\Gamma}_p \Vert}
                        \left[
                            \left(
                                    \boldsymbol{\Gamma}_p  \cdot \nabla
                                \right)
                                \mathbf{u} {\small (\mathbf{x}_p)}
                        \right]
                        \cdot \hat{\boldsymbol{\Gamma}}_p
                \\
                & \qquad
                    -
                    \left(
                        \frac{\f}{1 + 3\f}
                    \right)
                    \frac{\sigma_p}{\Vert \boldsymbol{\Gamma}_p \Vert}
                    \mathbf{M}_p \cdot \hat{\boldsymbol{\Gamma}}_p
            ,\end{align*}
        \end{subequations}
        and we substitute this into~\cref{eq:GE:Gamma:general} to get the equation governing the evolution of vortex strength,
        \begin{align} \label{eq:ge:rvpm:dGammadt}
            \begin{split}
                    \frac{\mathrm{d} }{\mathrm{d} t} \boldsymbol{\Gamma}_p
                =
                    \left(
                        \boldsymbol{\Gamma}_p \cdot \nabla
                    \right)
                    \mathbf{u} (\mathbf{x}_p)
                &   -
                    \frac{\g + \f}{\nicefrac{1}{3} + \f}
                    \left\{
                        \left[
                            \left(
                                    \boldsymbol{\Gamma}_p  \cdot \nabla
                                \right)
                                \mathbf{u} {\small (\mathbf{x}_p)}
                        \right]
                        \cdot \hat{\boldsymbol{\Gamma}}_p
                    \right\} \hat{\boldsymbol{\Gamma}}_p
                \\
                    +
                    \mathbf{M}_p
                &   -
                    \frac{\f}{\nicefrac{1}{3} + \f}
                    \left(
                        \mathbf{M}_p \cdot \hat{\boldsymbol{\Gamma}}_p
                    \right) \hat{\boldsymbol{\Gamma}}_p
            .\end{split}
        \end{align}
        \cref{eq:ge:rvpm:dsigmadt,eq:ge:rvpm:dGammadt} constitute the governing equations of the reformulated VPM, where the parameters $\f,\,\g$ are derived from a specific implementation of the conservations laws.
        In particular when ${\f = \g = 0}$,~\cref{eq:ge:rvpm:dsigmadt,eq:ge:rvpm:dGammadt} collapse back to the classic VPM equations, making the reformulated VPM a generalization of the classic method.
        Furthermore, notice that the reformulated equations do not require more computation than the classic method: When SFS and non-localized vorticity effects are neglected ($\mathbf{M}_p=0$), both $\frac{\mathrm{d} \sigma_p }{\mathrm{d}t}$ and $\frac{\mathrm{d} \boldsymbol{\Gamma}_p}{\mathrm{d} t}$ are calculated directly from vortex stretching, $\left( \boldsymbol{\Gamma}_p \cdot \nabla  \right) \mathbf{u} (\mathbf{x}_p)$.

\subsection{Formulation Analysis} \label{sec:reformulatedvpm:analysis}

    In~\cref{sec:reformulatedvpm:candidateformulations} we have shown several plausible formulations that can be derived from the conservation laws.
    We now cast each formulation into its $\f$--$\g$ form, as summarized in~\cref{table:fgform}, and contrast the physical implications of each formulation to help us choose one formulation to further explore.

    We wish to determine how each formulation may hinder or augment the effects of vortex stretching on both vortex strength and vortex re-orientation.
    First, we focus on vortex re-orientation.
    The rate of vortex re-orientation, denoted $\frac{\mathrm{d} \hat{\boldsymbol{\Gamma}}_p}{\mathrm{d} t}$, is constructed as
    \begin{align*}
            \frac{\mathrm{d} \hat{\boldsymbol{\Gamma}}_p}{\mathrm{d} t}
        =
            \frac{1}{\Vert \boldsymbol{\Gamma}_p \Vert}
            \frac{\mathrm{d} \boldsymbol{\Gamma}_p}{\mathrm{d} t}
            -
            \frac{1}{\Vert \boldsymbol{\Gamma}_p \Vert}
            \left(
                \frac{\mathrm{d} \boldsymbol{\Gamma}_p}{\mathrm{d} t} \cdot \hat{\boldsymbol{\Gamma}}_p
            \right)
            \hat{\boldsymbol{\Gamma}}_p
    ,\end{align*}
    which is the change in direction of vortex strength over time, as shown in~\cref{sec:app:der:reorientation}.
    Using~\cref{eq:analysis:stretching:fg} and~\cref{eq:ge:rvpm:dGammadt} and neglecting $\mathbf{M}$, this becomes
    \begin{align*}
            \frac{\mathrm{d} \hat{\boldsymbol{\Gamma}}_p}{\mathrm{d} t}
        =
            \left(
                \hat{\boldsymbol{\Gamma}}_p \cdot \nabla
            \right)
            \mathbf{u} (\mathbf{x}_p)
            -
            \left\{
                \left[
                    \left(
                            \hat{\boldsymbol{\Gamma}}_p  \cdot \nabla
                        \right)
                        \mathbf{u} {\small (\mathbf{x}_p)}
                \right]
                \cdot \hat{\boldsymbol{\Gamma}}_p
            \right\} \hat{\boldsymbol{\Gamma}}_p
    ,\end{align*}
    which is independent of $\f$ and $\g$.
    Hence, vortex re-orientation is not affected by the formulation, staying the same between the classic VPM and any $\f$--$\g$ formulation.

    \begin{table}[t]

        \caption{$\f$--$\g$ form of formulations derived in~\cref{sec:reformulatedvpm:candidateformulations}}
        \label{table:fgform}

        \begin{tabular}{ccccc}
            \hline
            \textbf{Formulation}                                                                             & \textbf{$\f$}                          & \textbf{$\g$}                          & \textbf{$h_\Gamma = \frac{1 - 3\g}{1 + 3\f}$} & \textbf{$h_\sigma = \frac{\g + \f}{1 + 3\f}$} \\ \hline
            \begin{tabular}[c]{@{}c@{}}Mass-conserving sphere\\ {\footnotesize  (Classic VPM) }\end{tabular} & {\color{fclr}{0}}                      & {\color{gclr}{0}}                      & 1                                             & 0                                             \\ [0.6cm]
            Mass-conserving tube                                                                             & ${\color{fclr}{\nicefrac{1}{2}}}$      & {\color{gclr}{0}}                      & 0.4                                           & 0.2                                           \\ [0.6cm]
            \begin{tabular}[c]{@{}c@{}}Momentum-conserving sphere\\ {\footnotesize  (Reformulated VPM) }\end{tabular}            & {\color{fclr}{0}}  & ${\color{gclr}{\nicefrac{1}{5}}}$      & 0.4                                           & 0.2                                           \\ [0.6cm]
            Momentum-conserving tube                                                                         & ${\color{fclr}{\nicefrac{1}{4}}}$      & ${\color{gclr}{\nicefrac{1}{4}}}$      & 0.143                                         & 0.286                                         \\ [0.6cm]
            \begin{tabular}[c]{@{}c@{}}Momentum-conservation \\ {\footnotesize   w/ assumed mass conservation}\end{tabular} & {\color{fclr}{0}}  & ${\color{gclr}{\nicefrac{1}{2}}}$      & -0.5                                          & 0.5                                           \\ [0.3cm] \hline
        \end{tabular}
    \end{table}

    Now, we shift our attention to the effects of vortex stretching on vortex strength.
    Neglecting SFS and non-localized vorticity effects ($\mathbf{M}_p=0$), from~\cref{eq:analysis:stretching:fg} we have
    \begin{align} \label{eq:analysis:stretchingmagitude}
            \frac{\mathrm{d} \boldsymbol{\Gamma}_p}{\mathrm{d} t} \cdot \hat{\boldsymbol{\Gamma}}_p
        =
            \frac{1 - 3\g}{1 + 3\f}
            \left[
                \left(
                        \boldsymbol{\Gamma}_p  \cdot \nabla
                    \right)
                    \mathbf{u} {\small (\mathbf{x}_p)}
            \right]
            \cdot \hat{\boldsymbol{\Gamma}}_p
    ,\end{align}
    where $\frac{\mathrm{d} \boldsymbol{\Gamma}_p}{\mathrm{d} t} \cdot \hat{\boldsymbol{\Gamma}}_p$ is the change in the magnitude of vortex strength.
    We introduce the parameters
    \begin{align*}
        h_\Gamma \equiv \frac{1 - 3\g}{1 + 3\f}
        ,\quad
        h_\sigma \equiv \frac{\g + \f}{1 + 3\f}
    \end{align*}
    and rewrite this as
    \begin{align*}
            \frac{\mathrm{d} \boldsymbol{\Gamma}_p}{\mathrm{d} t} \cdot \hat{\boldsymbol{\Gamma}}_p
        =
            h_\Gamma
            \left[
                \left(
                        \boldsymbol{\Gamma}_p  \cdot \nabla
                    \right)
                    \mathbf{u} {\small (\mathbf{x}_p)}
            \right]
            \cdot \hat{\boldsymbol{\Gamma}}_p
    ,\end{align*}
    while~\cref{eq:ge:rvpm:dsigmadt} becomes
    \begin{align*}
            \frac{\Vert \boldsymbol{\Gamma}_p \Vert}{\sigma_p}\frac{\mathrm{d} }{\mathrm{d}t} \sigma_p
        =
            - h_\sigma
                \left[
                    \left(
                            \boldsymbol{\Gamma}_p  \cdot \nabla
                        \right)
                        \mathbf{u} {\small (\mathbf{x}_p)}
                \right]
                \cdot \hat{\boldsymbol{\Gamma}}_p
    .\end{align*}

    It can be shown that $h_\Gamma$ and $h_\sigma$ must satisfy ${h_\Gamma + 3 h_\sigma = 1}$ in order for a formulation to be a solution to the vorticity Navier-Stokes equation.
    Hence, any $\f$--$\g$ formulation simply differs in how vortex stretching is distributed between the lengthening of vortex strength $\boldsymbol{\Gamma}_p$, and shrinking of particle size $\sigma_p$.
    The classic VPM results in $h_\Gamma=1$ and $h_\sigma=0$, meaning that vortex strength takes the full amount of vortex stretching.

\begin{figure}[t]
        \centering
        \includegraphics[width=\largerfigwidth]{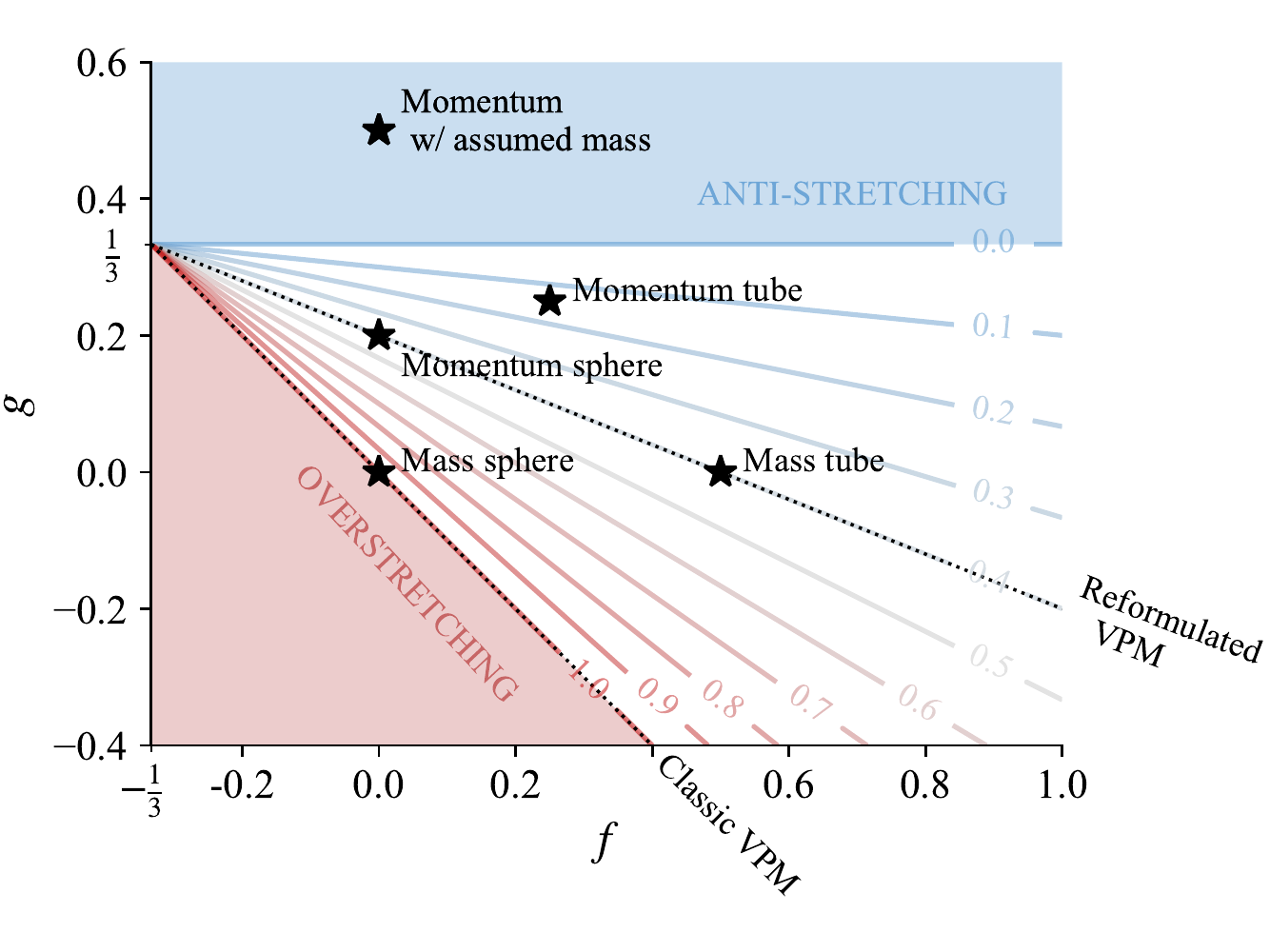}
        \caption{Space of possible $\f$-$\g$ formulations, with contour levels of $h_\Gamma = \frac{1 - 3\g}{1 + 3\f}$ that are solutions to the vorticity Navier-Stokes equation. Markers show each formulation in~\cref{table:fgform}.}
        \label{fig:fgspace}
    \end{figure}

    The space of all possible $\f$--$\g$ formulations is shown in~\cref{fig:fgspace}, along with the contours of $h_\Gamma$ that satisfy ${h_\Gamma + 3 h_\sigma = 1}$.
    Any formulation with $h_\Gamma > 1$ will result in a negative $h_\sigma$, meaning that the lengthening of vortex strength under stretching is so aggressive that $\sigma_p$ has to expand instead of shrink, which is unphysical.
    This is referred to as \textit{overstretching}.
    On the other extreme, $h_\sigma > \nicefrac{1}{3}$ results in a negative $h_\Gamma$, meaning that the shrinking of the particle is so aggressive that $\boldsymbol\Gamma$ has to compensate by shrinking when it ought to have lengthened with vortex stretching, which is also unphysical.
    This is referred to as \textit{anti-stretching}.
    The region between overstretching and anti-stretching is the space of VPM formulations that are physically plausible.

    There are four formulations that lay in the space of physically plausible formulations, shown in~\cref{fig:fgspace}.
    First, notice that the classic VPM (mass-conserving sphere) lies at the threshold of overstretching, which explains its tendency to be numerically unstable.
    We are then left with three possible formulations, two of which lay on the same $h_\Gamma$ contour line: the momentum-conserving sphere and the mass-conserving tube.
    Even though these two formulations stem from different physical principles resulting in different $\f$--$\g$ values, they share the same $h_\Gamma$ and $h_\sigma$ values, leading to the same governing equations\footnote{This is only true after neglecting $\mathbf{M}$. When SFS and non-localized vorticity effects are re-incorporated, the $\mathbf{M}$-term ends up being multiplied by a factor ${h_M \equiv \frac{1}{1+3\f}}$ as shown in~\cref{eq:analysis:stretching:fg}. Hence, SFS and non-localized vorticity effects are stronger in the momentum-conserving sphere formulation (${h_M = 1}$) than the mass-conserving tube formulation (${h_M = 0.4}$).}, as mentioned in~\cref{sec:reformulatedvpm:tubeformulation}.
    Thus, by choosing either set of $\f$--$\g$ values, we end up in a formulation that preserves angular momentum in spherical chunks of the flow, while at the same time conserving mass in tubes of vorticity.
    Here on, the momentum-conserving sphere formulation is referred to as the \textit{reformulated vortex particle method}, or rVPM.

\subsection{Reformulated Governing Equations} \label{sec:reformulatedvpm:GE}
    Summarizing the preceding sections, the governing equations of the reformulated VPM are

    \begin{align}
        \label{eq:rvpm:ge:dxdt}
            \frac{\text{d}}{\text{d}t}\mathbf{x}_p
        & =
            \mathbf{u}(\mathbf{x}_p)
        \\ \label{eq:rvpm:ge:dsigmadt}
            \frac{\mathrm{d} }{\mathrm{d} t} \sigma_p
        & =
            - \left(
                \frac{\g + \f}{1 + 3\f}
            \right)
            \frac{\sigma_p}{\Vert \boldsymbol{\Gamma}_p \Vert}
                \left[
                    \left(
                            \boldsymbol{\Gamma}_p  \cdot \nabla
                        \right)
                        \mathbf{u} {\small (\mathbf{x}_p)}
                \right]
                \cdot \hat{\boldsymbol{\Gamma}}_p
        \\ \label{eq:rvpm:ge:dGammadt}
            \frac{\mathrm{d} }{\mathrm{d} t} \boldsymbol{\Gamma}_p
        & =
            \left(
                \boldsymbol{\Gamma}_p \cdot \nabla
            \right)
            \mathbf{u} (\mathbf{x}_p)
            -
            \frac{\g + \f}{\nicefrac{1}{3} + \f}
            \left\{
                \left[
                    \left(
                            \boldsymbol{\Gamma}_p  \cdot \nabla
                        \right)
                        \mathbf{u} {\small (\mathbf{x}_p)}
                \right]
                \cdot \hat{\boldsymbol{\Gamma}}_p
            \right\} \hat{\boldsymbol{\Gamma}}_p
        \\ \label{eq:rvpm:ge:viscous}
            & \hspace{-10mm}
            \left(
                    \frac{\text{d} }{\text{d} t} \filter{ \boldsymbol\omega }
                \right)_\mathrm{viscous}
            =
                \nu \nabla^2 \filter{ \boldsymbol\omega }
    ,\end{align}
    where~\cref{eq:rvpm:ge:dxdt} resolves vorticity advection by convecting the particles,~\cref{eq:rvpm:ge:dsigmadt} governs the evolution of particle size, and~\cref{eq:rvpm:ge:dGammadt} governs the evolution of vortex strength.
    \cref{eq:rvpm:ge:dGammadt} in conjunction with~\cref{eq:rvpm:ge:dxdt,eq:rvpm:ge:dsigmadt} resolve the inviscid part of the LES-filtered vorticity Navier-Stokes equation, while the viscous part in~\cref{eq:rvpm:ge:viscous} is resolved by one of the schemes mentioned in~\cref{sec:fundamentals:splitting}.
    For simplicity, \cref{eq:rvpm:ge:dsigmadt,eq:rvpm:ge:dGammadt} omit the SFS contributions, but they are readily re-incorporated as shown in~\cref{eq:ge:rvpm:dsigmadt,eq:ge:rvpm:dGammadt}.

    The formulation used for the rest of this study implements $\f = 0$ and $\g = \nicefrac{1}{5}$, which ensures conservation of angular momentum in spherical chunks of the flow, while at the same time conserving mass in tubes of vorticity.
    As previously mentioned, the rVPM equations do not require more computation than the classic VPM since $\frac{\mathrm{d} \sigma_p }{\mathrm{d}t}$ and $\frac{\mathrm{d} \boldsymbol{\Gamma}_p}{\mathrm{d} t}$ are calculated directly and solely from vortex stretching, $\left( \boldsymbol{\Gamma}_p \cdot \nabla  \right) \mathbf{u} (\mathbf{x}_p)$.

    \section{Anisotropic Dynamic SFS Model} \label{sec:sfs}

    In the previous section we have formulated a numerical scheme for solving the large scales of the LES-filtered vorticity equation,
    \begin{align}  \tag{\ref{eq:filtered:angular:vector}}
        \frac{\text{d} }{\text{d} t} \filter{ \boldsymbol\omega }
        = \left( \filter{ \boldsymbol\omega } \cdot \nabla \right) \filter{ \mathbf{u} } +
        \nu \nabla^2 \filter{ \boldsymbol\omega }
        - \mathbf{E}_\mathrm{adv} - \mathbf{E}_\mathrm{str}
    .\end{align}
    We will now focus on the subfilter-scale (SFS) stresses associated with advection and vortex stretching, $\mathbf{E}_\mathrm{adv}$ and $\mathbf{E}_\mathrm{str}$, respectively.
    In~\cref{sec:sfs:existing} we briefly discuss existing SFS models relevant to meshless vortex methods.
    In~\cref{sec:sfs:newmodel} we introduce a novel anisotropic structural model of SFS vortex stretching that is suitable for turbulent flows where the predominant cascade mechanism is vortex stretching.
    In~\cref{sec:sfs:dynamic,sec:sfs:backscatter} we develop a dynamic procedure for computing the model coefficient while also providing the means for backscatter control.
    Finally, in~\cref{sec:sfs:meshcfd} we show how our model can be implemented in conventional mesh-based CFD with a pressure-velocity solver.

    \subsection{Existing Models} \label{sec:sfs:existing}
        Over the years, only a few models have been proposed that are suitable for meshless vortex methods.
        The most popular one is a variant of the Smagorinsky eddy-viscosity model formulated for the vorticity stress \cite{Winckelmans1995,Mansfield1998,Mansfield1996,Mansfield1999},
        \begin{align*}
            \mathbf{E} = - \nabla \times \left( \nu_\mathrm{SFS} \nabla \times \filter{\boldsymbol\omega} \right)
        ,\end{align*}
        where $\mathbf{E}\equiv \mathbf{E}_\mathrm{adv} + \mathbf{E}_\mathrm{str}$, $\nu_\mathrm{SFS} = C_d^2 \sigma^2 \sqrt{2 S_{mn}S_{mn}}$, $S_{mn}$ is the strain-rate tensor, $\sigma$ is the filter width, and $C_d$ is a model coefficient which is either prescribed or computed dynamically.
        This functional model and others alike were developed on the basis of homogeneous isotropic turbulence, which makes them overly diffusive in simulations with coherent vortical structures.
        In the latest developments of the vortex particle-mesh scheme\cite{Caprace2020,Caprace2021}, this drawback has been avoided with the variational multiscale method\cite{Hughes2000,Vreman2003,Jeanmart2007,Cocle2009}, however its applicability to a meshless scheme is not clear.

        In a different approach, Cottet\cite{Cottet1996,Cottet1998,cottet2003} developed an anisotropic structural model of the advective SFS term $\mathbf{E}_\mathrm{adv}$ as
        \begin{align*}
            &
                \mathbf{E}_\mathrm{adv} (\mathbf{x}_p)
            =
            \\
            & \quad
                C_d
                \sum\limits_q
                    \mathrm{Vol}_q \left( \boldsymbol{\omega}_p - \boldsymbol{\omega}_q \right)
                    \left[
                        \left(\mathbf{u} (\mathbf{x}_p) -\mathbf{u} (\mathbf{x}_q)\right)
                        \cdot
                        \nabla \zeta_{\sigma_q} (\mathbf{x}_p - \mathbf{x}_q)
                    \right]
        ,\end{align*}
        where the model coefficient $C_d$ is usually prescribed with \textit{ad hoc} calibration \cite{Mimeau2019}.
        This model is reportedly\cite{Stock2010,Mimeau2018} less dissipative than Smagorinsky-type models, however, it was developed on the basis of 2D flow which is absent of vortex stretching.
        Thus, this model neglects $\mathbf{E}_\mathrm{str}$ even though vortex stretching is known to be one of the main mechanisms for enstrophy production in the energy cascade in three dimensions.

        To address the need for a low-dissipation SFS model that captures vortex stretching as the physical mechanism for turbulence, in the following sections we develop an anisotropic structural model of SFS vortex stretching with a dynamic model coefficient.

    \subsection{SFS Vortex Stretching Model} \label{sec:sfs:newmodel}
        Starting from the definition of SFS vortex stretching,
        \begin{align} \label{eq:Estr}
                E_i^\mathrm{str}
            =
                - \frac{\partial T_{ij}}{\partial x_j}
            \equiv
                - \left(
                    \filter{ \omega_j \frac{\partial u_i}{\partial x_j} } -
                    \filter{ \omega_j } \frac{\partial \filter{ u_i }}{\partial x_j}
                \right)
        ,\end{align}
        we group the filter operator as
        \begin{align*}
                E_i^\mathrm{str} \left( \mathbf{x} \right)
            =
                - &
                    \int\limits
                        \omega_j \left( \mathbf{y} \right)
                        \left(
                            \frac{\partial u_i}{\partial x_j} {\small \left( \mathbf{y} \right)}
                            -
                            \frac{\partial \filter{u_i} }{\partial x_j} \left( \mathbf{x} \right)
                        \right)
                        \zeta_\sigma(\mathbf{x}-\mathbf{y})
                    \,\mathrm{d}\mathbf{y}
        .\end{align*}
        Using the singular particle approximation, $\boldsymbol{\omega}(\mathbf{x}) \approx \sum \limits_q \boldsymbol{\Gamma}_q \delta  (\mathbf{x} - \mathbf{x}_q)$, the integral collapses to
        \begin{align*}
                E_i^\mathrm{str} \left( \mathbf{x} \right)
            \approx
                -
                \sum\limits_q
                    \Gamma_j^q
                    \left(
                        \frac{\partial u_i}{\partial x_j} \left( \mathbf{x}_q \right)
                        - \frac{\partial \filter{u_i} }{\partial x_j} \left( \mathbf{x} \right)
                    \right)
                    \zeta_{\sigma}(\mathbf{x}-\mathbf{x}_q)
        .\end{align*}
        Assuming $\frac{\partial u_i}{\partial x_j} \approx \frac{\partial \filter{u_i} }{\partial x_j}$ and writing in vector notation, the model then becomes
        \begin{align*}
                \mathbf{E}_\mathrm{str} \left( \mathbf{x} \right)
            \approx
                \sum\limits_q
                    \zeta_{\sigma}(\mathbf{x}-\mathbf{x}_q)
                    \left(
                        \boldsymbol{\Gamma}_q \cdot \nabla
                    \right)
                    \left(
                        \filter{\mathbf{u}} \left( \mathbf{x} \right) - \filter{\mathbf{u}} \left( \mathbf{x}_q \right)
                    \right)
        .\end{align*}

        Interestingly, both Cottet's model and our model are strikingly similar to the non-localized vorticity terms $\mathbf{M}_p^1$ and $\mathbf{M}_p^2$, respectively (defined in~\cref{eq:M1,eq:M2}), which are neglected (or filtered out) by the localized-vorticity assumption.
        Thus, these models can be thought of as soft deconvolutions of the localized-vorticity assumption.
        Furthermore, introducing a model coefficient, $C_d$, they can approximate full deconvolutions encompassing the entire spectrum of subfilter scales.
        This model coefficient is computed dynamically as follows.

    \subsection{Dynamic Procedure} \label{sec:sfs:dynamic}

        \subsubsection{Derivative Balance}
            Since SFS vortex stretching needs to be modeled from resolved quantities, the initial equality given in~\cref{eq:Estr}, written here in vector notation
            \begin{align*}
                        \mathbf{E}_\mathrm{str}
                    = -
                        \left[
                            \filter{ (\boldsymbol{\omega} \cdot \nabla )\mathbf{u} }
                            -
                            (\filter{ \boldsymbol{\omega} } \cdot \nabla )\filter{ \mathbf{u} }
                    \right]
            ,\end{align*}
            becomes only an approximation,
            \begin{align*}
                    \mathbf{E}_\mathrm{str}
                \approx
                    - \left[
                        \filter{ (\boldsymbol{\omega} \cdot \nabla )\mathbf{u} }
                        -
                        (\filter{ \boldsymbol{\omega} } \cdot \nabla )\filter{ \mathbf{u} }
                    \right]
            \end{align*}
            once the $\mathbf{E}_\mathrm{str}$ model is introduced.
            To recover the equality (or at least improve the approximation), we introduce a dynamic model coefficient $C_d (\mathbf{x},\, t)$ satisfying
            \begin{align} \label{eq:Cd:intro}
                    C_d
                    \mathbf{E}_\mathrm{str}
                & =
                    - \left[
                        \filter{ (\boldsymbol{\omega} \cdot \nabla )\mathbf{u} }
                        - (\filter{ \boldsymbol{\omega} } \cdot \nabla )\filter{ \mathbf{u} }
                    \right]
            .\end{align}
            However, this equation is not useful \textit{per se} since, in order to determine $C_d$,~\cref{eq:Cd:intro} requires knowing the SFS quantity ${\filter{ (\boldsymbol{\omega} \cdot \nabla )\mathbf{u} } - (\filter{ \boldsymbol{\omega} } \cdot \nabla )\filter{ \mathbf{u} }}$, which is exactly what we are trying to model with $\mathbf{E}_\mathrm{str}$.
            However, assuming scale similarity (meaning that $C_d$ is independent of the filter width), we differentiate this equation with respect to the filter width $\sigma$ to obtain a more useful relation:
            \begin{align} \label{eq:Cd:derbalance}
                    C_d
                    \frac{\partial \mathbf{E}_\mathrm{str} }{\partial \sigma}
                =
                    - \frac{\partial }{\partial \sigma}
                    \left[
                        \filter{ (\boldsymbol{\omega} \cdot \nabla )\mathbf{u} }
                        -
                        (\filter{ \boldsymbol{\omega} } \cdot \nabla )\filter{ \mathbf{u} }
                    \right]
            .\end{align}

            Using the singular particle approximation, the right-hand side of~\cref{eq:Cd:derbalance} before differentiation becomes
            \begin{align*}
                &
                    \left.
                        \left[
                            \filter{ (\boldsymbol{\omega} \cdot \nabla )\mathbf{u} }
                            -
                            (\filter{ \boldsymbol{\omega} } \cdot \nabla )\filter{ \mathbf{u} }
                        \right]
                    \right\vert_\mathbf{x}
                \approx
                \\
                & \qquad \qquad
                    \sum\limits_q
                        \zeta_{\sigma}(\mathbf{x}-\mathbf{x}_q)
                        \left(
                            \boldsymbol{\Gamma}_q \cdot \nabla
                        \right)
                        \left(
                            \mathbf{u} \left( \mathbf{x}_q \right) - \filter{\mathbf{u}} \left( \mathbf{x} \right)
                        \right)
            ,\end{align*}
            and after differentiation,
            \begin{align*}
                &
                    \frac{\partial }{\partial \sigma}
                    \left.
                        \left[
                            \filter{ (\boldsymbol{\omega} \cdot \nabla )\mathbf{u} }
                            -
                            (\filter{ \boldsymbol{\omega} } \cdot \nabla )\filter{ \mathbf{u} }
                        \right]
                    \right\vert_\mathbf{x}
                \approx
                \\
                & \qquad \qquad
                        \sum\limits_q
                            \frac{\partial \zeta_{\sigma} }{\partial \sigma}(\mathbf{x}-\mathbf{x}_q)
                            \left(
                                \boldsymbol{\Gamma}_q \cdot \nabla
                            \right)
                            \left(
                                \mathbf{u} \left( \mathbf{x}_q \right) - \filter{\mathbf{u}} \left( \mathbf{x} \right)
                            \right)
                \\
                & \qquad \qquad
                        -
                        \sum\limits_q
                            \zeta_{\sigma}(\mathbf{x}-\mathbf{x}_q)
                            \left(
                                \boldsymbol{\Gamma}_q \cdot \nabla
                            \right)
                            \frac{\partial \filter{\mathbf{u}} }{\partial \sigma} \left( \mathbf{x} \right)
            .\end{align*}
            Recalling that $\zeta_\sigma$ is defined as $\zeta_\sigma (\mathbf{x}) = \frac{1}{\sigma^3} \zeta \left( \frac{\Vert \mathbf{x} \Vert}{\sigma} \right)$, its width-derivative is
            \begin{align*}
                    \frac{\partial \zeta_\sigma }{\partial \sigma} (\mathbf{x})
                & =
                    - %
                    \frac{3}{\sigma} \zeta_\sigma \left( \mathbf{x} \right)
                    - %
                    \frac{\Vert \mathbf{x} \Vert}{\sigma^5} \frac{\partial \zeta }{\partial r}
                    \left( \frac{\Vert \mathbf{x} \Vert}{\sigma} \right)
            ,\end{align*}
            and assuming that $\zeta_\sigma (\mathbf{x})$ reaches a maximum at $\mathbf{x} = \mathbf{0}$ (meaning, ${\frac{\partial \zeta}{\partial r} (0) = 0}$),
            \begin{align*}
                    \frac{\partial \zeta_\sigma }{\partial \sigma} (\mathbf{0})
                = - \frac{3}{\sigma} \zeta_\sigma \left( \mathbf{0} \right)
            .\end{align*}
            Then, evaluating at $\mathbf{x} = \mathbf{x}_p$, assuming $\zeta_\sigma \left( \mathbf{0} \right) \neq 0$, and using the localized-vorticity assumption to neglect all terms $q \neq p$,~\cref{eq:Cd:derbalance} becomes
            \begin{align*}
                &
                    C_d (\mathbf{x}_p)
                    \frac{1}{\zeta_\sigma \left( \mathbf{0} \right)}
                    \frac{\partial \mathbf{E}_\mathrm{str} }{\partial \sigma} (\mathbf{x}_p)
                =
                \\
                & \qquad \quad
                    \frac{3}{\sigma}
                    \left( \boldsymbol\Gamma_p \cdot \nabla \right)
                    \left(
                        \mathbf{u} (\mathbf{x}_p) - \filter{\mathbf{u}} (\mathbf{x}_p)
                    \right)
                    +
                    \left( \boldsymbol\Gamma_p \cdot \nabla \right)
                    \frac{\partial \filter{\mathbf{u}} }{\partial \sigma}  (\mathbf{x}_p)
            ,\end{align*}
            or
            \begin{align} \label{eq:Cd:ML}
                    C_d \mathbf{m} = \mathbf{L}
            ,\end{align}
            with
            \begin{align*}
                &
                    \mathbf{m} \equiv
                        \frac{\sigma^3}{\zeta(0)} \frac{\partial \mathbf{E}_\mathrm{str} }{\partial \sigma} (\mathbf{x}_p)
                \\
                &
                    \mathbf{L} \equiv
                        \frac{3}{\sigma}
                        \left( \boldsymbol\Gamma_p \cdot \nabla \right)
                        \left(
                            \mathbf{u} (\mathbf{x}_p) - \filter{\mathbf{u}} (\mathbf{x}_p)
                        \right)
                        +
                        \left( \boldsymbol\Gamma_p \cdot \nabla \right)
                        \frac{\partial \filter{\mathbf{u}} }{\partial \sigma}  (\mathbf{x}_p)
            .\end{align*}
            The relation in~\cref{eq:Cd:ML} will be the basis for our dynamic procedure.
            This procedure differs from the classic dynamic procedure developed by Germano \textit{et al.}\cite{Germano1991,MarioGenrmano1992} in that the former is motivated by the balance of derivatives between true and modeled SFS contributions given in~\cref{eq:Cd:derbalance}, while the latter is based on the Germano identity.

        \subsubsection{Enstrophy-Production Balance} \label{sec:sfs:dynamic:ensbalance}
            The procedure aims at calculating $C_d$ such as to impose the relation given in~\cref{eq:Cd:ML}, however, this is an overdetermined system as there are three equations (one for each spatial dimension) and only one unknown, $C_d$.
            Thus, the relation is now contracted by also imposing a balance of enstrophy production between true and modeled SFS contributions as follows.

            Enstrophy, a measure of the rotational kinetic energy of the flow, is defined locally as $\xi \equiv \frac{1}{2} \boldsymbol\omega \cdot \boldsymbol\omega$. The rate of local enstrophy production is then calculated as
            \begin{align*}
                    \frac{\mathrm{d} }{\mathrm{d} t} \xi
                = %
                    \boldsymbol\omega \cdot \left( \frac{\mathrm{d} }{\mathrm{d} t} \boldsymbol\omega \right)
            ,\end{align*}
            which can be decomposed between resolved and unresolved domains as
            \begin{align*}
                    \frac{\mathrm{d} }{\mathrm{d} t} \xi
                =
                    \boldsymbol\omega \cdot
                    \left(
                        \frac{\mathrm{d} }{\mathrm{d} t} \filter{\boldsymbol\omega}
                    \right)
                    +
                    \boldsymbol\omega \cdot
                    \left[
                        \frac{\mathrm{d} }{\mathrm{d} t}
                        \left( \boldsymbol\omega - \filter{\boldsymbol\omega} \right)
                    \right]
            ,\end{align*}
            The local enstrophy production in the resolved domain is then defined as
            \begin{align*}
                    \frac{\mathrm{d} }{\mathrm{d} t} \xi_r
                \equiv %
                    \boldsymbol\omega \cdot \left( \frac{\mathrm{d} }{\mathrm{d} t} \filter{\boldsymbol\omega} \right)
            ,\end{align*}
            and the global enstrophy production in the resolved domain is then calculated by integration,
            \begin{align*}
                    \int\limits_{-\infty}^\infty
                        \frac{\mathrm{d} }{\mathrm{d} t} \xi_r
                    \, \mathrm{d} \mathbf{y}
                = %
                    \int\limits_{-\infty}^\infty
                        \boldsymbol\omega \cdot \left( \frac{\mathrm{d} }{\mathrm{d} t} \filter{\boldsymbol\omega} \right)
                    \, \mathrm{d} \mathbf{y}
            .\end{align*}
            The SFS contribution to the rate of enstrophy production is isolated through operator splitting as
            \begin{align*}
                    \left(
                        \int\limits_{-\infty}^\infty
                            \frac{\mathrm{d} }{\mathrm{d} t} \xi_r
                        \, \mathrm{d} \mathbf{y}
                    \right)_\mathrm{SFS}
                \equiv %
                    \int\limits_{-\infty}^\infty
                        \boldsymbol\omega \cdot
                        \left(
                            \frac{\mathrm{d} }{\mathrm{d} t} \filter{\boldsymbol\omega}
                        \right)_\mathrm{SFS}
                    \, \mathrm{d} \mathbf{y}
            .\end{align*}

            Now, we constrain the SFS model to match the enstrophy production of the true SFS contribution,
            \begin{align*}
                    \int\limits_{-\infty}^\infty
                        \boldsymbol\omega \cdot
                        \left(
                            \frac{\mathrm{d} }{\mathrm{d} t} \filter{\boldsymbol\omega}
                        \right)_\mathrm{SFS\,model}
                    \, \mathrm{d} \mathbf{y}
                = %
                    \int\limits_{-\infty}^\infty
                        \boldsymbol\omega \cdot
                        \left(
                            \frac{\mathrm{d} }{\mathrm{d} t} \filter{\boldsymbol\omega}
                        \right)_\mathrm{true\,SFS}
                    \, \mathrm{d} \mathbf{y}
            .\end{align*}
            Using the singular particle approximation, $\boldsymbol\omega (\mathbf{x}) \approx \sum\limits_p \boldsymbol\Gamma_p \delta (\mathbf{x} - \mathbf{x}_p)$, the integrals collapse into
            \begin{align*}
                &
                    \sum\limits_p
                        \boldsymbol\Gamma_p
                        \cdot
                        \left(
                            \frac{\mathrm{d} }{\mathrm{d} t} \overline{\boldsymbol\omega}
                        \right)_{(\mathbf{x}_p)}^\mathrm{SFS\,model}
                = %
                    \sum\limits_p
                        \boldsymbol\Gamma_p
                        \cdot
                        \left(
                            \frac{\mathrm{d} }{\mathrm{d} t} \overline{\boldsymbol\omega}
                        \right)_{(\mathbf{x}_p)}^\mathrm{true\,SFS}
            .\end{align*}
            One instance that this equality is satisfied is when each term in the sum satisfies
            \begin{align*}
                &
                    \boldsymbol\Gamma_p
                    \cdot
                    \left(
                        \frac{\mathrm{d} }{\mathrm{d} t} \overline{\boldsymbol\omega}
                    \right)_{(\mathbf{x}_p)}^\mathrm{SFS\,model}
                = %
                    \boldsymbol\Gamma_p
                    \cdot
                    \left(
                        \frac{\mathrm{d} }{\mathrm{d} t} \overline{\boldsymbol\omega}
                    \right)_{(\mathbf{x}_p)}^\mathrm{true\,SFS}
            .\end{align*}
            Differentiating with respect to the filter width, we arrive to
            \begin{align} \label{eq:ensbalance}
                    \boldsymbol\Gamma_p
                    \cdot
                    \frac{\partial }{\partial \sigma}
                    \left(
                        \frac{\mathrm{d} }{\mathrm{d} t} \overline{\boldsymbol\omega}
                    \right)_{(\mathbf{x}_p)}^\mathrm{SFS\,model}
                = %
                    \boldsymbol\Gamma_p
                    \cdot
                    \frac{\partial }{\partial \sigma}
                    \left(
                        \frac{\mathrm{d} }{\mathrm{d} t} \overline{\boldsymbol\omega}
                    \right)_{(\mathbf{x}_p)}^\mathrm{true\,SFS}
            .\end{align}
            When SFS advection is neglected, the SFS model and the true SFS become
            \begin{align*}
                &
                    \left(
                        \frac{\mathrm{d} }{\mathrm{d} t} \overline{\boldsymbol\omega}
                    \right)^\mathrm{SFS\,model}
                =
                    -C_d \mathbf{E}_\mathrm{str}
                \\
                &
                    \left(
                        \frac{\mathrm{d} }{\mathrm{d} t} \overline{\boldsymbol\omega}
                    \right)^\mathrm{true\,SFS}
                =
                    {
                        \filter{ (\boldsymbol{\omega} \cdot \nabla )\mathbf{u} }
                        -
                        (\filter{ \boldsymbol{\omega} } \cdot \nabla )\filter{ \mathbf{u} }
                    }
            ,\end{align*}
            thus~\cref{eq:ensbalance} is simply the dot-product of~\cref{eq:Cd:derbalance} with $\boldsymbol\Gamma_p$,
            \begin{align*}
                    C_d
                    \boldsymbol\Gamma \cdot
                    \frac{\partial \mathbf{E}_\mathrm{str} }{\partial \sigma}
                =
                    -
                    \boldsymbol\Gamma \cdot
                    \frac{\partial }{\partial \sigma}
                    \left[
                        \filter{ (\boldsymbol{\omega} \cdot \nabla )\mathbf{u} }
                        -
                        (\filter{ \boldsymbol{\omega} } \cdot \nabla )\filter{ \mathbf{u} }
                    \right]
            .\end{align*}

            Finally, following the same steps that led from~\cref{eq:Cd:derbalance} to~\cref{eq:Cd:ML}, the enstrophy balance in~\cref{eq:ensbalance} becomes
            \begin{align*}
                    C_d \boldsymbol\Gamma_p \cdot \mathbf{m} = \boldsymbol\Gamma_p \cdot \mathbf{L}
            ,\end{align*}
            and $C_d$ is then calculated dynamically as
            \begin{align} \label{eq:Cd:GammaML}
                        C_d
                    =
                        \frac{
                            \boldsymbol\Gamma_p \cdot \mathbf{L}
                        }{
                            \boldsymbol\Gamma_p \cdot \mathbf{m}
                        }
            .\end{align}
            Thus, this $C_d$ calculated at the position of every particle is the coefficient that approximates the derivative balance while satisfying a local balance of enstrophy production between the model and the true SFS contribution.

        \subsubsection{Lagrangian Average}
            Similar to the classic procedure based on the Germano identity \cite{Germano1991}, \cref{eq:Cd:GammaML} poses numerical issues when the denominator is close to zero, leading to large fluctuations.
            In the classic procedure, Meneveau \textit{et al.} \cite{Meneveau1996} addressed this issue by integrating both numerator and denominator along Lagrangian trajectories (pathlines), effectively building ensemble averages.
            Applying this technique, our dynamic procedure becomes
            \begin{align} \label{eq:Cd:GammaMLave}
                    C_d
                =
                    \frac{
                        \ave{\boldsymbol\Gamma_p \cdot \mathbf{L}}
                    }{
                        \ave{\boldsymbol\Gamma_p \cdot \mathbf{m}}
                    }
            ,\end{align}
            where $\ave{\cdot}$ denotes the Lagrangian integration.
            As suggested by Meneveau \textit{et al.}, the integration is performed as a relaxation process at every time step of the form
            \begin{align*}
                    \ave{\phi}_\mathrm{new}
                =
                    (1 - \alpha)\ave{\phi}_\mathrm{old} + \alpha\phi
            ,\end{align*}
            where $\alpha$ is calculated as $\alpha = \Delta t / T \leq 1$, $\Delta t$ is the time step of the simulation, and $T$ is the time length of the ensemble average.

    \subsection{Backscatter Control} \label{sec:sfs:backscatter}
        For stability reasons, it is common in LES to control the amount of energy being backscattered from the unresolved scales into the resolved scales.
        The SFS term is purely-dissipative if its contributions to the enstrophy budget decrease the total enstrophy.
        This is,
        \begin{align*}
                \left(
                    \int\limits_{-\infty}^\infty
                        \frac{\mathrm{d} }{\mathrm{d} t} \xi_r
                    \, \mathrm{d} \mathbf{y}
                \right)_\mathrm{SFS}
            \leq
                0
        .\end{align*}
        Following the derivation in~\cref{sec:sfs:dynamic:ensbalance}, this condition is satisfied if
        \begin{align*}
                \boldsymbol\Gamma_p
                \cdot
                \left(
                    \frac{\mathrm{d} }{\mathrm{d} t} \overline{\boldsymbol\omega}
                \right)_{(\mathbf{x}_p)}^\mathrm{SFS\,model}
            \leq
                0
        \end{align*}
        for each particle, or equivalently,
        \begin{align*}
                C_d
                \boldsymbol\Gamma_p
                \cdot
                \mathbf{E}_\mathrm{str} (\mathbf{x}_p)
            \geq
                0
        \end{align*}
        Therefore, enstrophy backscatter can be filtered out at each particle by clipping the model coefficient as $C_d = 0$ whenever the condition shown above is not satisfied.

    \subsection{Usage in Conventional Mesh-Based CFD} \label{sec:sfs:meshcfd}
        Even though our SFS model is tailored for vortex methods as it only uses the primitive variables of the VPM, the model can also be readily applied to conventional mesh-based CFD as follows.
        The vortex strength is expressed from mesh quantities as $\boldsymbol\Gamma_q = \ave{\boldsymbol\omega}_q \mathrm{Vol}_q$, where $\ave{\boldsymbol\omega}_q$ and $\mathrm{Vol}_q$ are the average vorticity and volume associated to each element in the grid (\textit{i.e.}, cells in a finite volume method or nodes in a finite difference method).
        The model in the vorticity transport equation then becomes
        \begin{align*}
            &
                \frac{\partial T_{ij}}{\partial x_j} \left( \mathbf{x} \right)
            =
            \\
            &
                C_d \Delta^{-3}
                \sum\limits_q
                    \ave{\omega}_j^q \mathrm{Vol}_q
                    \left(
                        \frac{\partial \filter{u_i}}{\partial x_j} \left( \mathbf{x}_q \right)
                        - \frac{\partial \filter{u_i} }{\partial x_j} \left( \mathbf{x} \right)
                    \right)
                    \zeta \left( \frac{\Vert \mathbf{x}-\mathbf{x}_q \Vert}{\Delta} \right)
        ,\end{align*}
        where the index $q$ iterates over each element in the grid, $\Delta$ is the width of the grid filter, and $\zeta$ is the filter kernel.
        Furthermore, the model can be used in the pressure-velocity form of the momentum equation noticing that the SFS terms in the vorticity transport equation are simply the curl of the SFS term in the pressure-velocity equation, $\frac{\partial \tau_{ij}}{\partial x_j}$, as
        \begin{align*}
                \epsilon_{ijk} \frac{\partial }{\partial x_j} \frac{\partial \tau_{kl}}{\partial x_l}
            =
                \frac{\partial T'_{ij}}{\partial x_j}
                -
                \frac{\partial T_{ij}}{\partial x_j}
        ,\end{align*}
        where $\epsilon_{ijk}$ is the Levi-Civita tensor.
        Hence, the SFS models of the vorticity transport equation can be used to model $\frac{\partial \tau_{ij}}{\partial x_j}$ by ``un-curling'' the equation above.

    \section{Results} \label{sec:res}

  In the preceding sections we have developed a scheme for numerically solving the LES-filtered Navier-Stokes equations in their vorticity form.
  The proposed scheme uses a reformulation of the VPM and a novel model of SFS vortex stretching to achieve a meshless large eddy simulation.

  We now proceed to test and validate both the VPM reformulation and the vortex-stretching SFS model that comprise our meshless LES.
  Simulations are compared to results reported in the literature from experimental work, direct numerical simulation (DNS), the lattice Boltzmann method (LBM), and unsteady Reynolds-average Navier-Stokes (URANS) simulation.
  These validation cases also serve as examples on how to impose initial and boundary conditions in our meshless scheme.

  In~\cref{sec:res:isolatedring}, advection and viscous diffusion is validated on an isolated vortex ring, which is also used to test convergence of the reformulated VPM.
  In~\cref{sec:res:leapfrog}, vortex stretching is validated through the simulation of two leapfrogging vortex rings.
  Results with classic and reformulated VPM are first compared without SFS effects.
  The SFS model is then introduced to capture the subfilter energy cascade that leads the rings into turbulent breakdown.
  In~\cref{sec:res:jet}, a turbulent round jet is simulated to test the accuracy and numerical stability of the LES scheme.
  The evolution of the jet and predicted Reynolds stress are compared to experimental measurements, validating the scheme as an LES method accurately resolving large-scale features of turbulent flow.
  Finally, in~\cref{sec:res:rotor}, our meshless LES is compared to conventional mesh-based CFD in a real-world engineering problem: an aircraft rotor and its wake.

\subsection{Vortex Ring} \label{sec:res:isolatedring}

\begin{figure}[t!]
        \centering
        \includegraphics[width=\figwidth]{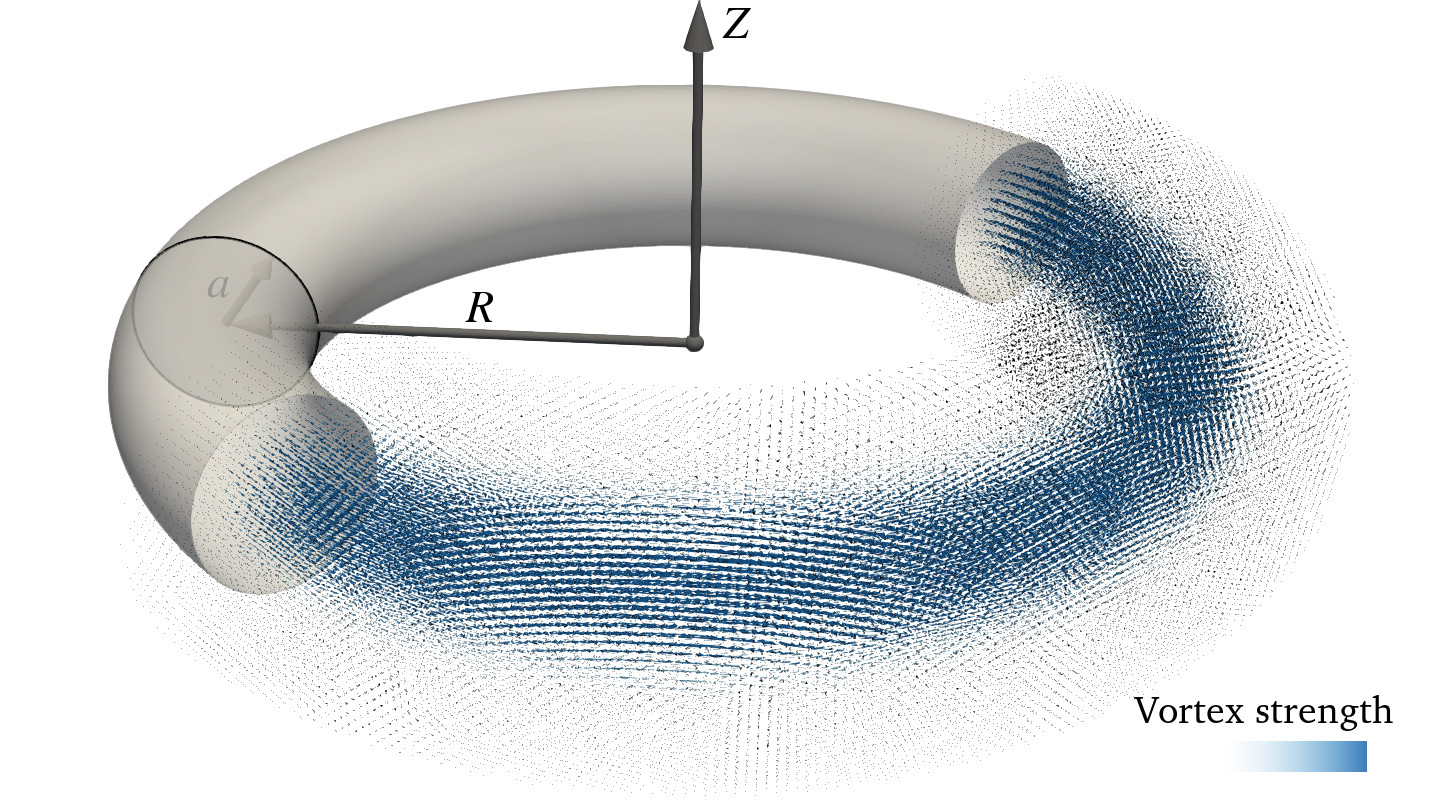}
        \caption{Vortex ring as discretized through particles with $\Delta x = 0.13a$.}
        \label{fig:vortexrings:visual}
    \end{figure}

    A vortex ring travels with a self-induced velocity that slowly decelerates as its vorticity spreads due to viscous diffusion.
    This poses a good validation case for both vorticity advection and viscous diffusion.
    The initial vorticity inside the ring core is assumed to follow a Gaussian distribution of the form ${\omega_\theta (r) = \frac{\Gamma_0}{\pi a^2} e^{-r^2/a^2}}$, where $\Gamma_0$ and $a$ are the ring's initial circulation and core size, respectively.
    The initial Reynolds number is defined as $\mathrm{Re}=\Gamma_0/\nu$.
    In all the following simulations, the rings were initially discretized by placing particles (evenly spaced by $\Delta x$) everywhere that the local vorticity was larger than 5\% of the peak vorticity $\omega_\theta(0)$, and then performing a radial-basis function (RBF) fit to the target vorticity $\omega_\theta$ to determine the initial vortex strength of the particles.
    The initial particle size $\sigma$ was set as to provide a particle overlap $\lambda=\frac{\sigma}{\Delta x}$ of 2.4.
    \cref{fig:vortexrings:visual} shows a discretized vortex ring along with a depiction of the ring's geometric parameters.

    A vortex ring at $\mathrm{Re}=7500$ and thickness $\frac{a}{R}=0.2$ was simulated and compared to direct-numerical simulation (DNS) reported by Archer \textit{et al.} \cite{Archer2008a}
    The DNS is a finite-difference method with second-order spatial accuracy solving the incompressible pressure-velocity form of the Navier Stokes equations, using over 100 million grid cells.

    \begin{figure}[t!]
        \raggedright
        \includegraphics[width=\figwidth]{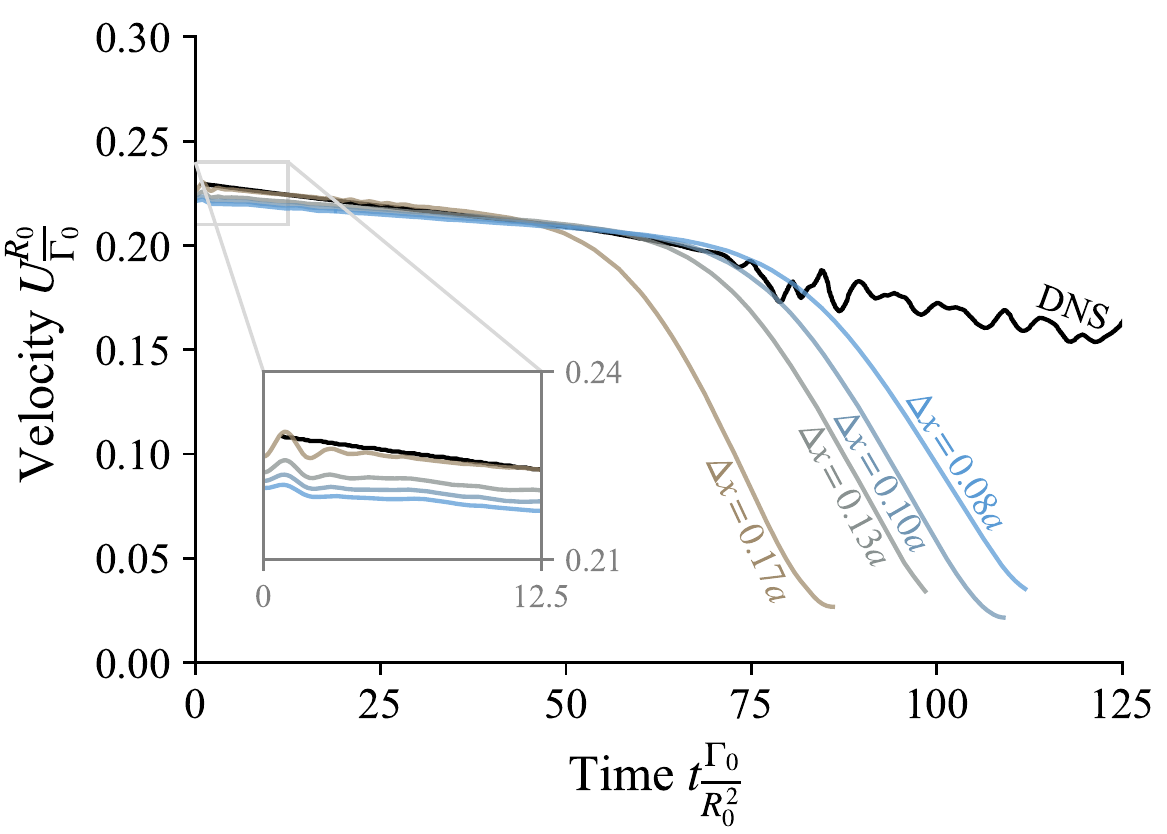}
        \caption{Spatial convergence of vortex ring simulated with reformulated VPM, compared to direct numerical simulation (DNS).}
        \label{fig:vortexrings:convergence}
    \end{figure}

    Convergence of the reformulated VPM was tested without the SFS model by increasing the spatial resolution from $\Delta x = 0.17a$ (resulting in 60k particles) to $\Delta x = 0.08a$ (resulting in 500k particles), shown in \cref{fig:vortexrings:convergence}.
    To minimize numerical noise, we suppressed the core-resetting step of the viscous diffusion scheme \cite{Barba2005a}, which leads the VPM simulation to become increasingly coarser as time goes by.
    For this reason, and without the SFS model, the VPM simulation is under-resolved and inaccurate in the turbulent regime
    This is observed in \cref{fig:vortexrings:convergence} as an abrupt turbulent breakdown after transition.
    In the laminar regime (nondimensional time between 0 and 50), the velocity shows little sensitivity to a spatial discretization finer than $\Delta x = 0.13a$, being between 4\% of the DNS velocity at time 0, as shown in the zoom-in box of~\cref{fig:vortexrings:convergence}.
    The DNS shows a slow transition from laminar to turbulent starting at nondimensional time 75, while the VPM also transitions around time 75 for $\Delta x \le 0.13a$, quickly leading the ring into turbulent breakdown.
    Hence, $\Delta x = 0.13a$ was used in all the simulations that follow, which results in 150k particles.
    These 150k vortex elements, contrasted with the 100 million grid elements used in the DNS, highlight the computational efficiency of our meshless LES, which enabled us to run our simulations in only a matter of hours on a desktop workstation.

\begin{figure}[t!]
        \raggedright
        \includegraphics[width=\figwidth]{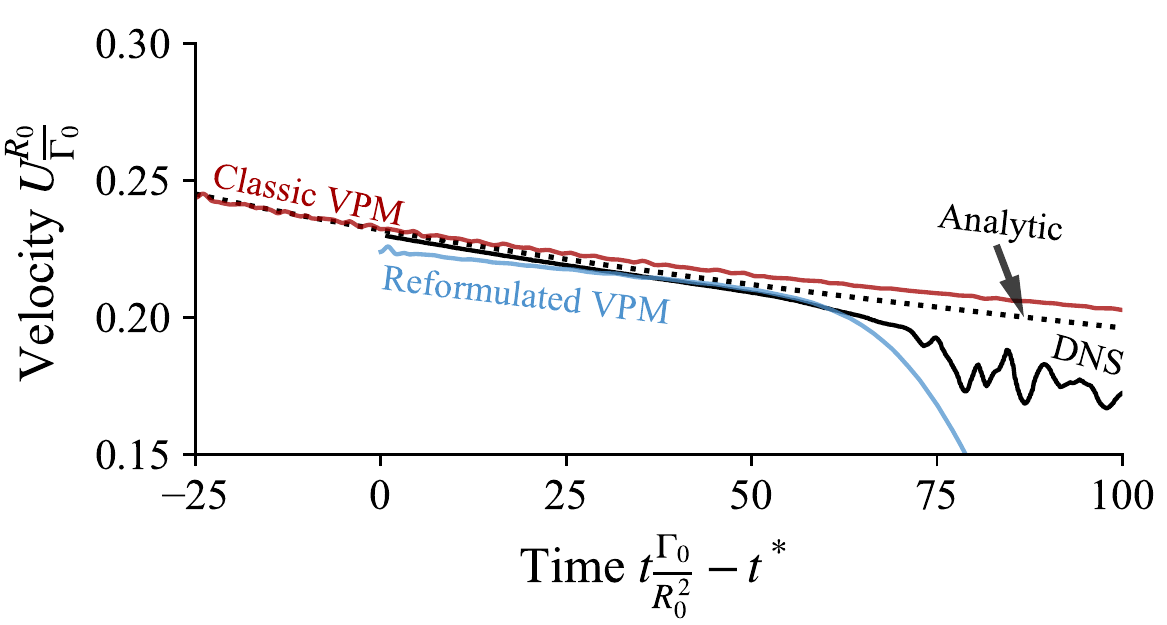}
        \caption{Vortex ring simulated with classic VPM ($t^*=25$) and reformulated VPM ($t^*=0$), compared to direct numerical simulation (DNS, $t^*=25$) and analytic solution ($t^*=25$).}
        \label{fig:vortexrings:Ucomparison}
    \end{figure}

    \cref{fig:vortexrings:Ucomparison} compares the DNS to classic and reformulated VPM simulations, along with the analytic velocity of a viscous Gaussian vortex ring (derived in~\cref{sec:app:vortexring}).
    Archer \textit{et al.} \cite{Archer2008a} identified that the initial Gaussian distribution goes through a transient phase until achieving a steady skewed distribution.
    They denoted the time that the simulation takes to achieve this steady distribution as $t^*$, reporting $t^*=25$ for their DNS.
    Similarly, the VPM simulations were also observed to have a phase of transient vorticity distribution.
    In \cref{fig:vortexrings:Ucomparison} both the classic VPM simulation and the analytic solution were shifted by $t^*=25$, showing good agreement with the DNS velocity at time 0. However, the classic VPM slowly drifts away from both the analytic and DNS velocity after time 0.
    In contrast, the reformulated VPM starts off with a velocity between 4\% of the DNS velocity at time 0 and maintains good agreement up to turbulent breakdown.
    This confirms that the reformulated VPM accurately resolves vorticity advection and viscous diffusion in the laminar regime.

\subsection{Leapfrogging Vortex Rings} \label{sec:res:leapfrog}

    Two vortex rings that travel in the same direction can repeatedly leapfrog, stretching and contracting as they pass one another.
    This poses a good validation case for vortex stretching.
    Two identical vortex rings with initial radii $R_0=1$, thickness $\frac{a}{R_0}=0.1$, and separation $\Delta Z=R_0$ were simulated at $\mathrm{Re}=3000$ and compared to the simulation through lattice Boltzmann method (LBM) reported by Cheng \textit{et al.} \cite{Cheng2015}
    The LBM simulation is a finite-difference DNS with second order spatial accuracy solving the generalized lattice-Boltzmann equation, reportedly using 15 billion cells.\cite{Cheng2010}
    In contrast, the VPM simulations used only 600k particles, which again highlights the computational efficiency of our meshless LES.

\begin{figure}[t]
        \raggedright
        \includegraphics[width=\midlargerfigwidth]{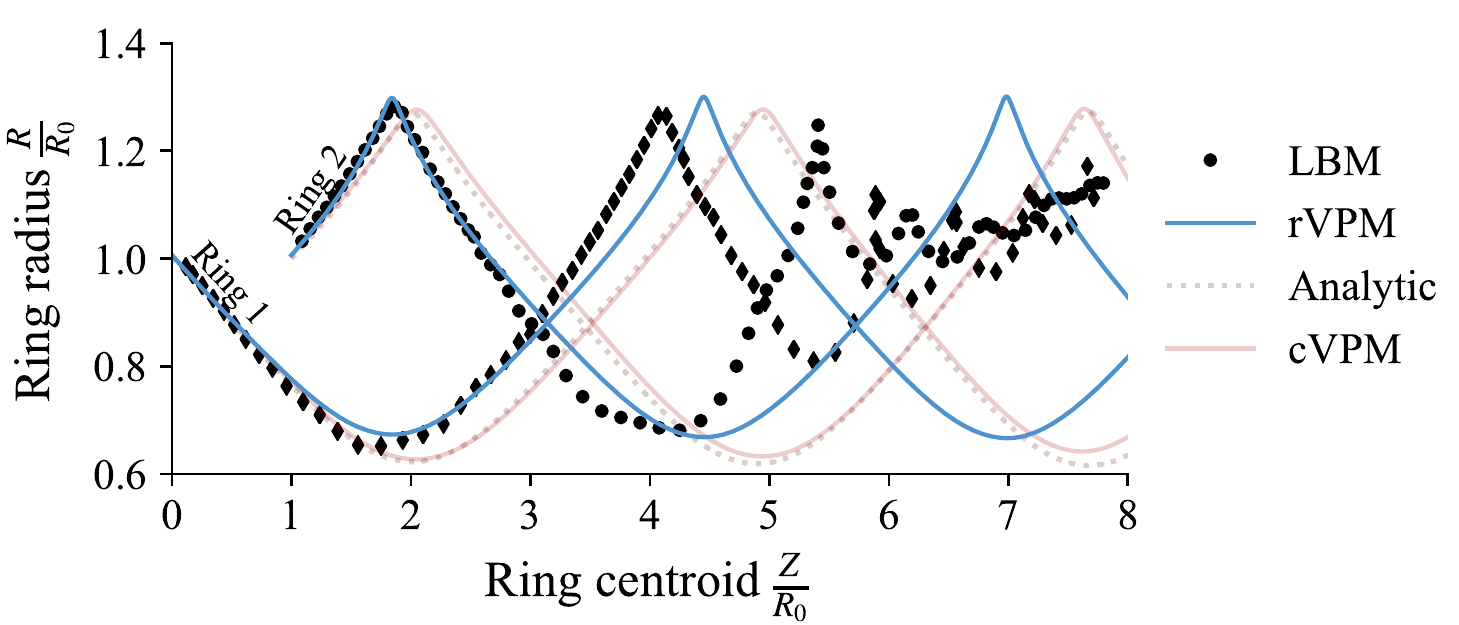}
        \caption{Leapfrogging vortex rings simulated with classic VPM (cVPM) and reformulated VPM (rVPM) without SFS model, compared to lattice Boltzmann method (LBM) and analytic solution.}
        \label{fig:leapfrog:RZcomparison:formulation}
    \end{figure}

    First, the simulation was performed on both classic and reformulated VPM without the SFS model, shown in \cref{fig:leapfrog:RZcomparison:formulation}.
    Reformulated VPM and LBM show good agreement in the first leapfrog ($Z/R_0 \le 3$), but they deviate in subsequent cycles ($Z/R_0 > 3$) once turbulence starts to arise in the LBM causing the rings to slow down and run into each other.
    This is not captured in the reformulated VPM as the simulation is under-resolved (it does not resolve all the scales of turbulent motion) and leads to a laminar behavior, which indicates the need for an SFS model.
    Before incorporating the SFS model, however, we conclude that the reformulated VPM accurately resolves vortex stretching in the laminar regime.

    \cref{fig:leapfrog:RZcomparison:formulation} also shows the classic VPM simulation and compares it to an analytic solution derived in~\cref{sec:app:vortexring}.
    The classic VPM leads to dynamics that are substantially different from both reformulated VPM and LBM,
    but it matches the analytic solution.
    As described in~\cref{sec:app:vortexring}, the analytic solution assumes that  (1) the core size $a$ is not affected by the stretching of the ring and that (2) the core size does not affect the interactions between rings.
    The former clearly violates conservation of mass, while the latter is suspected to violate conservation of angular momentum.
    Such good agreement between the classic VPM and this unphysical analytic solution supports our previous claim in~\cref{sec:reformulatedvpm:physicalimplications:implications} that the classic VPM may lead to unphysical results, which may also cause the method to be numerically unstable.

    \begin{figure*}[]

        \vspace{3mm}

        \centering
        \includegraphics[width=0.83\textwidth]{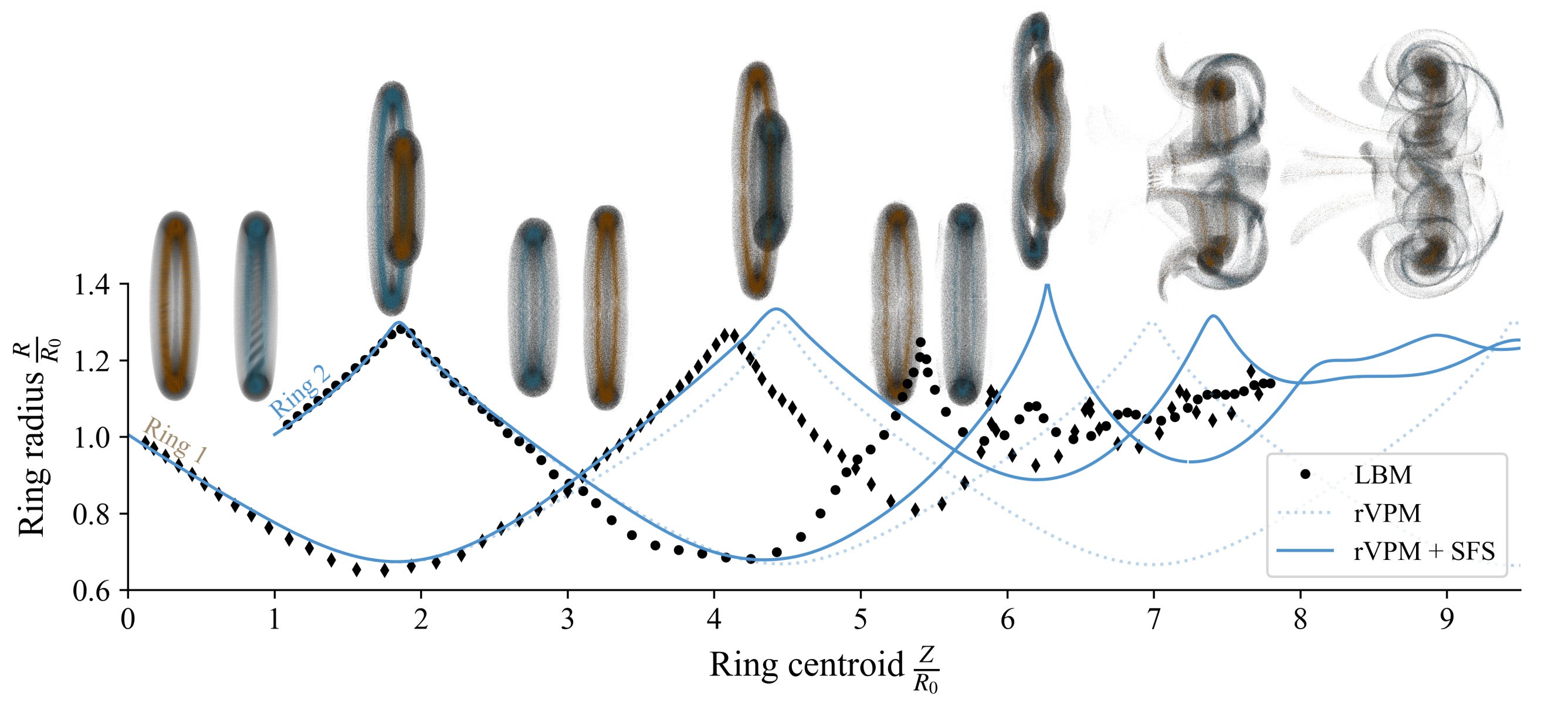}
        \caption{Leapfrogging of vortex rings using reformulated VPM with and without SFS model, compared to lattice Boltzmann method (LBM). Visualization of rVPM+SFS simulation overlaid on top.}
        \label{fig:leapfrog:RZcomparison:SFS}
    \end{figure*}

        \begin{figure*}[]

            \vspace{3mm}

                \begin{subfigure}{0.06\textwidth} \centering
                    \includegraphics[width=\linewidth]{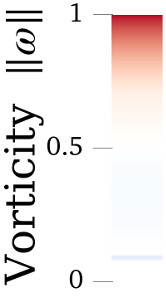}
                \end{subfigure}
                \begin{subfigure}{0.15\textwidth} \centering
                    \includegraphics[width=\linewidth]{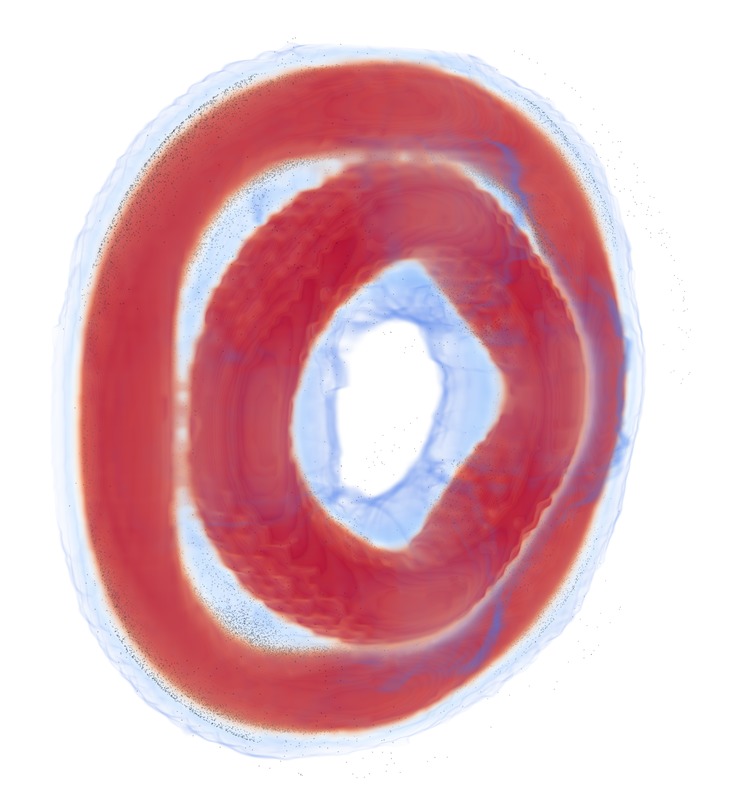}
                \end{subfigure}
                \begin{subfigure}{0.15\textwidth} \centering
                    \includegraphics[width=\linewidth]{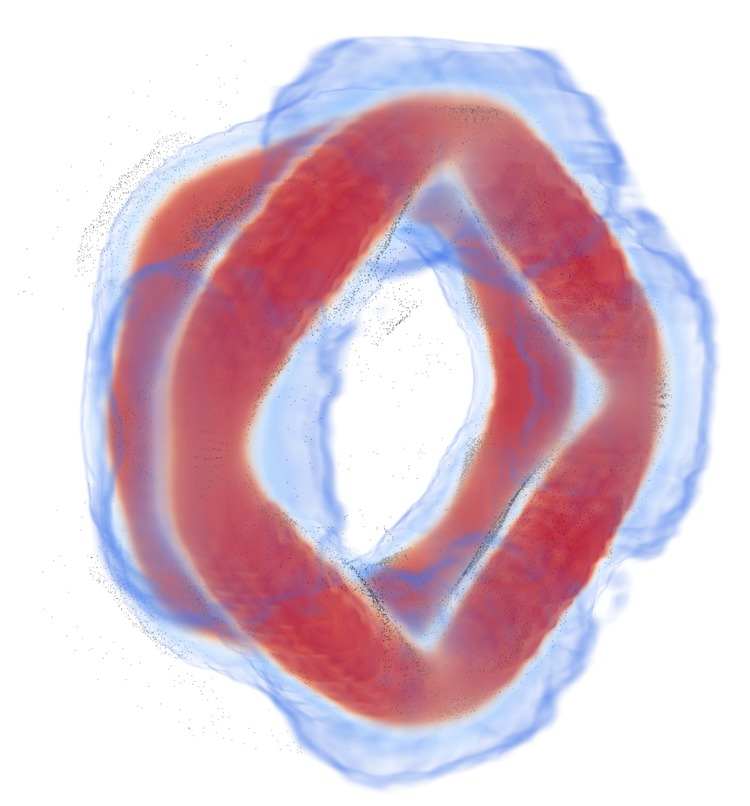}
                \end{subfigure}
                \begin{subfigure}{0.15\textwidth} \centering
                    \includegraphics[width=\linewidth]{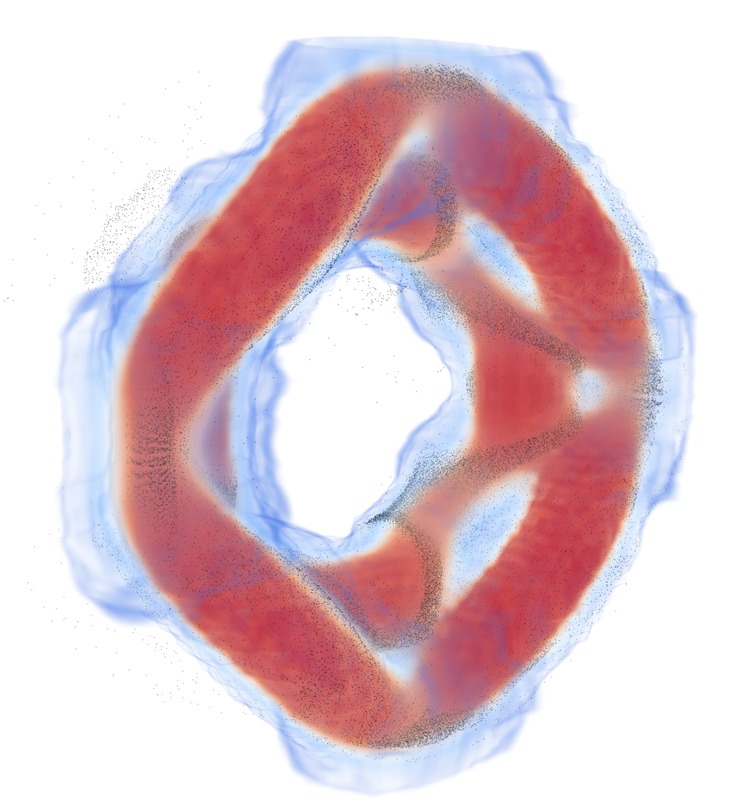}
                \end{subfigure}
                \begin{subfigure}{0.15\textwidth} \centering
                    \includegraphics[width=\linewidth]{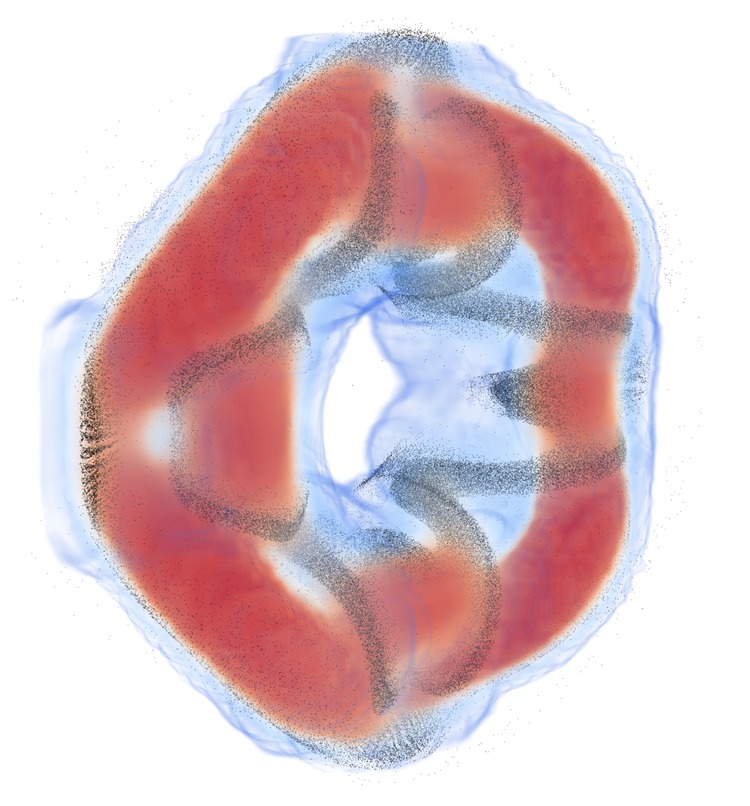}
                \end{subfigure}
                \begin{subfigure}{0.15\textwidth} \centering
                    \includegraphics[width=\linewidth]{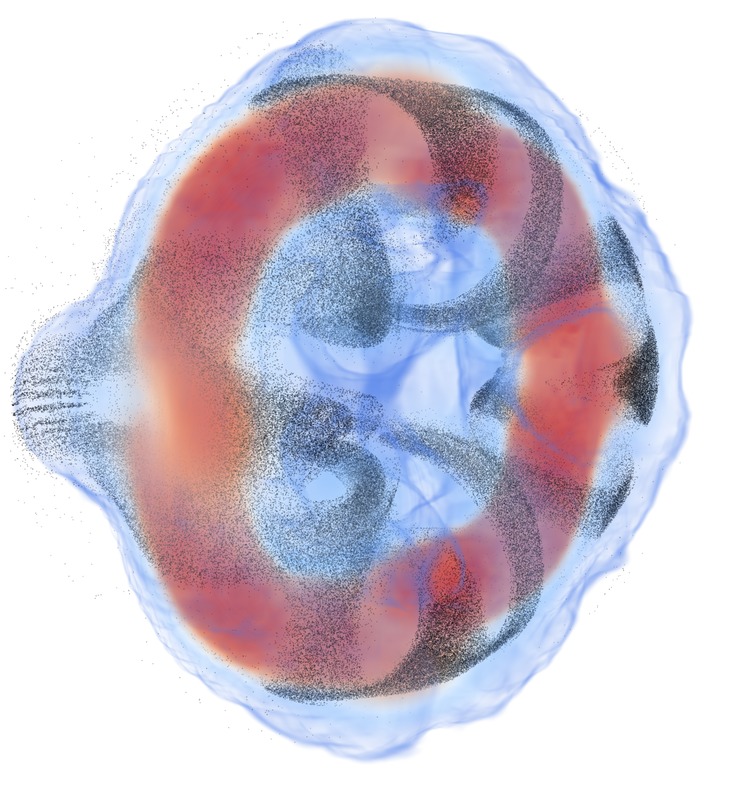}
                \end{subfigure}
                \begin{subfigure}{0.15\textwidth} \centering
                    \includegraphics[width=\linewidth]{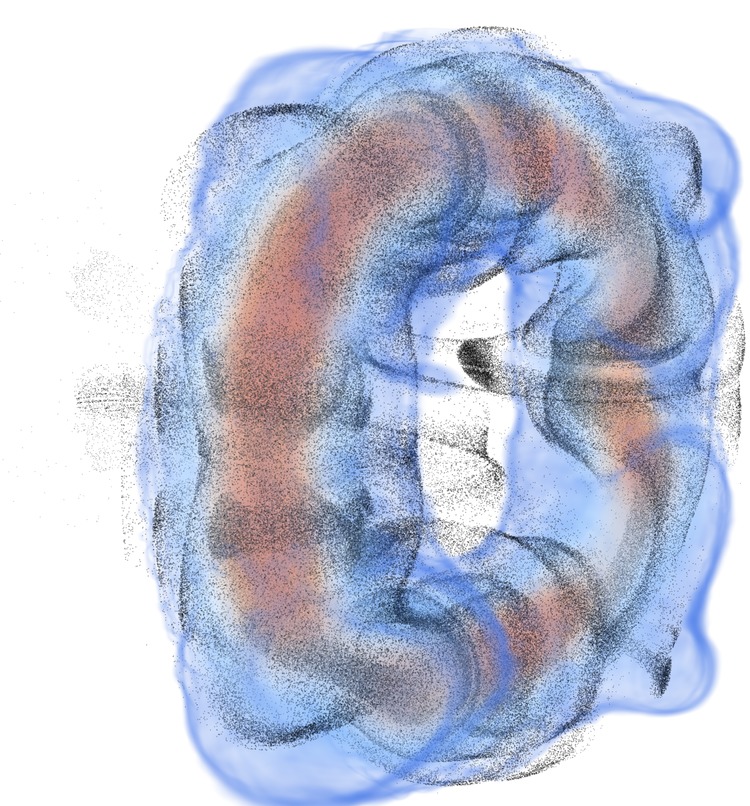}
                \end{subfigure}
            \caption{Volume rendering of vorticity field as the leapfrogging rings merge. Particles shown as black points. Video available at \href{https://youtu.be/viE-i0KzVOw}{https://youtu.be/viE-i0KzVOw} in the preprint version of this paper.}
            \label{fig:leapfrog:fdom}


        \end{figure*}

    Finally, the leapfrog simulation was repeated with the SFS model added to the reformulated VPM, shown in~\cref{fig:leapfrog:RZcomparison:SFS}.
    Notice that the SFS model only affects the dynamics after the first leapfrog ($Z/R_0 > 3$) once turbulence starts to develop.
    This confirms that the dynamic procedure succeeds at auto-regulating the SFS model, suppressing it in the laminar regime while activating it in the turbulent regime.
    The diffusivity of the SFS model slows the rings down, eventually running into each other and mixing after three leapfrogs at $Z/R_0 > 6$ (shown in more detail in~\cref{fig:leapfrog:fdom}).
    As seen in~\cref{fig:leapfrog:RZcomparison:SFS}, the process of slowing down and running into each other happens at a different pace between the rVPM+SFS simulation and the LBM.
    It is unclear whether this discrepancy is due to our vortex-stretching SFS model lacking the full range of turbulent diffusion (requiring the addition of a vorticity-advection SFS model), or due to excessive numerical dissipation in the LBM.
    However, they are both consistent in that the rings mix after three leapfrog cycles, showing the ability of the SFS model to capture diffusion associated with vortex stretching in the turbulent regime.

    In order to estimate the computational cost of the VPM reformulation and the SFS model, \cref{table:leapfrog:benchmark} compares the computational time of each simulation to the classic VPM.
    The VPM reformulation runs as fast as the classic VPM, adding no significant overhead.
    The SFS model adds a computational overhead of 8\% when a constant model coefficient is prescribed and 43\% with the dynamic procedure.
    Comparing this to a benchmark study by Chapelier \textit{et al.} \cite{Chapelier2018}, we conclude that the overhead of our SFS model is comparable to a Smagorinsky eddy-viscosity model, while the dynamic procedure adds an overhead comparable to a Germano-identity dynamic model.
    The work of Chapelier \textit{et al.} suggests that further speedup could be achieved by implementing a sensor function, but this is left for future work.

    \begin{table}[]

        \centering
        \caption{Computational time of the leapfrog VPM simulations.}
        \label{table:leapfrog:benchmark}
        \begin{tabular}{cccc}
            \hline
            \textbf{Formulation} & \,\,\textbf{SFS Model} & \,\,\textbf{CPU Time}      & \,\,\textbf{Overhead} \\ \hline
            Classic          & None               & $t_\mathrm{ref}$       & --                              \\
            Reformulated     & None               & $1.01\,t_\mathrm{ref}$ & $<$1\%                            \\
            Reformulated     & Constant $C_d$     & $1.08\,t_\mathrm{ref}$ & +8\%                            \\
            Reformulated     & Dynamic $C_d$      & $1.43\,t_\mathrm{ref}$  & +43\%                           \\ \hline
        \end{tabular}

    \end{table}

\subsection{Turbulent Round Jet} \label{sec:res:jet}

    {

        \begin{figure}[t!]

            \vspace{2mm}

            \centering
            \includegraphics[width=\figwidth]{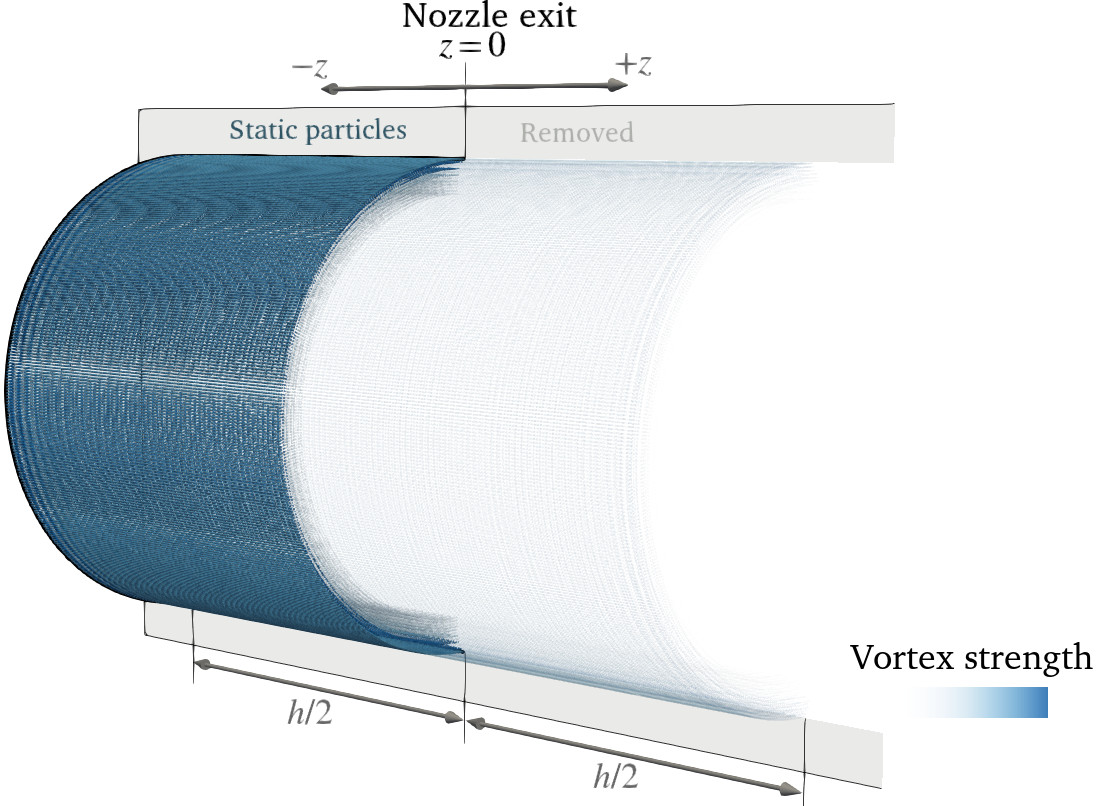}
            \caption{Boundary condition of round jet simulation defined with particles.}
            \label{fig:roundjet:visual}
        \end{figure}

        \begin{figure}[t!]

            \vspace{2mm}
            
            \raggedright
            \includegraphics[width=\figwidth]{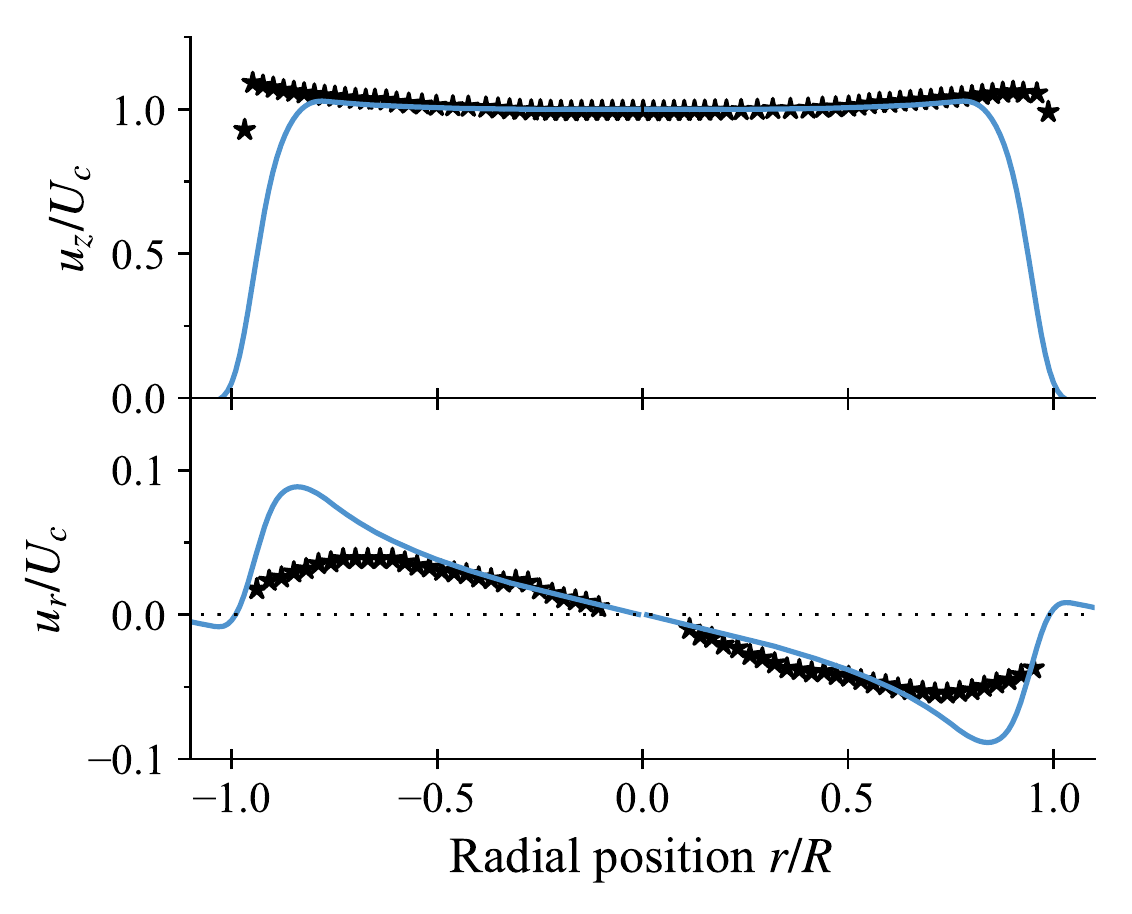}
            \caption{Streamwise (top) and radial (bottom) velocity close to nozzle exit in experiment by Quinn and Militzer \cite{Quinn1989} (markers), compared to boundary condition of simulation (solid line) probed at $z/d=0.1$.}
            \label{fig:roundjet:boundaryconditions}
        \end{figure}
    
        \begin{figure}[t!]

            \vspace{2mm}
            
            \centering
            \includegraphics[width=\figwidth]{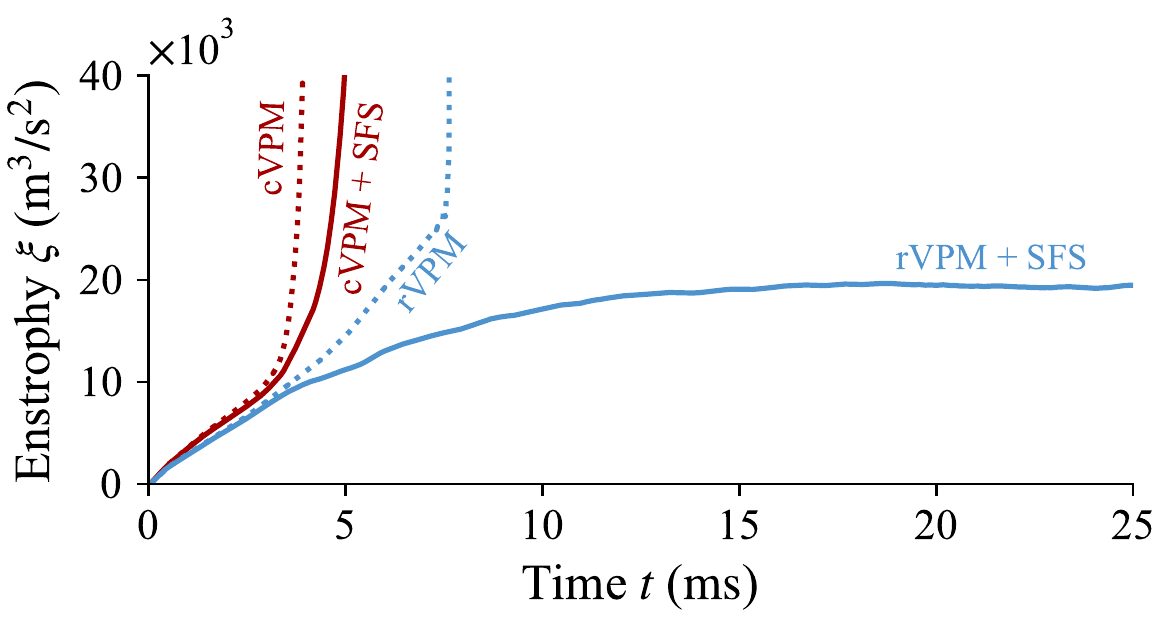}
            \caption{Global enstrophy of turbulent round jet with classic VPM (cVPM) and reformulated VPM (rVPM).}
            \label{fig:roundjet:enstrophy}
        \end{figure}
    }

    A jet discharging into a quiescent environment poses a canonical case for the study of turbulence, encompassing  a laminar region near the nozzle that breaks down into turbulence away from the nozzle.
    Experimental measurements on a round jet were used to test that the reformulated VPM is able to resolve the mean and fluctuating components of turbulent flow, while resolving Reynolds stress directly.
    The simulation replicated the experiment by Quinn and Militzer \cite{Quinn1989} which used a contoured nozzle with an exit diameter $d$ of ${45.4\,\mathrm{mm}}$, discharging air into stagnant ambient air with a centerline velocity $U_c$ of ${60\,\mathrm{m/s}}$.
    This corresponds to a Mach number of 0.18 and a diameter-based Reynolds number $\mathrm{Re} = U_c d / \nu$ of ${2 \times 10^5}$.

    The simulation assumed a top-hat velocity profile with smooth edges at the nozzle exit given by
    \begin{align*}
        u_e(r) = U_c \tanh\left( \frac{\frac{d}{2} - r}{\theta} \right)
    ,\end{align*}
    where $\theta$ is the momentum thickness of the shear layer, assumed to be $\theta=0.025d$.
    As depicted in~\cref{fig:roundjet:visual}, this velocity profile was imposed as a boundary condition at $z=0$ by defining a volumetric cylinder of length $h$ with particles spanning from $z=-h/2$ to $z=h/2$ and performing an RBF fit to the vorticity profile $\omega_\theta = \frac{\mathrm{d} u_e}{\mathrm{d} r}$.
    This computes the vortex strengths that induce the velocity profile $u_e$ inside the cylinder.
    The $+z$ half of the cylinder was then removed, the $-z$ half was kept in the computational domain as static particles throughout the simulation, while the set of particles computed at $z=0$ were injected at each time step as free particles.
    The resulting boundary condition showed sensitivity to the cylinder length $h$, leading to a straight jet when $h \ge 2d$, a contracting jet when $h < 2d $, and an actuator disk when $h \rightarrow 0$.
    Since Quinn and Militzer reported a jet that was slightly contracting, the length of the boundary-condition cylinder was tailored to match the streamwise and radial velocity that they measured at the nozzle exit plane shown in~\cref{fig:roundjet:boundaryconditions}, finding sufficient agreement when $h = 1.7d$.

    \begin{figure}[b!]
        \centering
        \includegraphics[width=\figwidth]{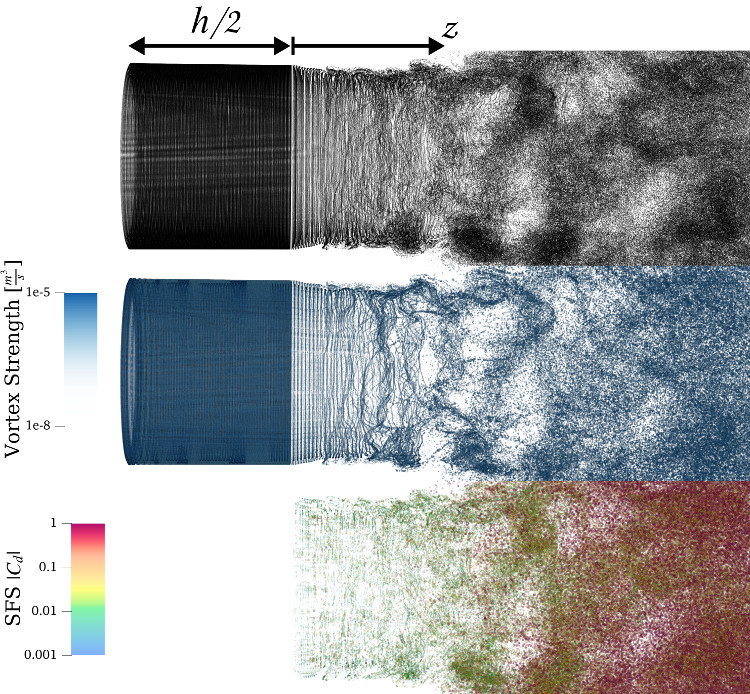}
        \caption{Vortex particles (top), vortex strength (middle), and SFS model coefficient (bottom) close to jet nozzle at $t=48\,\mathrm{ms}$.}
        \label{fig:roundjet:allviews}
    \end{figure}

    \begin{figure*}[]

        \vspace{3mm}

        \centering
        \includegraphics[width=0.83\textwidth]{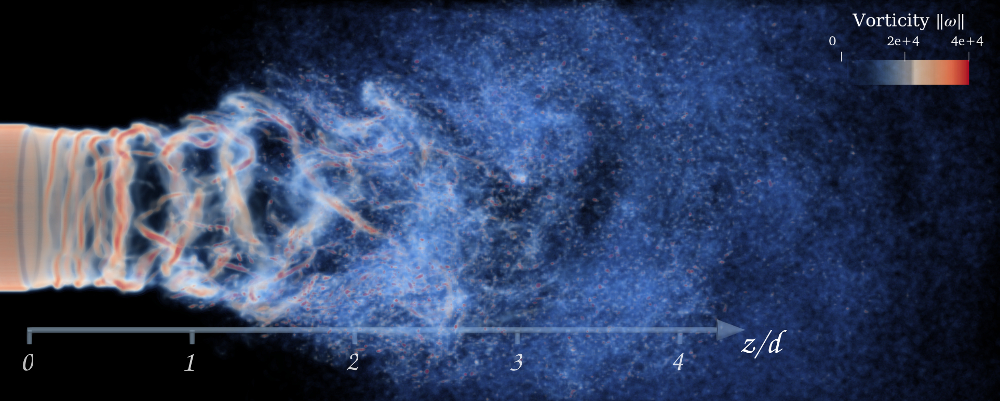}
        \caption{Volume rendering of vorticity field in turbulent jet at $t=48\,\mathrm{ms}$ showing distinct flow features: Coherent structures form in the initial region ($z<1d$) that mix and break down by $z>3d$.}
        \label{fig:roundjet:vorticity}

    \end{figure*}

    \begin{figure*}[]

        \vspace{1mm}

        \raggedright
        $t={\color{white}{0}}2\,\mathrm{ms}$ \hspace{0.4mm}
        \begin{minipage}{0.41\textwidth}
            \includegraphics[width=\linewidth]{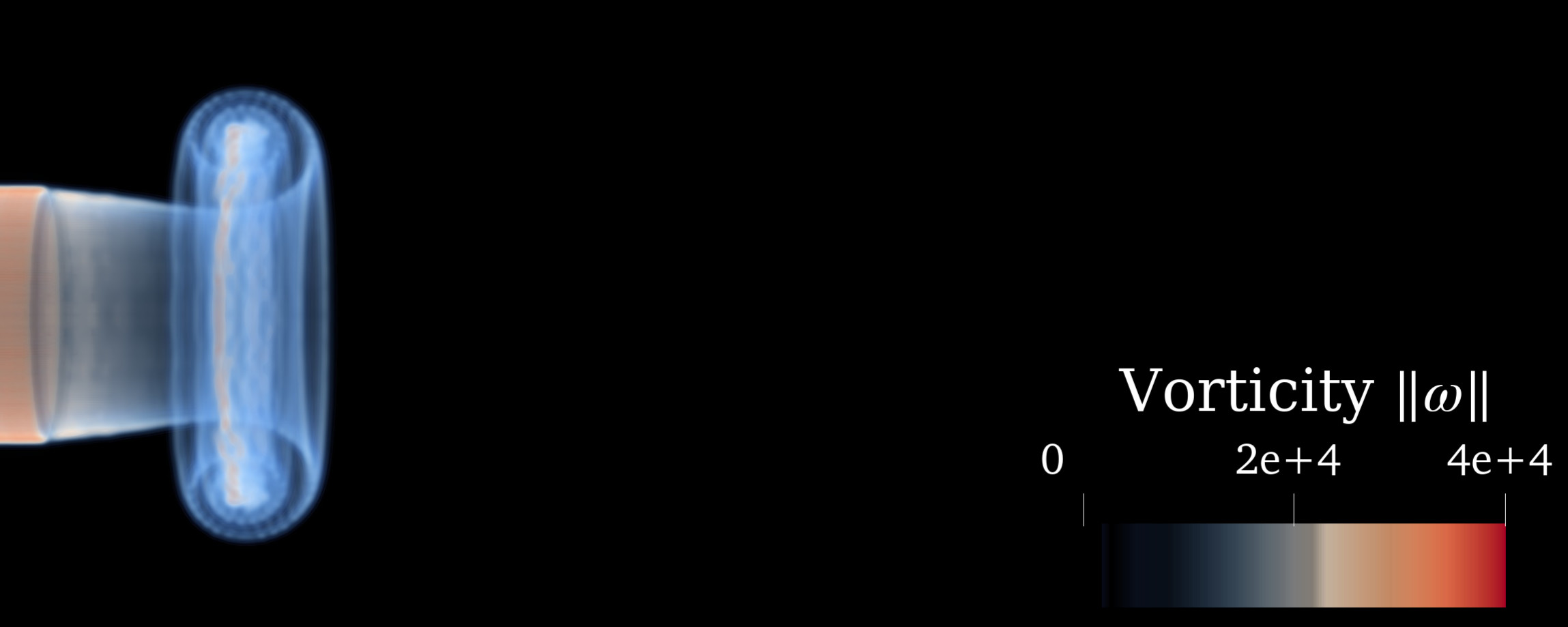}
        \end{minipage}
        \begin{minipage}{0.41\textwidth}
            \includegraphics[width=\linewidth]{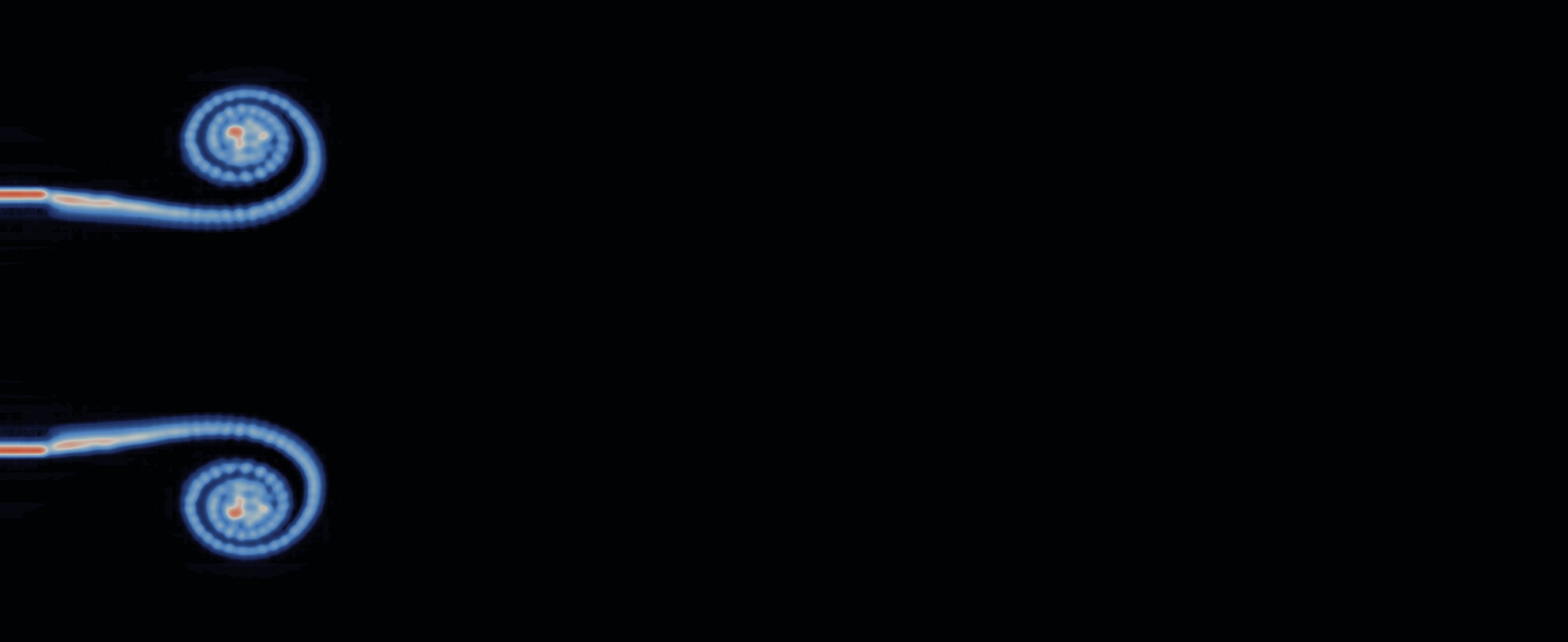}
        \end{minipage}

        \raggedright
        $t={\color{white}{0}}4\,\mathrm{ms}$ \hspace{0.4mm}
        \begin{minipage}{0.41\textwidth}
            \includegraphics[width=\linewidth]{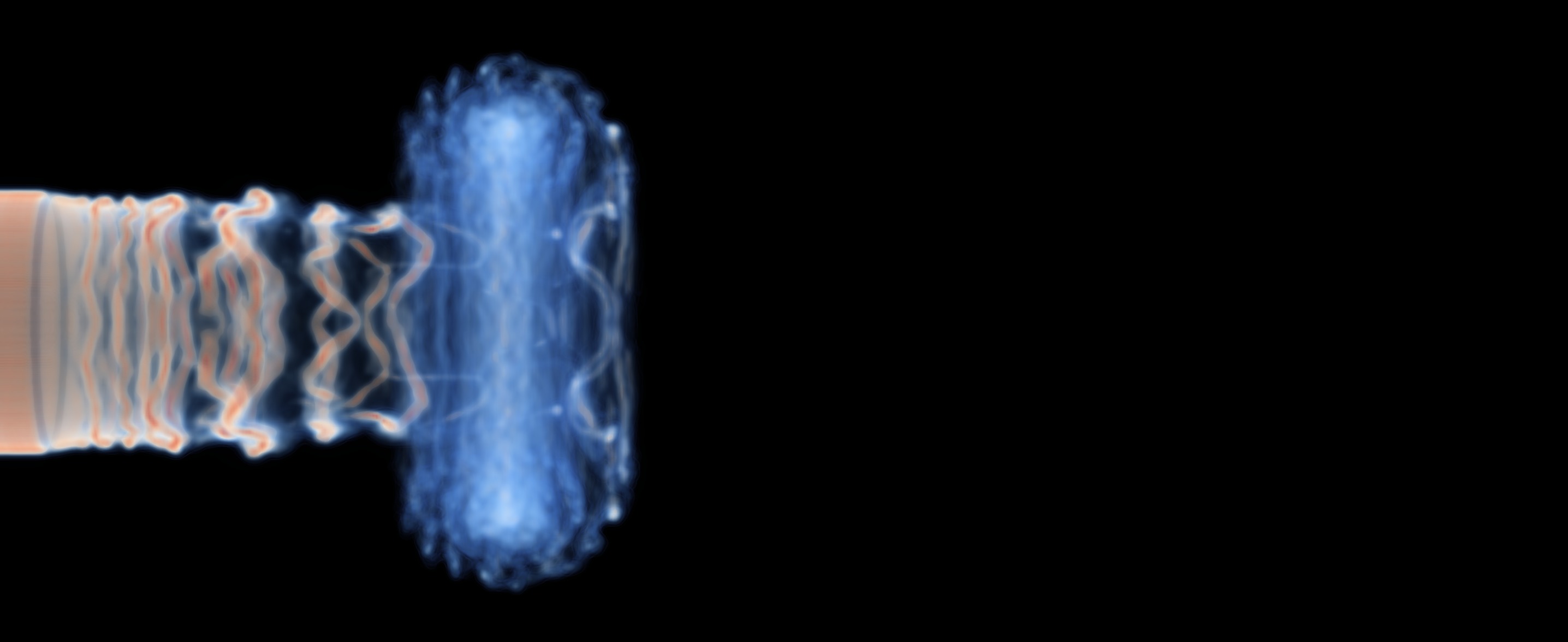}
        \end{minipage}
        \begin{minipage}{0.41\textwidth}
            \includegraphics[width=\linewidth]{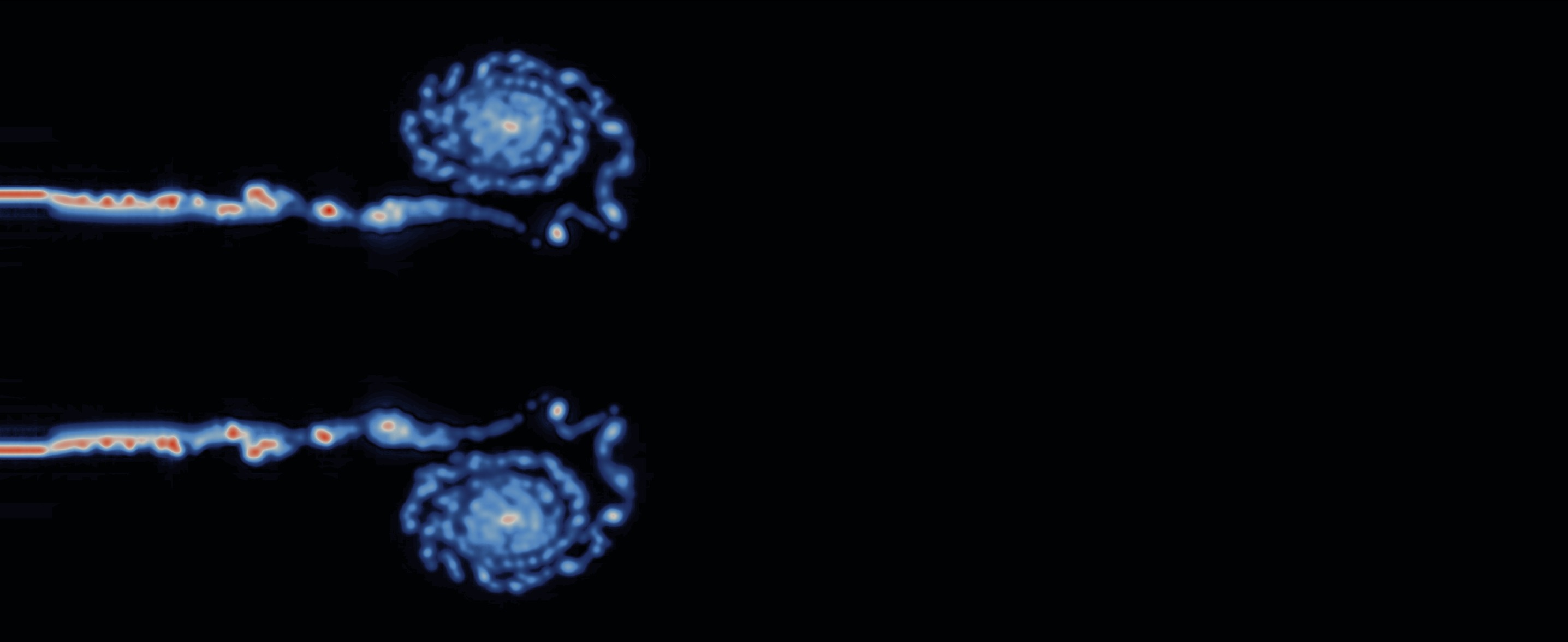}
        \end{minipage}

        \raggedright
        $t={\color{white}{0}}8\,\mathrm{ms}$ \hspace{0.4mm}
        \begin{minipage}{0.41\textwidth}
            \includegraphics[width=\linewidth]{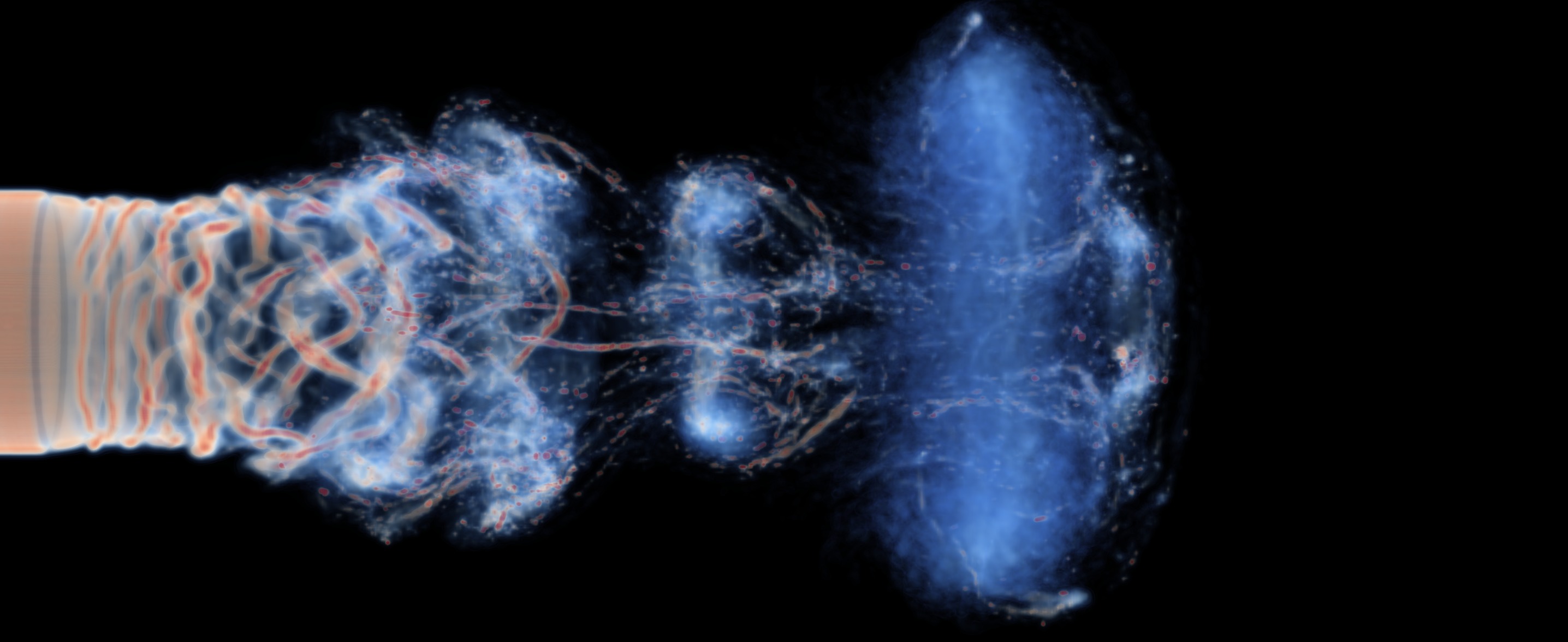}
        \end{minipage}
        \begin{minipage}{0.41\textwidth}
            \includegraphics[width=\linewidth]{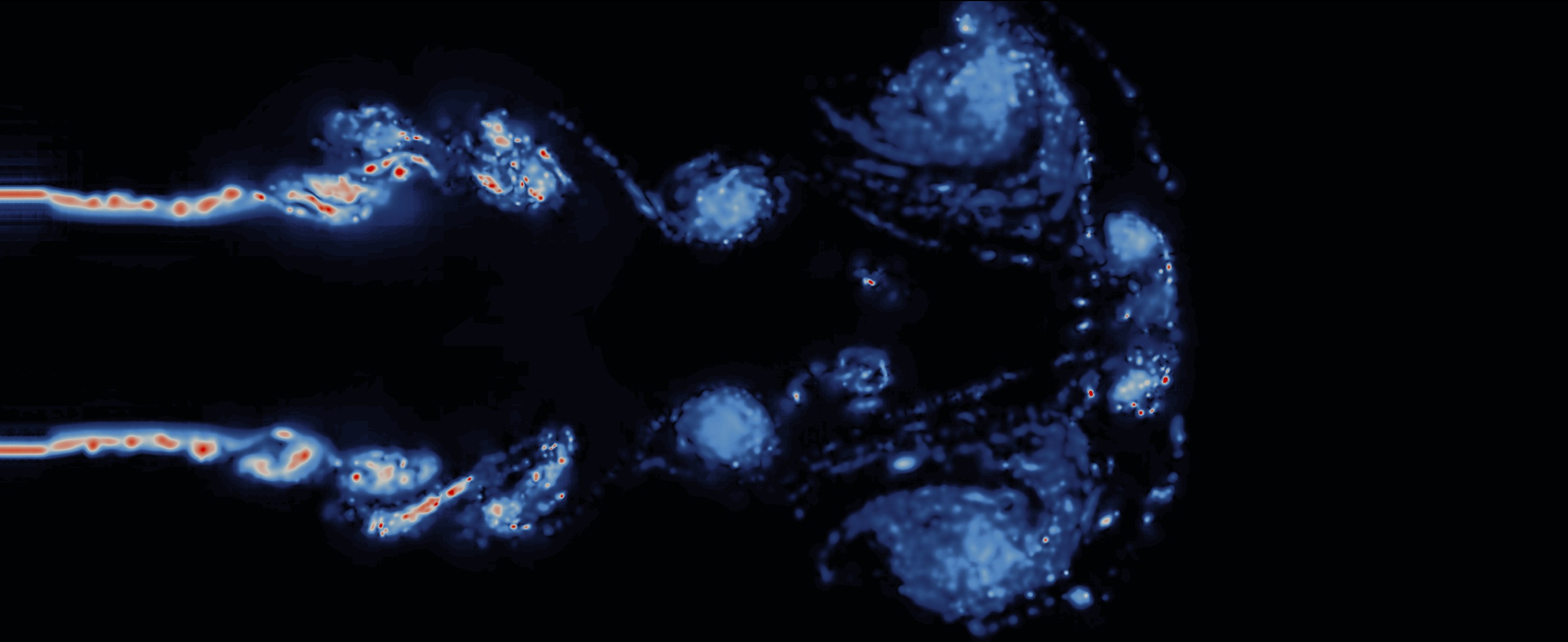}
        \end{minipage}

        \raggedright
        $t=10\,\mathrm{ms}$ \hspace{0.4mm}
        \begin{minipage}{0.41\textwidth}
            \includegraphics[width=\linewidth]{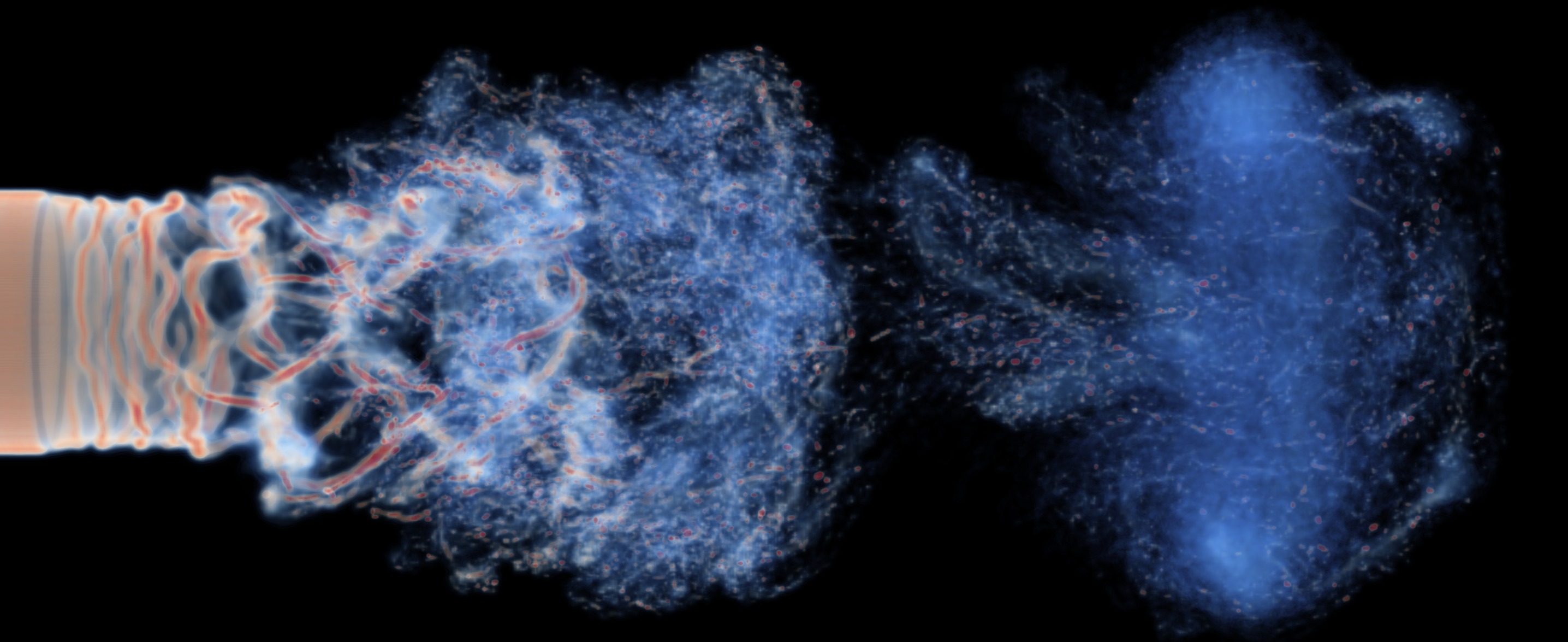}
        \end{minipage}
        \begin{minipage}{0.41\textwidth}
            \includegraphics[width=\linewidth]{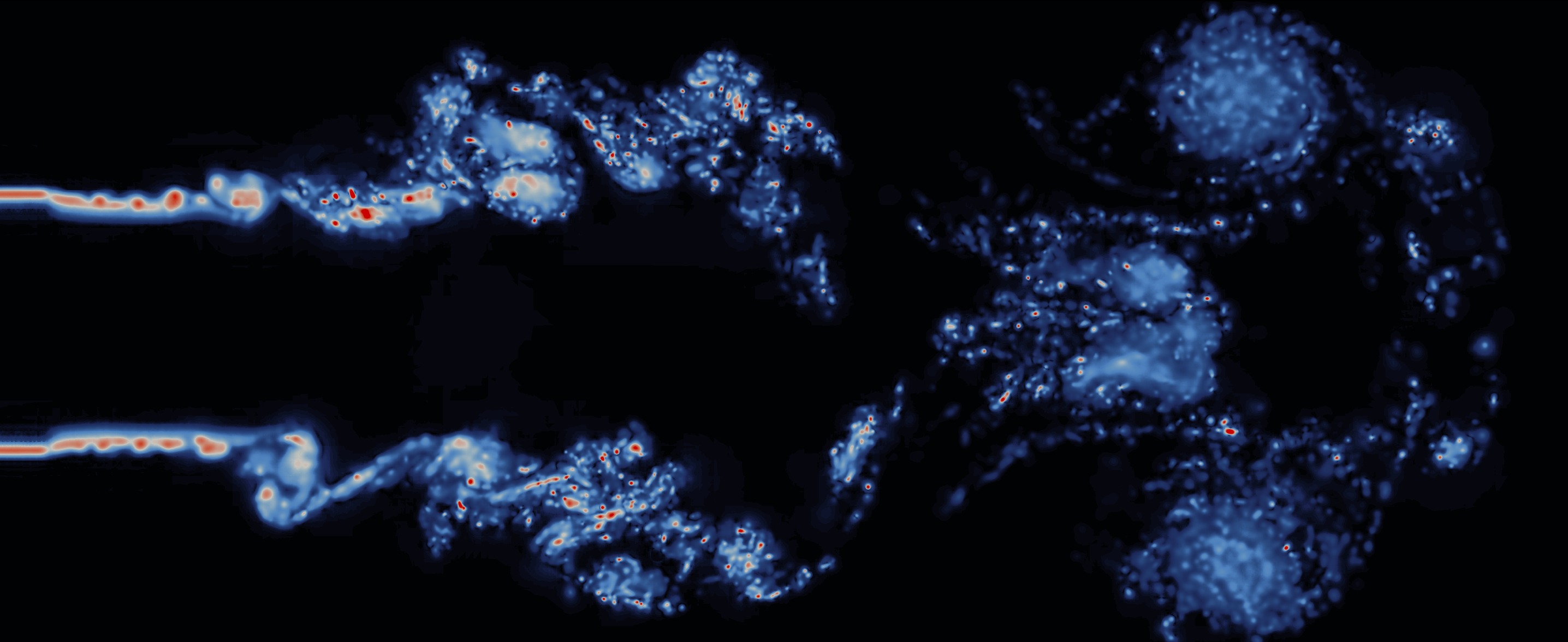}
        \end{minipage}

        \caption{Evolution of turbulent jet simulation: (left) volume rendering and (right) slice of vorticity field.}

    \end{figure*}



    \begin{figure*}[] \ContinuedFloat

        \raggedright
        $t=22\,\mathrm{ms}$ \hspace{0.4mm}
        \begin{minipage}{0.41\textwidth}
            \includegraphics[width=\linewidth]{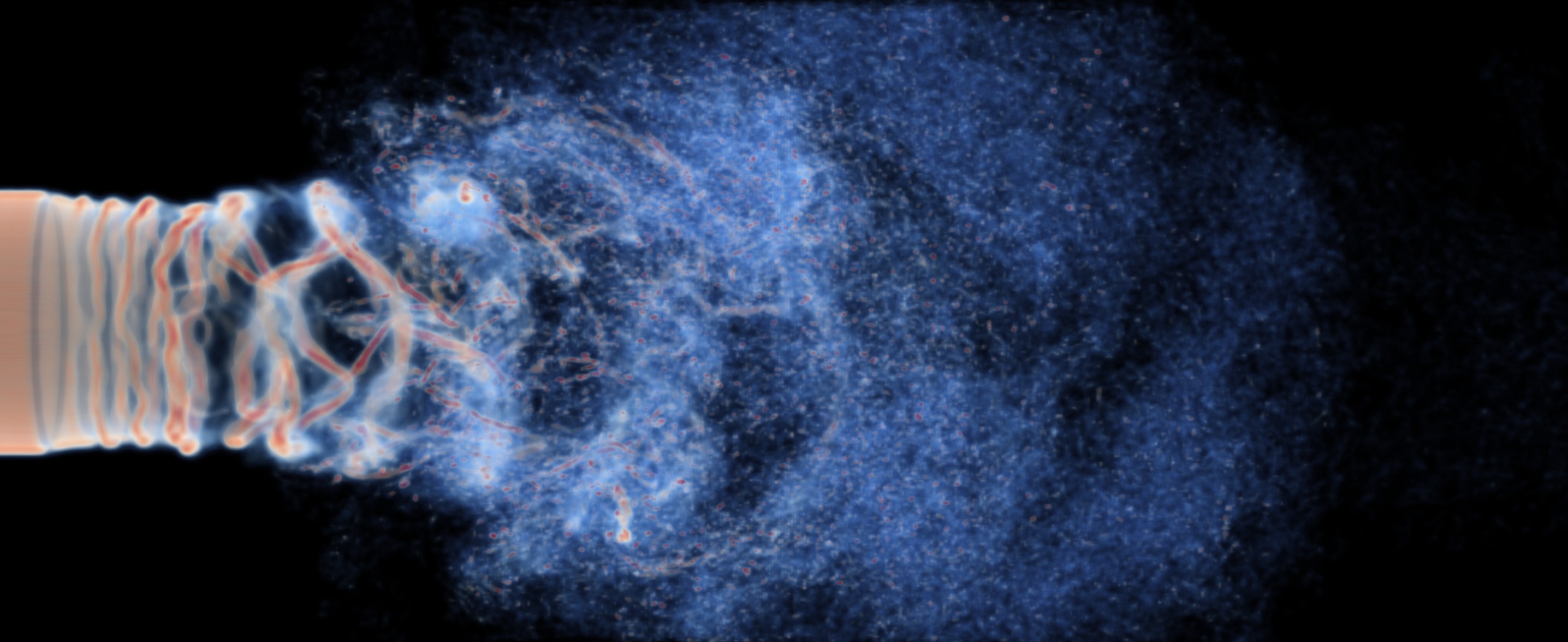}
        \end{minipage}
        \begin{minipage}{0.41\textwidth}
            \includegraphics[width=\linewidth]{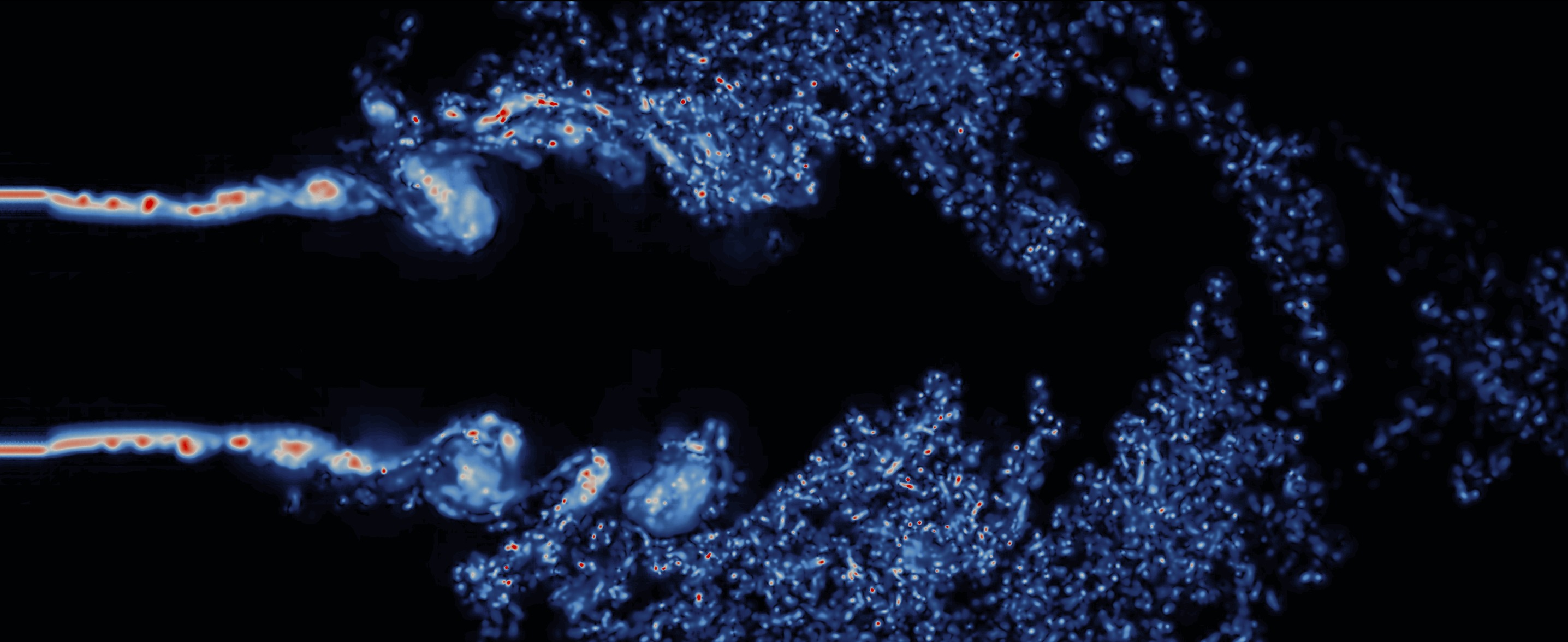}
        \end{minipage}

        \raggedright
        $t=45\,\mathrm{ms}$ \hspace{0.4mm}
        \begin{minipage}{0.41\textwidth}
            \includegraphics[width=\linewidth]{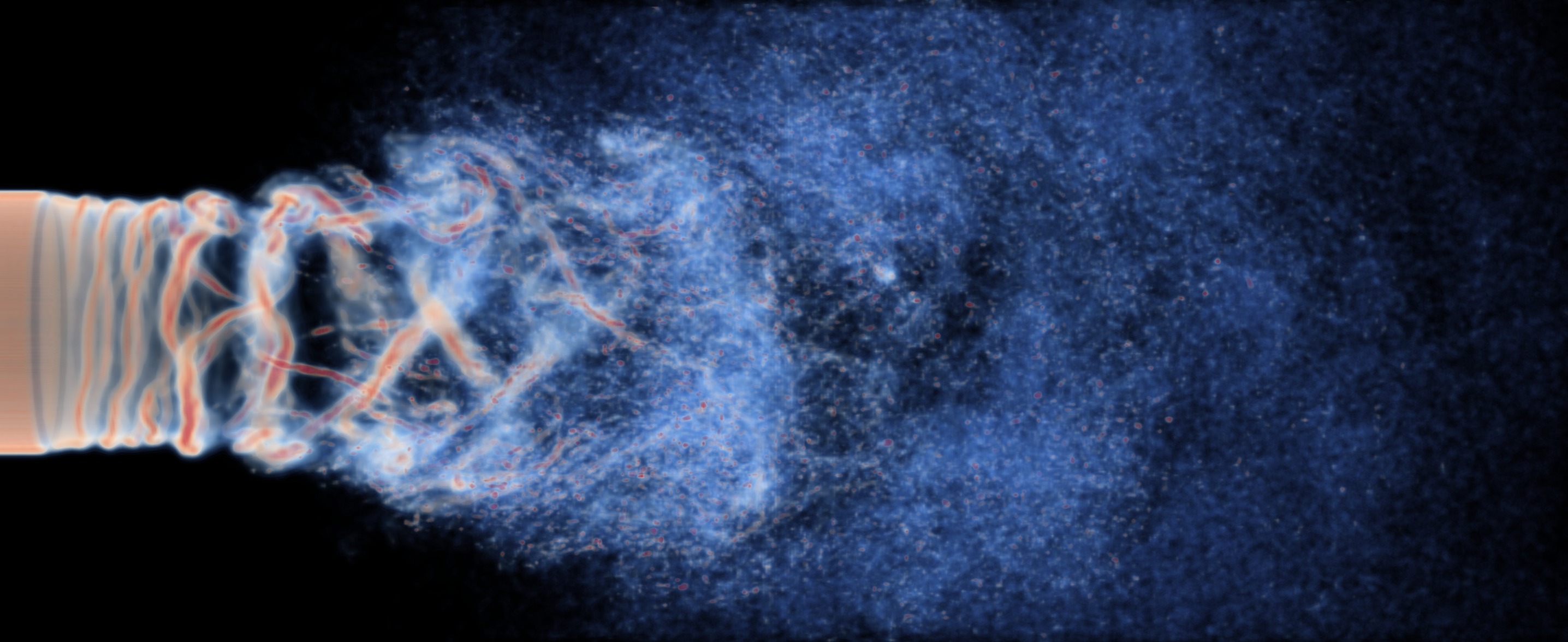}
        \end{minipage}
        \begin{minipage}{0.41\textwidth}
            \includegraphics[width=\linewidth]{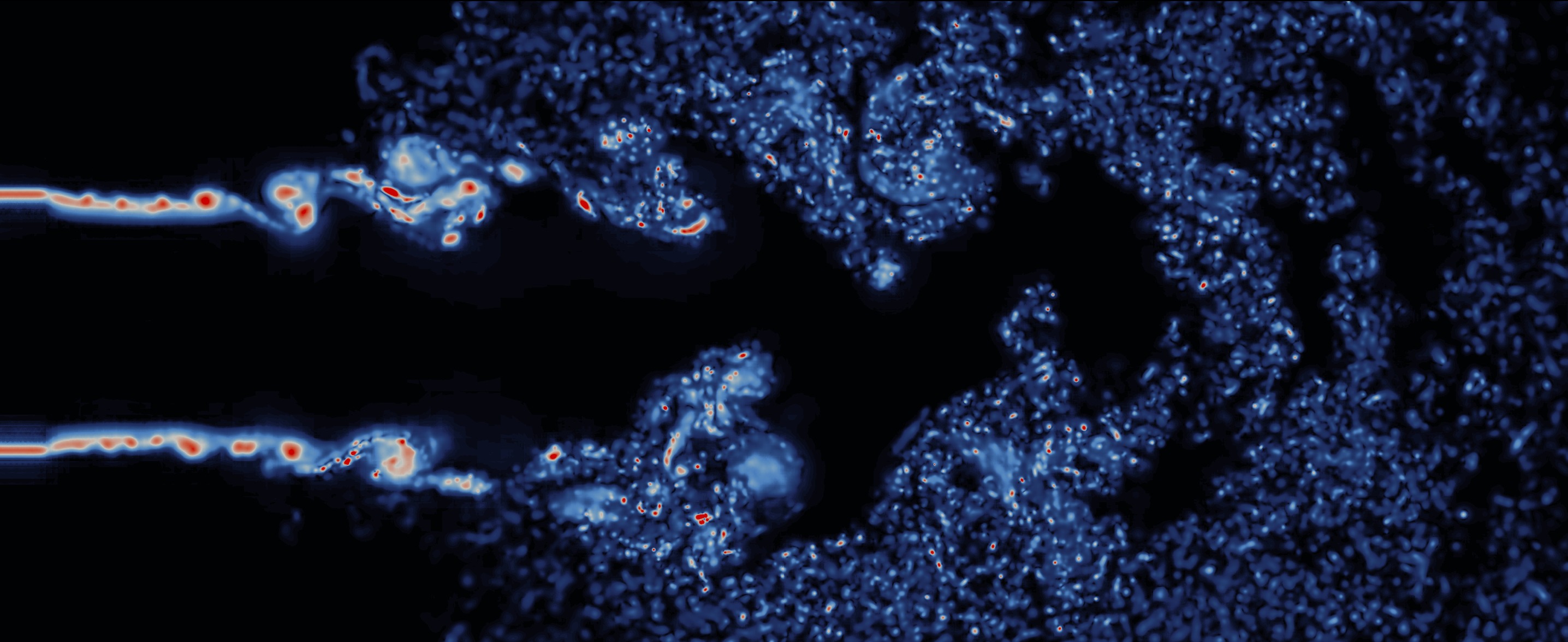}
        \end{minipage}

        \raggedright
        $t=50\,\mathrm{ms}$ \hspace{0.4mm}
        \begin{minipage}{0.41\textwidth}
            \includegraphics[width=\linewidth]{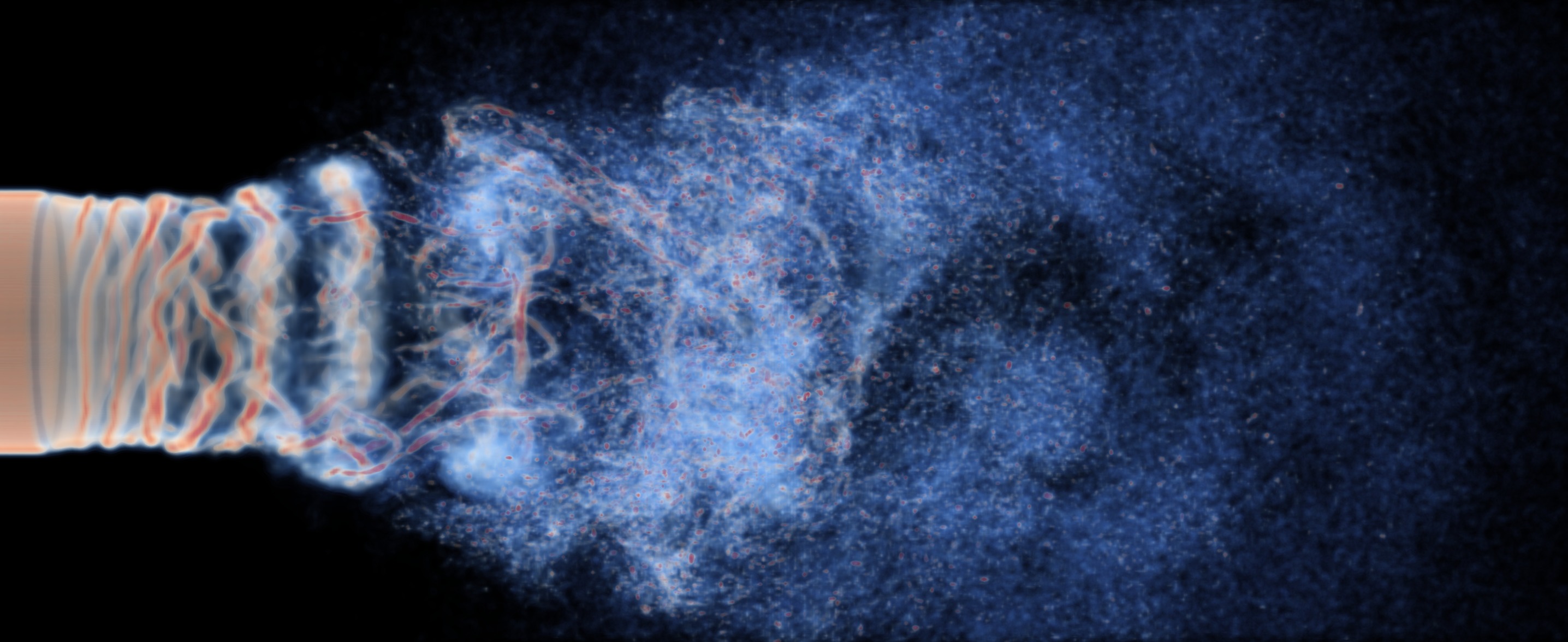}
        \end{minipage}
        \begin{minipage}{0.41\textwidth}
            \includegraphics[width=\linewidth]{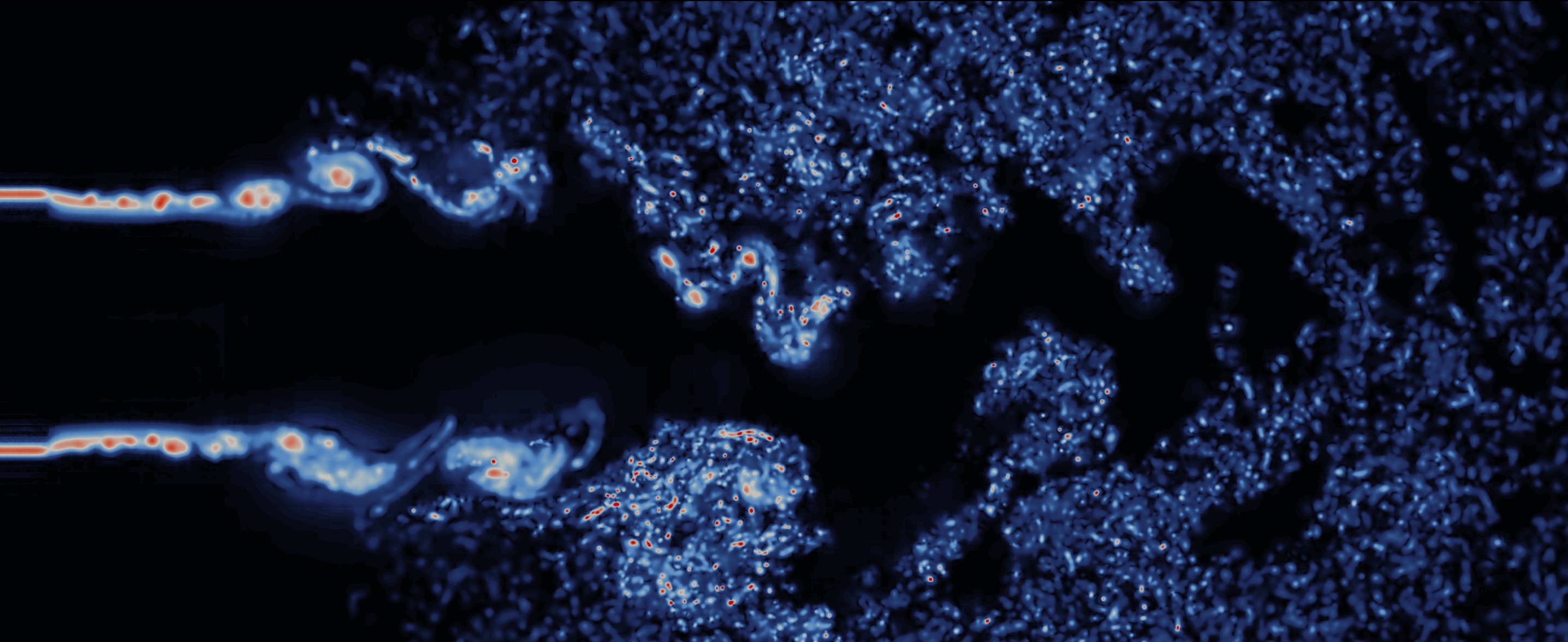}
        \end{minipage}

        \caption{Evolution of turbulent jet simulation (continued).}
        \label{fig:roundjet:vorticityevolution}
    \end{figure*}

    The simulation was first attempted with the classic VPM, but it quickly ended in numerical blow up at the initial stage of the jet.
    This is shown in~\cref{fig:roundjet:enstrophy} through an abrupt jump in global enstrophy.
    The reformulated VPM proved to be numerically stable in the initial and transition stages of the jet, however, the simulation becomes unstable in the fully-developed turbulent regime as enstrophy builds up in the absence of SFS turbulent diffusion.
    Introducing the SFS model, the enstrophy production of the inlet balances out with the forward-scatter of the fully-developed region and the rVPM simulation becomes indefinitely stable, as shown in~\cref{fig:roundjet:enstrophy}.
    Interestingly, the SFS model did not stabilize the classic VPM simulation, which indicates that both the reformulated VPM and the SFS model are needed to achieve numerical stability.

    \begin{figure}[t]

        \vspace{5mm}
        
        \centering
        \includegraphics[width=\figwidth]{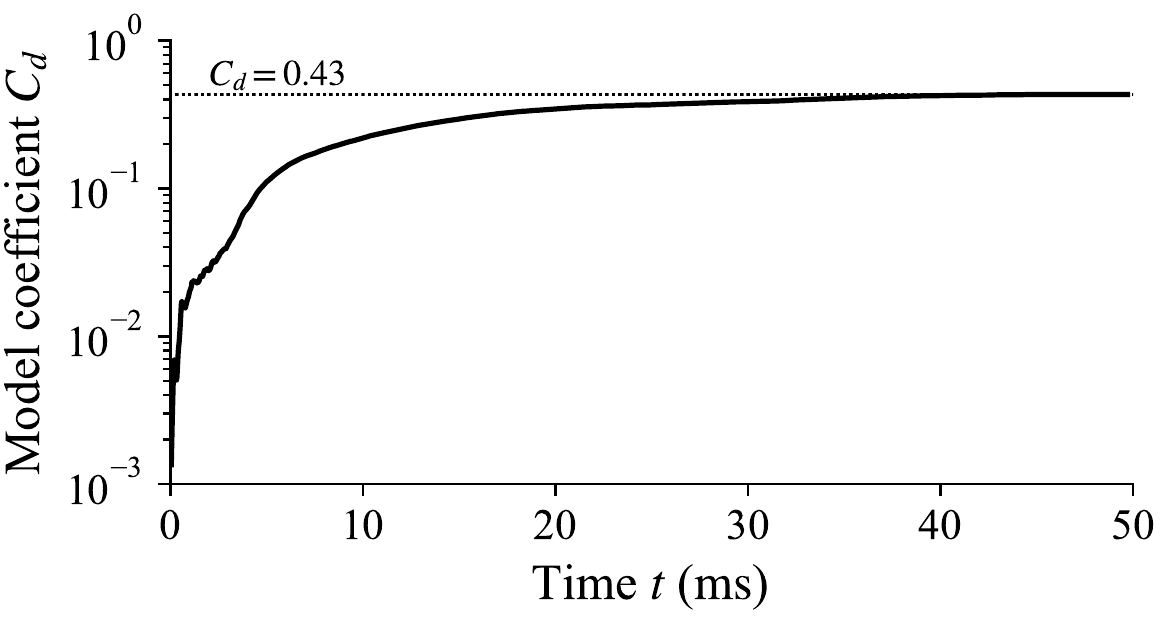}
        \caption{Average SFS model coefficient in turbulent jet simulation as computed through the dynamic procedure.}
        \label{fig:roundjet:Cd}
    \end{figure}
    
    \begin{figure*}[t!]

        \vspace{5mm}

        \begin{subfigure}{0.9\textwidth} \centering
            \includegraphics[width=\linewidth]{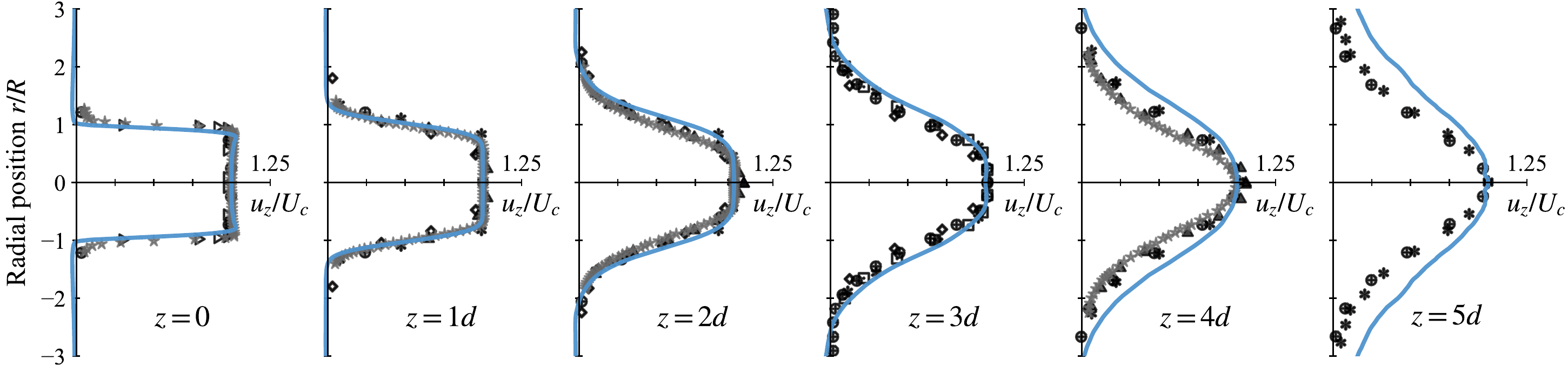}
        \end{subfigure}

        \caption{Development of streamwise mean velocity profile along jet. Simulation: {\LARGE \color{vpm}-} rVPM. Experimental:
            {\color{gray}$\star$} Quinn and Militzer \cite{Quinn1989}; $\oplus$ Fellouah and Pollard \cite{fellouah-jet}; $\triangleright$ Mi \textit{et al.}, contraction nozzle \cite{mi-jet}; $\ast$ Iqbal and Thomas \cite{iqbal-jet}; $\diamond$ Romano \cite{romano-jet}; $\boxdot$ Xu and Antonia, contraction nozzle \cite{xu-jet}
        .}
        \label{fig:roundjet:Umean}
    \end{figure*}

    \begin{figure*}[t!]

        \vspace{5mm}
        
        \begin{subfigure}{0.9\textwidth} \centering
            \includegraphics[width=\linewidth]{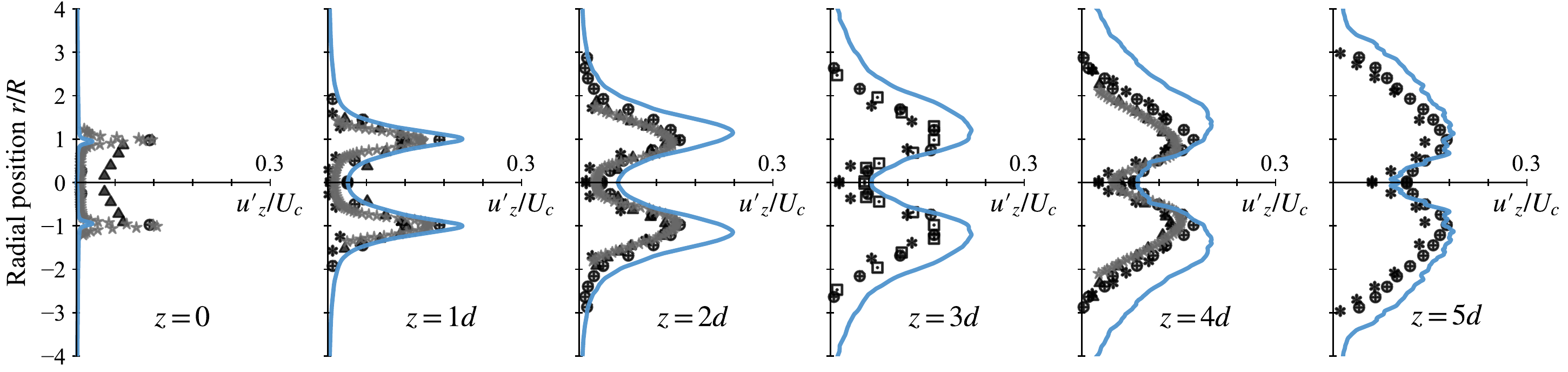}
        \end{subfigure}

        \caption{Development of fluctuating component of the streamwise velocity along jet. Simulation: {\LARGE \color{vpm}-} rVPM. Experimental:
            {\color{gray}$\star$} Quinn and Militzer \cite{Quinn1989}; $\oplus$ Fellouah and Pollard \cite{fellouah-jet}; $\ast$ Iqbal and Thomas \cite{iqbal-jet}; $\boxdot$ Xu and Antonia, contraction nozzle \cite{xu-jet}
        .}
        \label{fig:roundjet:Ufluctuating}
    \end{figure*}

    \begin{figure*}[t!]

        \vspace{5mm}
        
        \begin{subfigure}{0.9\textwidth} \centering
            \includegraphics[width=\linewidth]{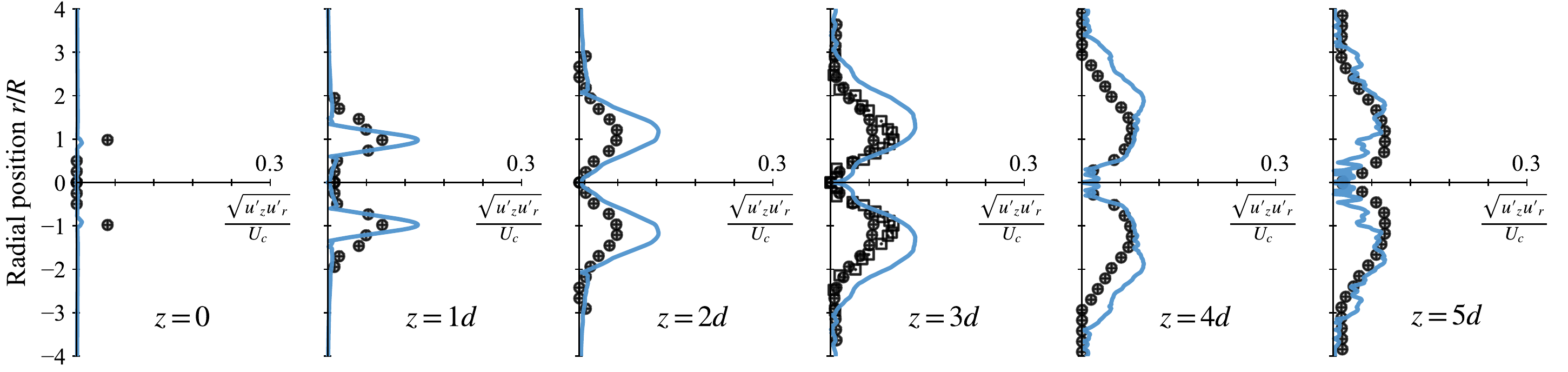}
        \end{subfigure}

        \caption{Development of Reynolds stress along jet. Simulation: {\LARGE \color{vpm}-} rVPM. Experimental:
            $\oplus$ Fellouah and Pollard \cite{fellouah-jet}; $\boxdot$ Xu and Antonia, contraction nozzle \cite{xu-jet}
        .}
        \label{fig:roundjet:reynoldsstress}
    \end{figure*}

    Once proven stable, the simulation was run for $50$ ms with time step $\Delta t = 0.02\,\mathrm{ms}$ until achieving a fully-developed region that was statistically stationary.
    This was run on a single node with 128 CPU cores (dual 64-core AMD EPYC 7702, 2.0 GHz) resulting on a wall-clock time of two and a half days using up to $3\times 10^6$ particles.
    ~\cref{fig:roundjet:allviews} shows the particle field, vortex strengths, and the SFS model coefficient $C_d$ close to the nozzle at $t=48\,\mathrm{ms}$, while~\cref{fig:roundjet:vorticity} shows the vorticity of the entire field.
    As seen in~\cref{fig:roundjet:vorticity}, coherent vortical structures form in the initial region ($z<1d$), which leapfrog and mix transitioning to fully-developed turbulent flow by $z>3d$.
    Notice in~\cref{fig:roundjet:allviews} that the model coefficient is negligibly small in the initial region and increases to $0.1 < C_d < 1$ in the fully-developed regime.
    This shows that the dynamic procedure succeeds at automatically calibrating $C_d$, hindering SFS turbulent diffusion in the laminar regime while facilitating it in turbulent regions.

    The formation and time evolution of the jet are shown in~\cref{fig:roundjet:vorticityevolution} and available as a video in the supplemental content\footnote{Available at \href{https://youtu.be/V9hthE7m1d4}{https://youtu.be/V9hthE7m1d4} in the preprint version of this paper.}.
    Initially, a vortex ring forms at the head of the jet.
    As the ring travels downstream, a vortex sheet is deployed forming the shear layer, seen at $t=2\,\mathrm{ms}$.
    The vortex sheet is stretched as the flow develops, eventually rolling up and forming filaments, seen at $t=4\,\mathrm{ms}$.
    The filaments pair up, leapfrog, merge, and breakdown (seen at $t=8\,\mathrm{ms}$) eventually forming the fully-developed turbulent region at $t>20\,\mathrm{ms}$.
    \cref{fig:roundjet:Cd} shows the history of $C_d$ throughout this process, calculated at each time step as the average $\vert C_d \vert$ over all the particles where $C_d \neq 0$.
    Here we see that the average $C_d$ is negligibly small in the initial development of the jet, but it ramps up and converges to a value of 0.43 as the flow becomes fully developed, again confirming the ability of the dynamic procedure to automatically calibrate the SFS model at each flow regime.

    To validate the dynamics predicted by the simulation, the velocity was probed at five stations along the jet and statistical properties were compared to the experimental measurements reported by Quinn and Militzer \cite{Quinn1989}.
    The data was also supplemented with other experiments of similar round jets compiled by Ball \textit{et al}. \cite{Ball2012}
    Statistical properties were calculated through temporal and spatial ensemble averages after the flow became statistically stationary in the region $0 \leq z \leq 5d$.
    Averaging was performed over the time interval $40\,\mathrm{ms} \le t < 50\,\mathrm{ms}$.

    \cref{fig:roundjet:Umean} shows the mean component of the streamwise velocity profile, $u_z$, normalized by the centerline velocity at each station.
    The mean component shows excellent agreement with the experiments in the initial and transition regions ($0 \leq z \leq 3d$) and reasonable agreement in the fully-developed turbulent region ($z \ge 4d$) though slightly overexpanded.
    \cref{fig:roundjet:Ufluctuating} shows the fluctuating component of the streamwise velocity, $u^\prime_z$, or standard deviation, while~\cref{fig:roundjet:reynoldsstress} shows the Reynolds stress between streamwise and radial velocity, defined as the square root of the covariance, $\sqrt{u^\prime_z u^\prime_r}$.
    At $z = 0$, the simulation shows only a small fluctuation and Reynolds stress as the flow is dominated by the boundary condition.
    Away from the nozzle exit plane ($z > 0$), fluctuations and Reynolds stress concentrate at the shear layer ($r/R = 1$) in the laminar region ($z \approx 1d$), and gradually spread as the jet breaks down in the turbulent regime ($z \ge 4d$).
    Fluctuations are overpredicted (and, as a consequence, also the Reynolds stress), however this is in within reasonable agreement with the experiments.
    These predictions can be further improved in future work with the addition of an SFS model of vorticity advection increasing the turbulent diffusion (and damping out fluctuations), or implementing a spatial adaptation strategy to better resolve small scales.
    However, the current agreement with the experiments suffices to confirm that our scheme is an LES able to resolve mean and fluctuating large-scale features of the flow.
    Also, its ability to directly resolve Reynolds stress renders the reformulated VPM a higher fidelity approach than Reynolds-average approaches like RANS and URANS (where Reynolds stresses are rather modeled), while being completely meshless.

	\subsection{Aircraft Rotor in Hover} \label{sec:res:rotor}

    The rotation of blades in static air drives a strong axial flow caused by the shedding of tip vortices.
    This is a challenging case to simulate since, in the absence of a freestream, the wake quickly becomes fully turbulent as tip vortices leapfrog and mix close to the rotor.
    Thus, a rotor in hover is a good engineering application to showcase the accuracy, numerical stability, and computational efficiency of the reformulated VPM.

    In this test case, we simulated the experiment by Zawodny \textit{et al.} \cite{Zawodny2016b} consisting of a DJI 9443 rotor in hover at 5400 RPM.
    This two-bladed rotor is 9.4 inches in diameter, resulting in a tip Mach number of 0.20 and chord and diameter-based Reynolds numbers at 70\% of the blade span of $6 \times 10^4$ and $7 \times 10^5$, respectively.
    Our simulations were also compared to unsteady Reynolds-average Navier-Stokes (URANS) results reported by Schenk, \cite{Schenk2020} obtained with the commercial software \text{STAR-CCM+}.
    The URANS is an unsteady compressible solver with an SST $k$--$\omega$ turbulence model, resolving the blades with an all-$y^+$ wall treatment on a rotating mesh surrounding the rotor.
    It used an unstructured mesh with 14 million cells, mesh refinement down to a $y^+$ of only 30, and time steps equivalent to $3^\circ$ of rotation on a first-order time integration scheme, which is a rather coarse simulation, but requiring very low computational resources.

\begin{figure}[t!]
        \vspace{10mm}
        \centering
        \includegraphics[width=\figwidth]{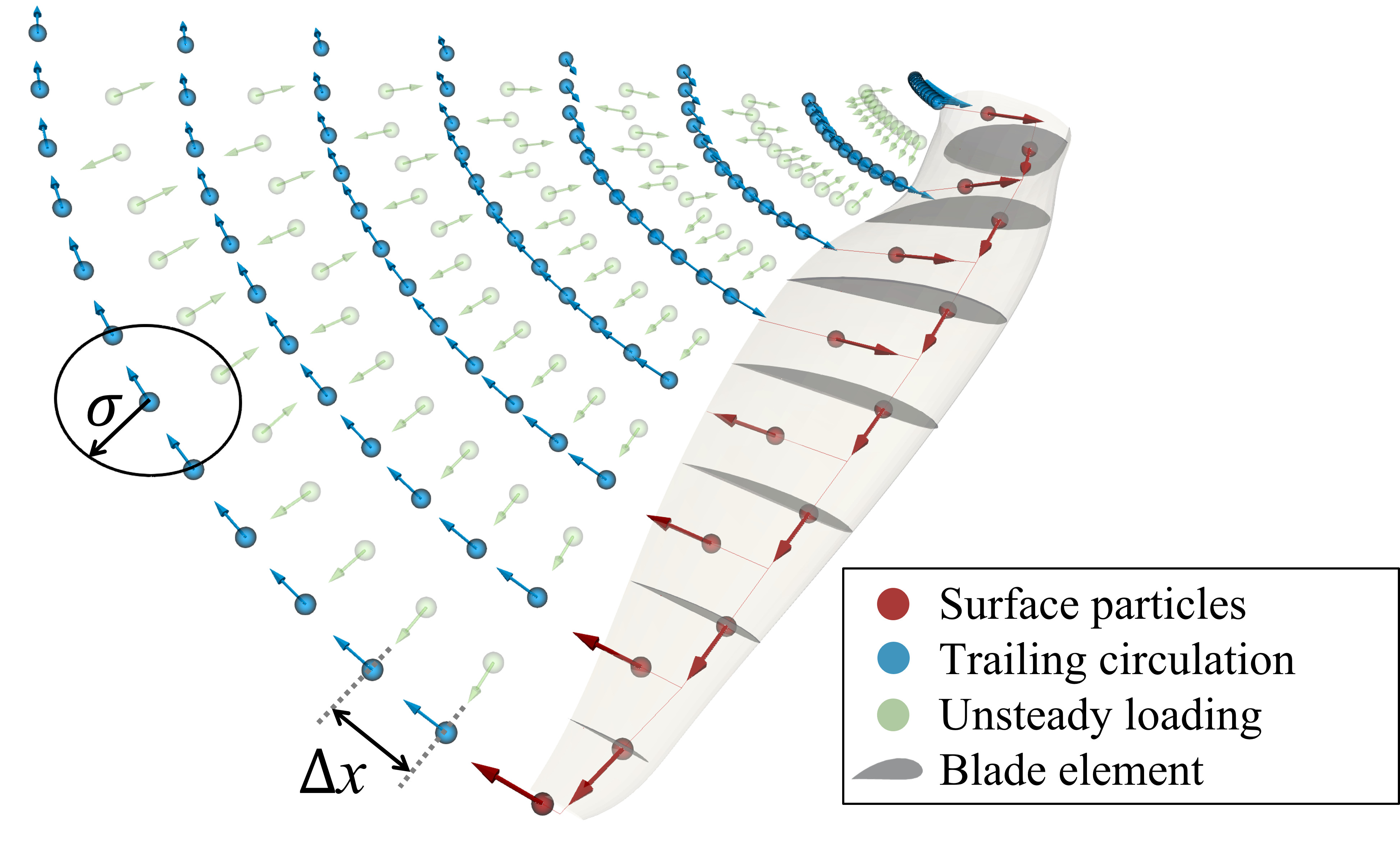}
        \caption{Actuator line model in rotor simulation. Particles colored by their source of vorticity; arrows indicate direction of vortex strength.}
        \label{fig:rotor:visual}
    \end{figure}

    The rotating blades are computed in our VPM through an actuator line model (ALM), which is a common practice in LES\cite{Churchfield2017}.
    Our ALM discretizes the geometry into blade elements, using a two-dimensional viscous panel method to compute forces and circulation along each blade cross section as the blades move, as described in previous work \cite{Alvarez2020}.
    The vorticity of each blade is introduced in the fluid domain by embedding static particles along the surface that capture the blade's circulation distribution, while shedding free particles at the trailing edge associated with unsteady loading and trailing circulation, as shown in~\cref{fig:rotor:visual}.
    The frequency of particle shedding per revolution determines the initial spacing $\Delta x$ in between particles, which, along with the core size $\sigma$, determines the spatial resolution at which the wake is being resolved.
    The initial particle size $\sigma$ was set as to provide a particle overlap $\lambda=\frac{\sigma}{\Delta x}$ of 2.125 at the blade tip.
    The VPM simulations used a time step equivalent to a rotation of $1^\circ$ while shedding particles every $0.5^\circ$ with 50 blade elements.

\begin{figure}[t!]
        \vspace{4mm}
        \centering
        \includegraphics[width=\figwidth]{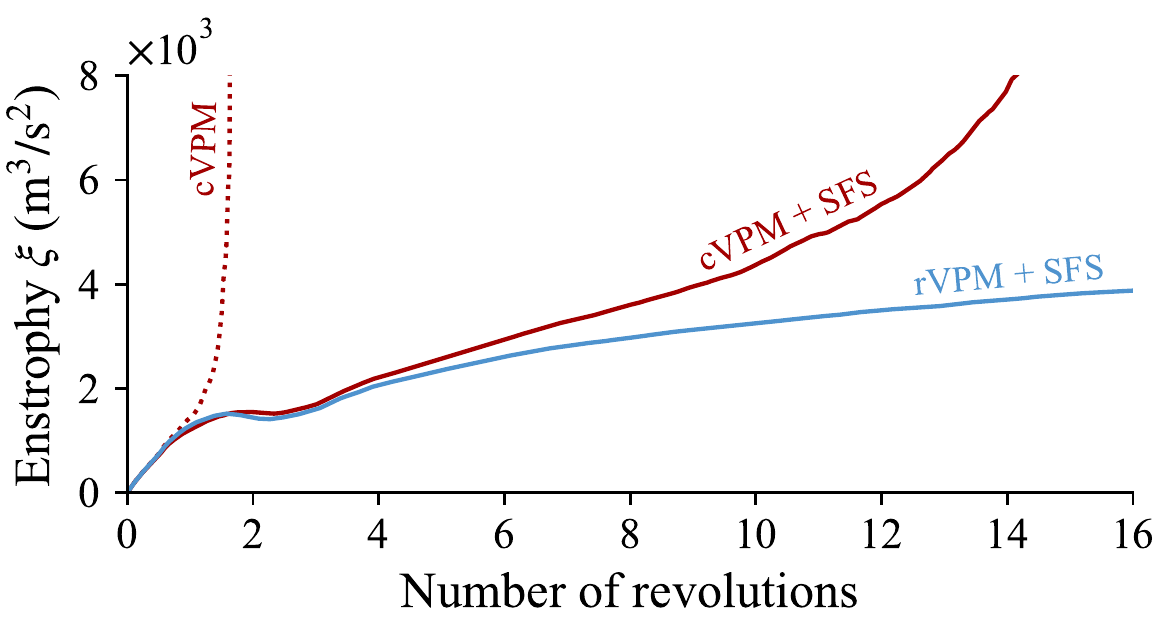}
        \caption{Global enstrophy in VPM rotor simulations.}
        \label{fig:rotor:enstrophy}
    \end{figure}

    First, the simulation was attempted on the classic VPM without the SFS model, which quickly ended in numerical blow up after one revolution, shown in~\cref{fig:rotor:enstrophy}.
    Introducing the SFS model made the classic VPM noticeably more stable, however, the simulation still blew up before the wake became fully developed.
    Introducing the reformulated VPM made the simulation completely stable, using up to $1\times10^6$ particles after 16 revolutions.
    As shown in~\cref{fig:rotor:enstrophy}, the rate of enstrophy produced by the rotor eventually balances out with the forward scatter of the SFS model.
    This further asserts the numerical stability gained with the reformulated VPM and SFS model.

\begin{figure}[t!]
        \vspace{4mm}
        \centering
        \includegraphics[width=\figwidth]{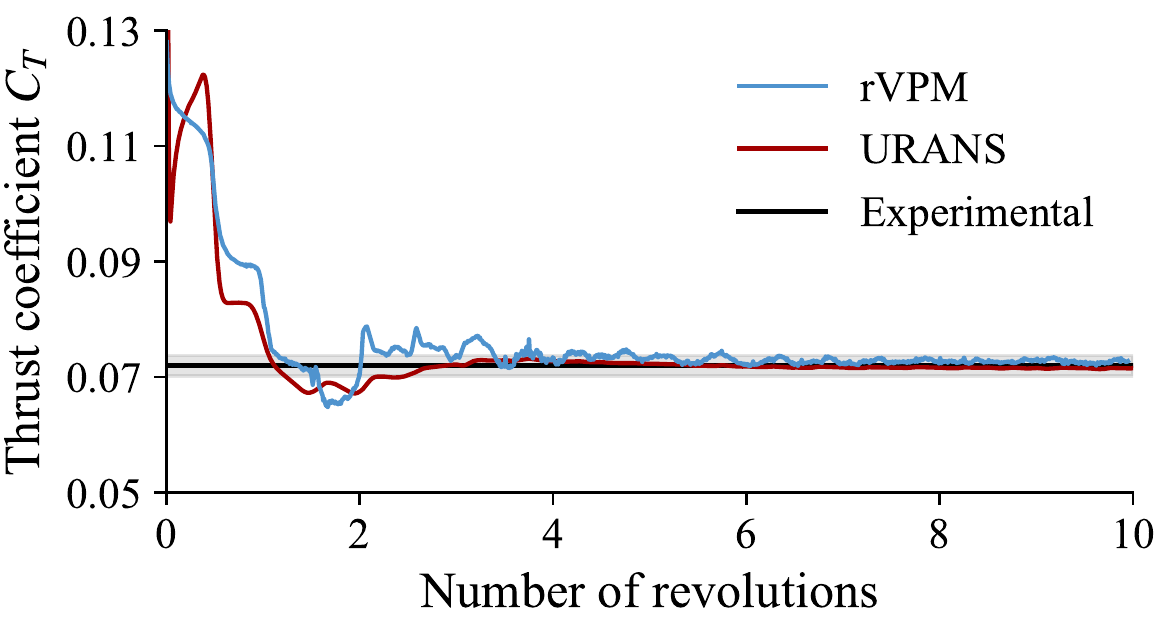}
        \caption{Thrust history in rotor simulations compared to experimental mean $C_T$. Shaded region encompasses the 95\%-confidence interval of the experiment.}
        \label{fig:rotor:CT}
    \end{figure}

    Next, the thrust coefficient $C_T$ predicted with rVPM was compared to the experimental coefficient reported by Zawodny \textit{et al.} \cite{Zawodny2016b}
    We define $C_T$ in the propeller convention as $C_T = \frac{T}{\rho n^2 d^4}$, where $T$ is the dimensional thrust, $\rho$ is air density, and $n$ is rotations per seconds.
    The experiment reported\footnote{The experiment reported the thrust coefficient defined as ${C_T=\frac{T}{\rho \pi R^2 (\Omega R)^2}}$ which is typical in the rotorcraft community, while here we have converted their measurement to $C_T$ as defined in the propeller convention.} a mean $C_T$ of $0.072$.
    In our simulation, the thrust is calculated integrating the force computed by the ALM at each blade element that is immersed in the fluid domain.
    As shown in~\cref{fig:rotor:CT}, the VPM simulation shows excellent agreement with the experiment, predicting a mean value within 2\% of the experimental mean value.
    This illustrates the capacity of our method to provide accurate predictions in a real engineering application.

    \begin{figure}[t]
        \centering
        \includegraphics[width=\figwidth]{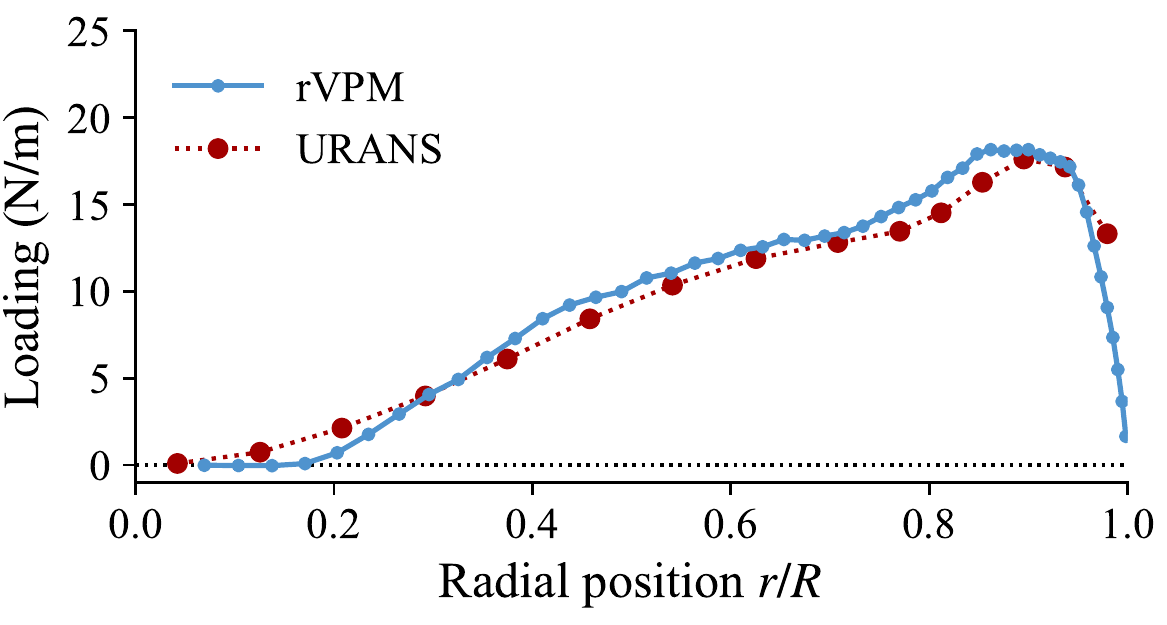}
        \caption{Time-average blade loading distribution in rotor simulations.}
        \label{fig:rotor:loading}
        \vspace{-4mm}
    \end{figure}

\begin{figure*}

        \vspace{2mm}

        \begin{minipage}{0.49\textwidth}
            \includegraphics[width=\linewidth]{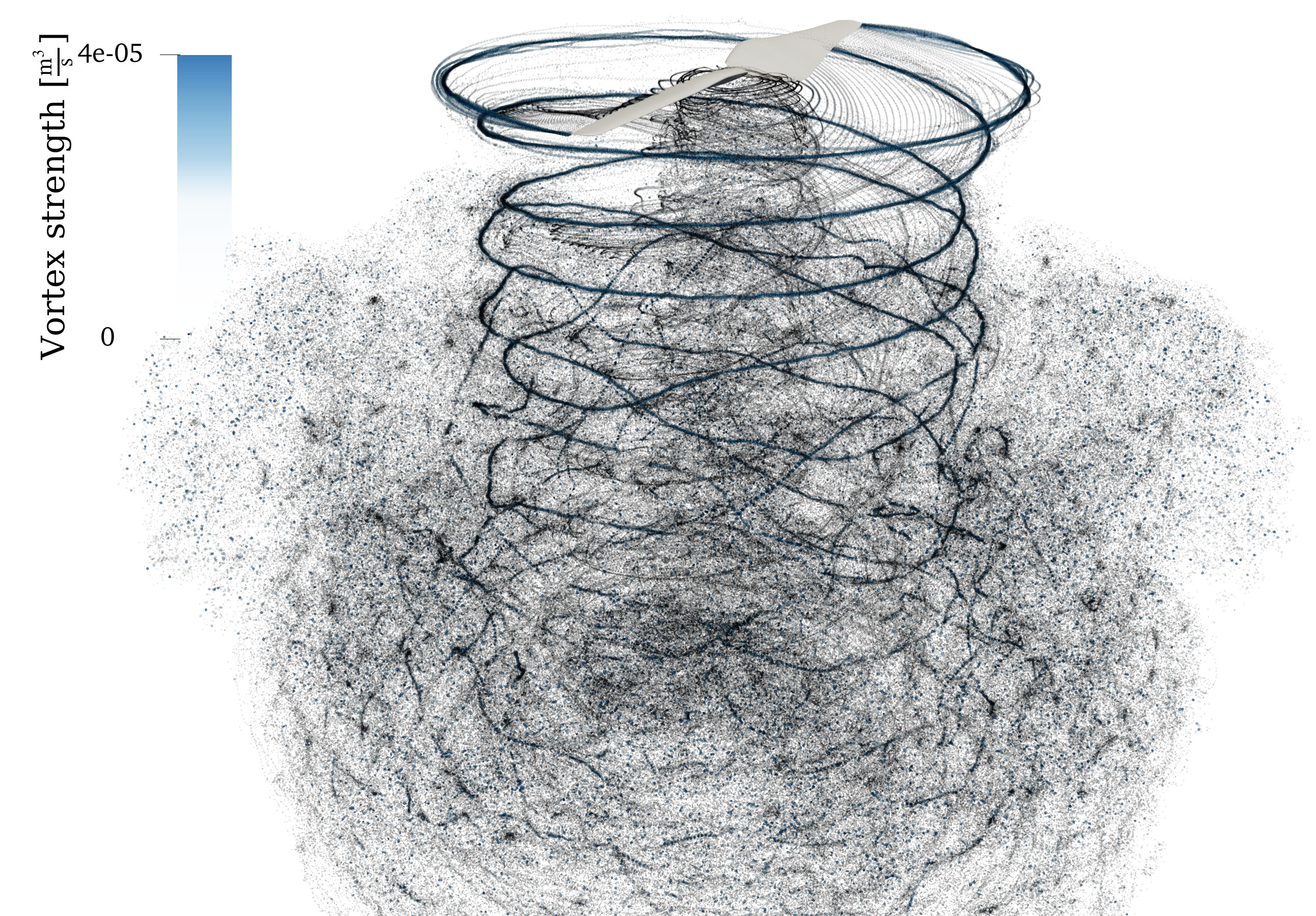}
        \end{minipage}
        \begin{minipage}{0.49\textwidth}
            \includegraphics[width=\linewidth]{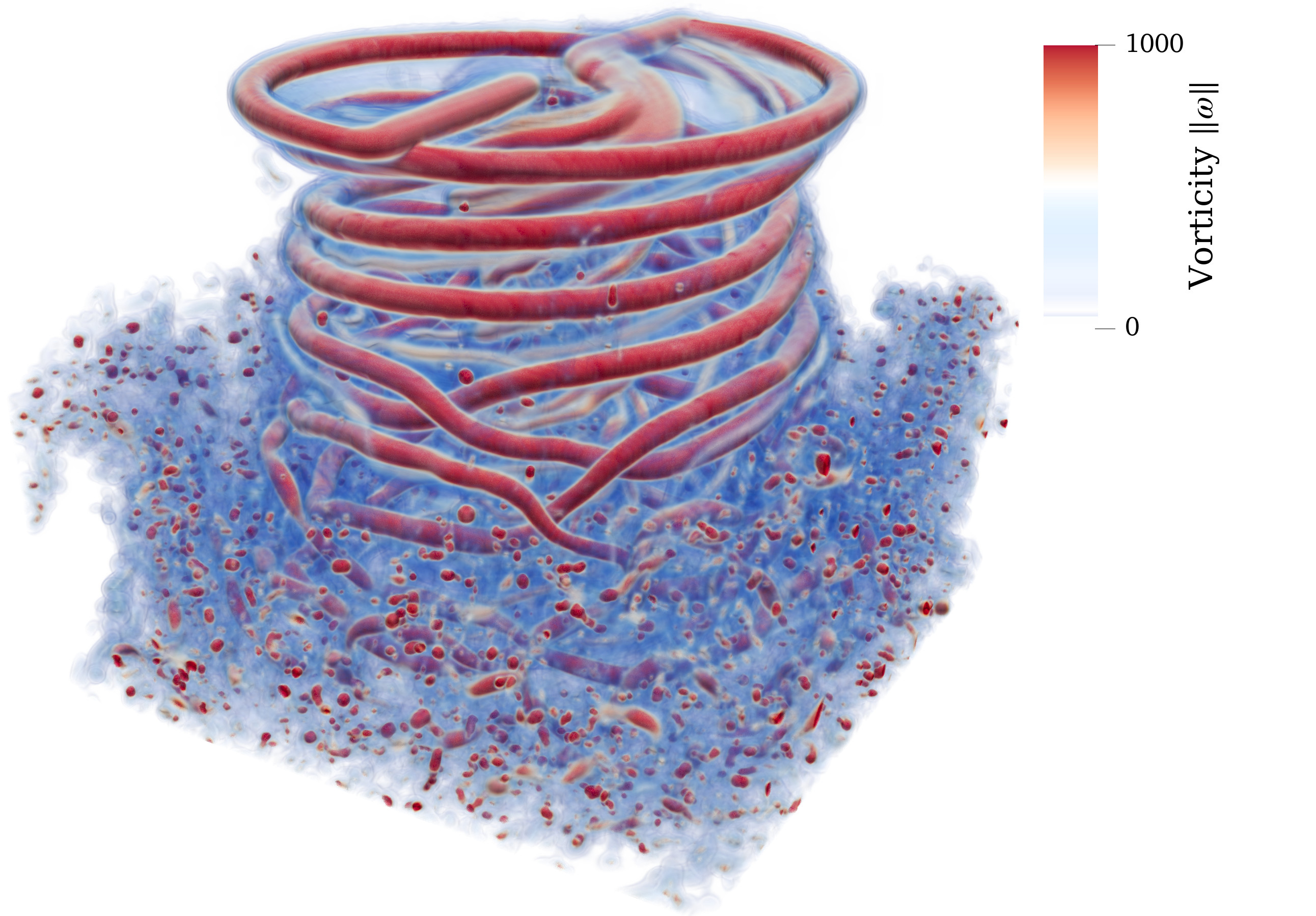}
        \end{minipage}

        \caption{Meshless LES of rotor in hover after 15 revolutions using rVPM: (left) computational elements (vortex particles and strength), and (right) volume rendering of vorticity field.  Video available at \href{https://youtu.be/u9SgYbYhPpU}{https://youtu.be/u9SgYbYhPpU} in the preprint version of this paper.}
        \label{fig:rotor:vpmsim}

    \end{figure*}

    In order to illustrate the low computational cost of our meshless LES, we now compare our simulation to the URANS simulation by Schenk.
    As previously pointed out, for the sake of minimizing computational effort, the URANS simulation is both spatially and temporally coarse (using only 14 million cells down to a $y^+$ of only 30, with time steps of $3^\circ$), while using a first-order time integration.
    Thus, this URANS simulation represents the minimum computational effort that is plausible for a simulation of this kind while still being accurate (the $C_T$ predicted by URANS is still within 2\% of the experimental mean value, as shown in~\cref{fig:rotor:CT}).
    \cref{fig:rotor:loading} compares the time-average loading distribution along the blade as predicted with rVPM and URANS, showing that both methods can resolve the blade loading with similar accuracy.
    However, the rVPM has the advantage of being able to accurately preserve the vortical structure of the wake, shown in~\cref{fig:rotor:vpmsim}, with minimal computational effort.
    The URANS simulation reportedly took about 10 wall-clock hours to resolve 10 revolutions, using 192 CPU cores across 12 nodes on the BYU Fulton supercomputer.
    This is equivalent to about 1800 core-hours.
    On the other hand, the rVPM simulation took about 4 wall-clock hours to resolve 10 revolutions on a single node of the BYU Fulton supercomputer with 32 CPU cores, equivalent to about 140 core-hours.
    Hence, rVPM is 13 times faster (or one order of magnitude faster) than this coarse URANS simulation, while providing LES accuracy.

    While Schenk's URANS simulation represents the low-fidelity end of mesh-based CFD, Zawodny \textit{et al.} \cite{Zawodny2016b} reported a high-fidelity mesh-based simulation using OVERFLOW2.
    OVERFLOW2 is a dettached-eddy simulation (DES) code using a URANS solver with the Spalart-Allmaras turbulence model near solid surfaces, while switching to a subgrid scale formulation in regions fine enough for large eddy simulation.
    Their simulation of the DJI 9443 rotor used 260 million grid cells with time steps corresponding to $0.25^\circ$ on a second-order time integration scheme, reportedly predicting a mean $C_T$ within 2.5\% of the experimental value.
    One rotor revolution reportedly took approximately 30 wall-clock hours using 1008 CPU cores on the NAS Pleiades supercomputer.
    Extrapolating this to 10 revolutions, the DES takes 300 wall-clock hours and 300k core-hours.
    Thus, recalling that rVPM took only 140 core-hours, the rVPM is 2200x faster (or three orders of magnitude faster) than this DES simulation.

\begin{figure*}


        \hspace{-20mm}
        \begin{minipage}{0.35\textwidth}
            \includegraphics[width=\linewidth]{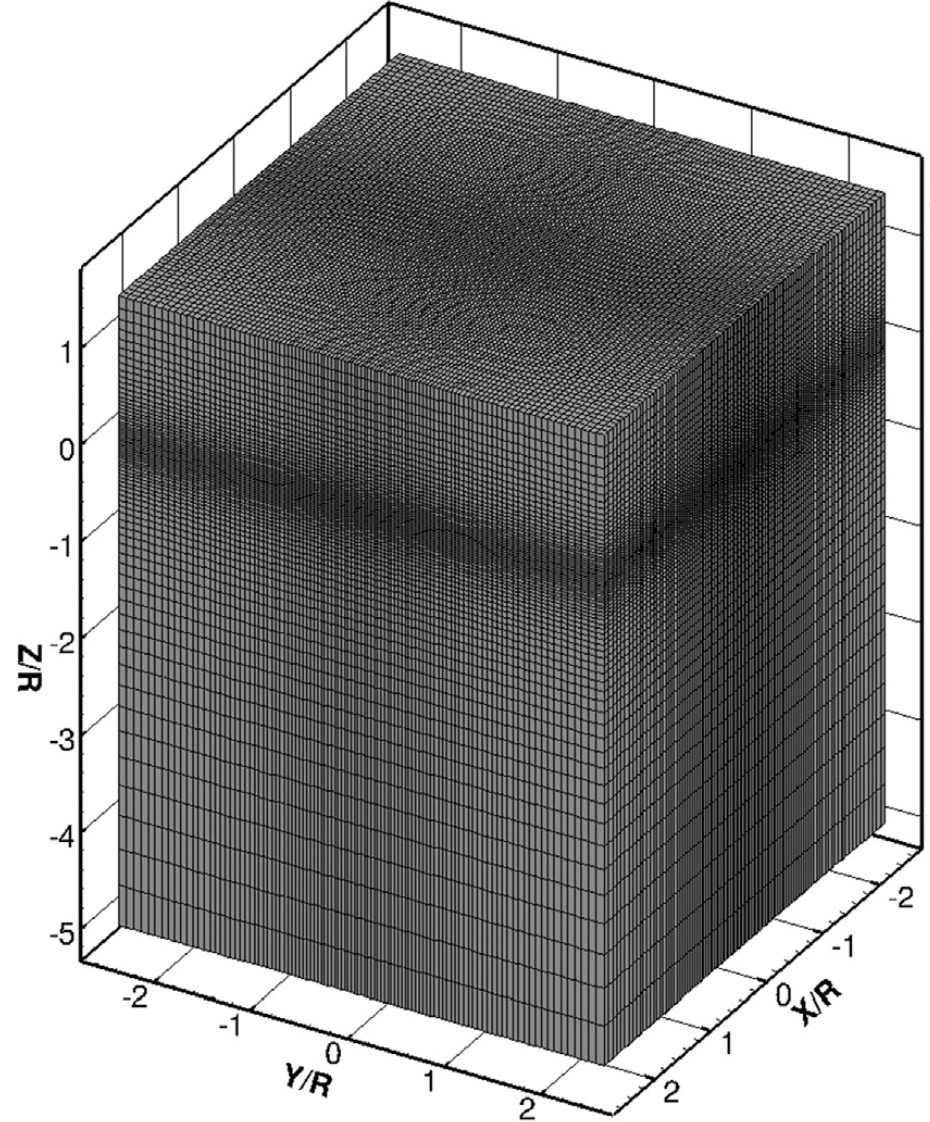}
        \end{minipage}
        \hspace{10mm}
        \begin{minipage}{0.28\textwidth}
            \includegraphics[width=\linewidth]{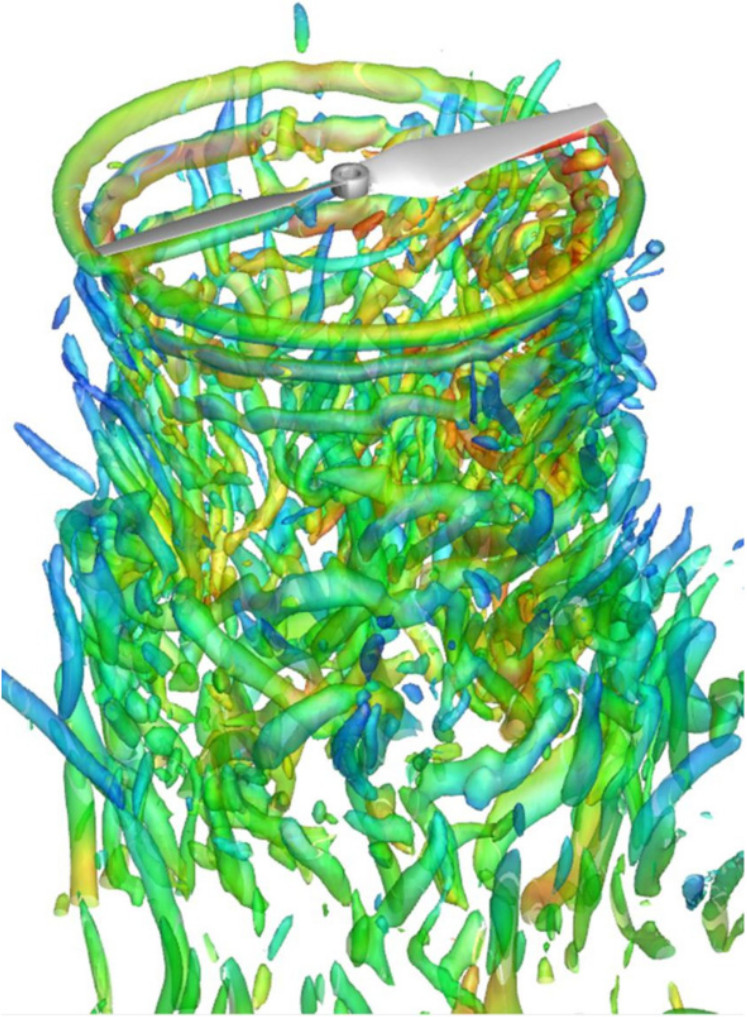}
        \end{minipage}

        \caption{Mesh-based LES of rotor in hover reported by Delorme \textit{et al.}: (left) computational elements, and (right) iso-surface of Q-criterion colored by velocity magnitude. Reprinted from Aerospace Science and Technology, Vol 108, Delorme \textit{et al}, \textit{Application of Actuator Line Model for Large Eddy Simulation of Rotor Noise Control}, Copyright 2020 Elsevier Masson SAS, with permission from Elsevier.}
        \label{fig:rotor:meshsim}

    \end{figure*}

    While Schenk's URANS and Zawodny's DES represent the low and high end of fidelity in mesh-based CFD, respectively, Delorme \textit{et al.} \cite{Delorme2021} reported two LES that lay somewhere in between both ends.
    One LES used an actuator line model (ALM) similar to our meshless LES, while the other used an immersed boundary method (IBM).
    All solvers previously mentioned are summarized in~\cref{table:rotor:solvers} and their computational costs are compared to our meshless LES in~\cref{table:rotor:benchmark}.
    The LES by Delorme \textit{et al.}, shown in~\cref{fig:rotor:meshsim}, is an implicit LES that relies on numerical dissipation to approximate subgrid-scale turbulent diffusion, reportedly predicting a mean $C_T$ within 2\% of the experimental value.
    Out of the aforementioned mesh-based simulations, their LES-ALM simulation is the most akin to our meshless LES since both use an ALM while resolving vortical structures with comparable fidelity, as shown in~\cref{fig:rotor:vpmsim} (right) and~\cref{fig:rotor:meshsim} (right).
    The main computational advantage of our meshless LES lays in that computational elements are only placed where vorticity is originated and are automatically convected by the flow field, as shown in~\cref{fig:rotor:vpmsim} (left), while mesh-based LES requires meshing the entire space, as shown in~\cref{fig:rotor:meshsim} (left).
    Their LES-ALM simulation of a modified DJI 9443 rotor reportedly took a wall-clock time of 24 hours per every 10 rotor revolutions using 845 CPUs, equivalent to 20k core-hours.
    Compared to the 140 core-hours of the rVPM, our meshless LES is 145x faster (or two orders of magnitude faster) than this mesh-based LES.

    In summary, as shown in~\cref{table:rotor:benchmark}, our meshless LES appears to be two orders of magnitude faster than a mesh-based LES with similar fidelity, while being one order of magnitude faster than a low-fidelity URANS simulation and three orders of magnitude than high-fidelity DES.
    It is difficult to justify an exact comparison between the computational cost of each simulation since each solver used different order-of-convergence schemes, spatial and temporal resolution, and computing hardware.
    However, the comparison is still qualitatively insightful: our meshless LES can be 10x to 1000x faster than conventional mesh-based CFD approaches.

\makeatletter\onecolumngrid@push\makeatother

    \begin{table*}[t]

        \vspace{5mm}

        \caption{Description of CFD solvers in rotor benchmark.}
        \label{table:rotor:solvers}

        \hspace{-10mm}
        \begin{tabular}{ccccc}
            \hline
            \textbf{Simulation} & \,\,\textbf{Software} & \,\,\textbf{Blade Scheme} & \,\,\textbf{Turbulence Model} & \,\,\textbf{Computational Elements}   \\ \hline \vspace{-2mm} \\

            \makecell{rVPM \vspace{-1mm}\\ {\tiny (meshless LES)}}                & FLOWVPM               & Actuator line model       &  Anisotropic dynamic SFS      &  1M vortex elements                  \vspace{1mm} \\

            URANS               & STAR-CCM+             & \makecell{Blade resolved \vspace{-1mm}\\ {\tiny ($y^+=30$, all-$y^+$ wall treament)}}  &  SST $k$--$\omega$    & 14M grid cells   \\

            LES-ALM             & MIRACLES              & Actuator line model       &  \makecell{None \vspace{-1mm}\\ {\tiny (numerical dissipation)}} & 50M grid points     \\

            LES-IBM             & MIRACLES              & Immersed boundary method  &  \makecell{None \vspace{-1mm}\\ {\tiny (numerical dissipation)}} & 216M grid points   \vspace{1mm} \\

            DES                 & OVERFLOW2             & Blade resolved            &  Spalart-Allmaras DES         & 260M grid points                      \vspace{1.5mm} \\ \hline
        \end{tabular}
    \end{table*}

    \begin{table*}[]

        \vspace{5mm}

        \caption{Rotor benchmark resolving 10 rotor revolutions.
        }
        \label{table:rotor:benchmark}

        \centering
        \begin{tabular}{ccccc}
            \hline
            \textbf{Simulation} & \,\,\textbf{CPU Cores} & \,\,\textbf{Wall-Clock} & \,\,\textbf{Core-Hours} & \,\,\textbf{rVPM Speedup} \\ \hline
            rVPM            & 32                & 4.3 hours  & 140   & --           \\
            URANS           & 192               & 9.6 hours  & 1.8k  & \raisebox{-.8ex}{\~{}}10x faster   \\
            LES-ALM         & 845               & 24  hours  & 20k  & \raisebox{-.8ex}{\~{}}100x faster  \\
            LES-IBM         & 1000              & 96  hours  & 96k  & \raisebox{-.8ex}{\~{}}500x faster   \\
            DES             & 1008              & 300 hours  & 300k  & \raisebox{-.8ex}{\~{}}1000x faster \\ \hline
        \end{tabular}
    \end{table*}

    \FloatBarrier


    \section{Conclusions}
        In this study we have developed a meshless scheme for large eddy simulation that is accurate, stable, and computationally efficient.
        We started off by deriving a general set of VPM governing equations that solve the LES-filtered Navier-Stokes equations.
        The classic VPM turns out to be one of the formulations arising from these general equations, which seems to locally violate both conservation of mass and angular momentum, thus explaining its tendency to be numerically unstable.
        We then derived a new VPM formulation that uses the particle shape to reinforce local conservation of mass and angular momentum.

        In addition to the reformulated VPM, we developed a low-dissipation anisotropic SFS model that uses vortex stretching as the physical mechanism for turbulence.
        We proposed a novel dynamic procedure for calculating the model coefficient and a strategy for backscatter control.
        The dynamic procedure is based on a simultaneous balance of enstrophy-production and derivatives between true and modeled SFS contributions.
        This SFS model is apt for both meshless and conventional mesh-based CFD, and is well suited for flows with coherent vortical structures where the predominant cascade mechanism is vortex stretching.

        Extensive validation was presented, asserting the scheme comprised of the reformulated VPM and SFS model as a meshless LES accurately resolving large-scale turbulent dynamics.
        Advection, viscous diffusion, and vortex stretching were validated through the simulation of vortex rings, showing good agreement with DNS and LBM.
        It was shown that the reformulated VPM runs as fast as the classic VPM, adding no significant overhead, while the SFS model adds a computational overhead of 8\% when a constant model coefficient is prescribed and 43\% with the dynamic procedure.
        In the simulation of a turbulent round jet, a favorable agreement with experiments showed that our scheme is able to resolve the mean and fluctuating components of turbulent flow, while also resolving Reynolds stress directly.

        Finally, computational efficiency was demonstrated in an engineering application through the simulation of an aircraft rotor in hover.
        Our meshless LES showed to be two orders of magnitude faster than a mesh-based LES with similar fidelity, while being one order of magnitude faster than a low-fidelity URANS simulation and three orders of magnitude than a high-fidelity DES.

        This study validates the reformulated VPM as an LES that is both numerically stable and meshless, while able to accurately resolve mean and fluctuating large-scale features of turbulent flow with minimal computational effort.
        In future work, a wider variety of boundary conditions can be explored, especially aiming at introducing solid boundaries without a mesh.
        This study limited its computation to conventional CPU paradigms, but previous work \cite{Yokota2013a,Hu2014} shows a strong potential for massive parallelization and speed up in heterogenous GPU architectures.

    \begin{acknowledgments}

        This material is based upon work supported by the National Science Foundation under Grant No.~2006219. Any opinions, findings, and conclusions or recommendations expressed in this material are those of the author(s) and do not necessarily reflect the views of the National Science Foundation.

        FLOWVPM uses a modified version of the open-source code ExaFMM \cite{Wang2021} originally developed by Lorena Barba and Rio Yokota.
        FLOWVPM is implemented in the Julia programming language \cite{Bezanson2017} and integrates the open-source software ParaView \cite{Ahrens2005,Ayachit2015} as its visualization engine.
        The authors wish to thank Prof. Adrin Gharakhani for the insightful conversations on vortex methods.
    \end{acknowledgments}

    \appendix
    
\section{Derivation of $\frac{\mathrm{d}}{\mathrm{d} t} \left(\zeta_{\sigma_p} ({\mathbf x} - {\mathbf x}_p) \right)$}
\label{sec:app:der:dzetadt}

    Let us define two auxiliary functions $G$ and $g$ such that
    \begin{align*}
            G\left( \mathbf{x},\, t \right)
        =
            g\left( \mathbf{x},\, \mathbf{y}(t),\, \sigma(t) \right)
    \end{align*}
    and
    \begin{align*}
            g\left( \mathbf{x},\, \mathbf{y}(t),\, \sigma(t) \right)
        =
            \zeta_\sigma \left( \mathbf{x} - \mathbf{y} \right)
    ,\end{align*}
    resulting in the following properties.

    \vspace{10mm}
    \noindent \textrm{Property 1}
    \begin{align*}
            \frac{\mathrm{d} }{\mathrm{d} t} G\left( \mathbf{x},\, t \right)
        & =
            \frac{\partial G}{\partial t} \left( \mathbf{x},\, t \right)
            +
            \left( \frac{\partial \mathbf{x}}{\partial t} \cdot \nabla \right)
            G\left( \mathbf{x},\, t \right)
    ,\end{align*}
    by definition.

    \vspace{10mm}
    \noindent \textrm{Property 2}
    \begin{align*}
            \frac{\partial G}{\partial t} \left( \mathbf{x},\, t \right)
        & =
            \frac{\partial }{\partial t}
            \left(
                g \left( \mathbf{x},\, \mathbf{y}(t),\, \sigma(t) \right)
            \right)
        \\
        & =
            \frac{\partial g}{\partial \sigma}\frac{\partial \sigma}{\partial t}
            +
            \sum\limits_i
                \frac{\partial g}{\partial y_i}
                \frac{\partial y_i}{\partial t}
        \\
        & =
            \frac{\partial g}{\partial \sigma}\frac{\partial \sigma}{\partial t}
            +
            \left( \frac{\partial \mathbf{y}}{\partial t} \cdot \nabla_y \right) g
    \end{align*}

    \noindent \textrm{Property 3}
    \begin{align*}
            \left( \frac{\partial \mathbf{y}}{\partial t} \cdot \nabla_y \right)
            g \left( \mathbf{x},\, \mathbf{y},\, \sigma \right)
        & =
            \left( \frac{\partial \mathbf{y}}{\partial t} \cdot \nabla_y \right)
            \zeta_\sigma \left( \mathbf{x} - \mathbf{y} \right)
        \\
        & =
            - \left( \frac{\partial \mathbf{y}}{\partial t} \cdot \nabla \right)
            \zeta_\sigma \left( \mathbf{x} - \mathbf{y} \right)
    ,\end{align*}
    since
    \begin{align*}
            \frac{\partial }{\partial y_i} \left( \zeta_\sigma \left( \mathbf{x} - \mathbf{y} \right) \right)
        =
            - \frac{\partial }{\partial x_i} \left( \zeta_\sigma \left( \mathbf{x} - \mathbf{y} \right) \right)
    \end{align*}

    \vspace{10mm}
    Now, using $\frac{\partial \zeta_\sigma}{\partial t} = \frac{\partial \zeta_\sigma}{\partial \sigma}\frac{\partial \sigma}{\partial t}$ and replacing Property 3 in Property 2 we get
    \begin{align*}
                \frac{\partial G}{\partial t} \left( \mathbf{x},\, t \right)
            & =
                \frac{\partial \zeta_\sigma }{\partial t} \left( \mathbf{x} - \mathbf{y} \right)
                -
                \left( \frac{\partial \mathbf{y}}{\partial t} \cdot \nabla \right)
                \zeta_\sigma \left( \mathbf{x} - \mathbf{y} \right)
    ,\end{align*}
    and using this in Property 1,
    \begin{align*}
            \frac{\mathrm{d} }{\mathrm{d} t} G\left( \mathbf{x},\, t \right)
        & =
            \frac{\partial \zeta_\sigma }{\partial t} \left( \mathbf{x} - \mathbf{y} \right)
            -
            \left( \frac{\partial \mathbf{y}}{\partial t} \cdot \nabla \right)
            \zeta_\sigma \left( \mathbf{x} - \mathbf{y} \right)
        \\
        & \qquad \qquad
            +
            \left( \frac{\partial \mathbf{x}}{\partial t} \cdot \nabla \right)
            \zeta_\sigma \left( \mathbf{x} - \mathbf{y} \right)
        \\
        & =
            \frac{\partial \zeta_\sigma }{\partial t} \left( \mathbf{x} - \mathbf{y} \right)
            +
            \left[
                \left(
                    \frac{\partial \mathbf{x}}{\partial t}
                    -
                    \frac{\partial \mathbf{y}}{\partial t}
                \right)
                \cdot \nabla
            \right]
            \zeta_\sigma \left( \mathbf{x} - \mathbf{y} \right)
    .\end{align*}
    Defining $\mathbf{u}\left( \mathbf{x} \right) = \frac{\partial \mathbf{x}}{\partial t}$ and assuming $\frac{\partial \mathbf{y}}{\partial t} = \mathbf{u}\left( \mathbf{y} \right)$,
    \begin{align*}
            \frac{\mathrm{d} }{\mathrm{d} t} G\left( \mathbf{x},\, t \right)
        & =
            \frac{\partial \zeta_\sigma }{\partial t} \left( \mathbf{x} - \mathbf{y} \right)
        \\
        & \qquad \qquad
            +
            \left[
                \left(
                    \mathbf{u}\left( \mathbf{x} \right)
                    -
                    \mathbf{u}\left( \mathbf{y} \right)
                \right)
                \cdot \nabla
            \right]
            \zeta_\sigma \left( \mathbf{x} - \mathbf{y} \right)
    .\end{align*}
    Finally, given that
    \begin{align*}
            \frac{\mathrm{d} }{\mathrm{d} t} G\left( \mathbf{x},\, t \right)
        & =
            \frac{\mathrm{d} }{\mathrm{d} t} \zeta_\sigma \left( \mathbf{x} - \mathbf{y} \right)
    \end{align*}
    by definition, we conclude that
    \begin{align*}
            \frac{\mathrm{d} }{\mathrm{d} t} \zeta_\sigma \left( \mathbf{x} - \mathbf{y} \right)
        & =
            \frac{\partial \zeta_\sigma }{\partial t} \left( \mathbf{x} - \mathbf{y} \right)
        \\
        & \qquad \qquad
            +
            \left[
                \left(
                    \mathbf{u}\left( \mathbf{x} \right)
                    -
                    \mathbf{u}\left( \mathbf{y} \right)
                \right)
                \cdot \nabla
            \right]
            \zeta_\sigma \left( \mathbf{x} - \mathbf{y} \right)
    .\end{align*}

\section{Derivation of $\frac{\mathrm{d} \hat{\boldsymbol{\Gamma}}}{\mathrm{d} t}$} \label{sec:app:der:reorientation}

    Decomposing the vortex strength $\boldsymbol\Gamma$ into its magnitude $\Gamma \equiv \Vert \boldsymbol\Gamma \Vert $ and unit vector $\hat{\boldsymbol\Gamma} \equiv \frac{\boldsymbol\Gamma}{\Vert \boldsymbol\Gamma \Vert}$ as $\boldsymbol\Gamma  =\Gamma \hat{\boldsymbol\Gamma}$, we calculate the following
    \begin{align*}
            \frac{\mathrm{d} \boldsymbol{\Gamma}}{\mathrm{d} t}
            -
            \left(
                \frac{\mathrm{d} \boldsymbol{\Gamma}}{\mathrm{d} t} \cdot \hat{\boldsymbol{\Gamma}}
            \right)
            \hat{\boldsymbol{\Gamma}}
        & =
            \frac{\mathrm{d} }{\mathrm{d} t}
            \left(
                \Gamma \hat{\boldsymbol\Gamma}
            \right)
            -
            \left\{
                \left[
                    \frac{\mathrm{d} }{\mathrm{d} t}
                    \left(
                    \Gamma \hat{\boldsymbol\Gamma}
                    \right)
                \right]
                    \cdot \hat{\boldsymbol{\Gamma}}
            \right\}
            \hat{\boldsymbol{\Gamma}}
        \\
        & =
            \frac{\mathrm{d}\Gamma }{\mathrm{d} t}
            \hat{\boldsymbol\Gamma}
            +
            \Gamma
            \frac{\mathrm{d} \hat{\boldsymbol\Gamma} }{\mathrm{d} t}
            -
            \left(
                \frac{\mathrm{d}\Gamma }{\mathrm{d} t}
                \cancelto{1}{\hat{\boldsymbol\Gamma} \cdot \hat{\boldsymbol{\Gamma}}}
                +
                \Gamma
                \frac{\mathrm{d} \hat{\boldsymbol\Gamma} }{\mathrm{d} t}
                \cdot \hat{\boldsymbol{\Gamma}}
            \right)
            \hat{\boldsymbol{\Gamma}}
        \\
        & =
            \Gamma
            \frac{\mathrm{d} \hat{\boldsymbol\Gamma} }{\mathrm{d} t}
            -
            \left(
                \Gamma
                \frac{\mathrm{d} \hat{\boldsymbol\Gamma} }{\mathrm{d} t}
                \cdot \hat{\boldsymbol{\Gamma}}
            \right)
            \hat{\boldsymbol{\Gamma}}
        \\
        & =
            \Gamma
            \frac{\mathrm{d} \hat{\boldsymbol\Gamma} }{\mathrm{d} t}
            -
            \left[
                \frac{\Gamma}{2}
                \frac{\mathrm{d} }{\mathrm{d} t}
                \left(
                    \cancelto{1}{ \hat{\boldsymbol{\Gamma}} \cdot \hat{\boldsymbol{\Gamma}} }
                \right)
            \right]
            \hat{\boldsymbol{\Gamma}}
        \\
        & =
            \Gamma
            \frac{\mathrm{d} \hat{\boldsymbol\Gamma} }{\mathrm{d} t}
    \end{align*}
    Hence, it follows that
    \begin{align*}
            \frac{\mathrm{d} \hat{\boldsymbol\Gamma} }{\mathrm{d} t}
        =
            \frac{1}{\Gamma}
            \frac{\mathrm{d} \boldsymbol{\Gamma}}{\mathrm{d} t}
            -
            \frac{1}{\Gamma}
            \left(
                \frac{\mathrm{d} \boldsymbol{\Gamma}}{\mathrm{d} t} \cdot \hat{\boldsymbol{\Gamma}}
            \right)
            \hat{\boldsymbol{\Gamma}}
    .\end{align*}

\section{Analytical Dynamics of Vortex Rings} \label{sec:app:vortexring}
    Given a system of $N$ collinear vortex rings with circulation $\Gamma_i$, centroid $Z_i$, and radius $R_i$ for $i=1,...,N$, let us assume a singular vorticity along each ring's centerline,
    \begin{align*}
            \omega(r, z, t)
        =
            \sum\limits_{i=1}^{N}
                \Gamma_i \delta\left( z - Z_i(t) \right) \delta\left( r - R_i(t) \right)
    \end{align*}
    where $\delta$ is the Dirac delta, and $r$ and $z$ are the radial and axial coordinates of a cylindrical system.
    As described by Borisov \textit{et al.} \cite{Borisov2013}, the interactions between singular rings lead to dynamics dictated by
    \begin{align} \label{eq:ring:dRdt}
            \frac{\mathrm{d} R_i}{\mathrm{d} t}
        =
            - \frac{1}{R_i} \frac{\partial }{\partial Z_i}
            \sum\limits_{j \neq i}
                \Gamma_j G\left( R_i, Z_i, R_j, Z_j \right)
    \end{align}
    and
    \begin{align} \label{eq:ring:dZdt}
            \frac{\mathrm{d} Z_i}{\mathrm{d} t}
        =
            V_i
            +
            \frac{1}{R_i} \frac{\partial }{\partial R_i}
            \sum\limits_{j \neq i}
                \Gamma_j G\left( R_i, Z_i, R_j, Z_j \right)
    \end{align}
    where $V_i$ is the self-induced velocity of the ring and $G$ is the streamfunction of an infinitely thin circular vortex filament of unit intensity.
    $G$ is given by
    \begin{align*}
            G(r, z, R, Z)
        =
            \frac{\sqrt{r R}}{2\pi}
            \left[
                \left(
                    \frac{2}{k} - k
                \right)
                K(k^2)
                -
                \frac{2}{k} E(k^2)
            \right]
    ,\end{align*}
    where
    \begin{align*}
            k
        =
            \sqrt{
                \frac{4rR}{\left( z - Z\right)^2 + \left( r + R \right)^2}
            }
    ,\end{align*}
    and $K(m)$ and $E(m)$ are the complete elliptic integrals of the first and second kind, respectively:
    \begin{align*}
            K (m)
        =
            \int\limits_0^{\frac{\pi}{2}}
                \frac{1}{\sqrt{1 - m \sin^2 \theta}}
                \, \mathrm{d}\theta
    \end{align*}
    and
    \begin{align*}
            E (m)
        =
            \int\limits_0^{\frac{\pi}{2}}
                \sqrt{1 - m \sin^2 \theta}
                \, \mathrm{d}\theta
    .\end{align*}

    Since the velocity induced by a singular filament on itself is infinite, a regularized vorticity distribution must be assumed when calculating $V_i$.
    Assuming a thin toroidal vortex ring, Saffman \cite{Saffman1970,Saffman1985} obtained
    \begin{align} \label{eq:ring:V}
            V
        =
            \frac{\Gamma}{4\pi R}
            \left(
                \ln\frac{8}{a/R} + C\left( a/R \right)
            \right)
    ,\end{align}
    where the function $C$ is determined from the vorticity distribution inside the ring core.
    For thin Gaussian rings with $a \ll R$, $C(\epsilon)$ becomes $C = -0.558$.
    For thick Gaussian rings, Archer \textit{et al.} \cite{Archer2008a} observed through direct numerical simulation that the initial Gaussian distribution quickly becomes skewed and leads to
    \begin{align} \label{eq:ring:C}
        C(\epsilon) \approx -0.558 - 1.12 \epsilon^2 - 5.0 \epsilon^4
    ,\end{align}
    where $\epsilon \equiv a/R$.
    Thus, the self-induced velocity depends on the current core size $a$ of the ring.
    At the same time, viscous diffusion constantly spreads the vorticity distribution according to $\frac{\mathrm{d} \boldsymbol\omega}{\mathrm{d} t} = \nu \nabla^2 \boldsymbol\omega$.
    In a vortex filament with a Gaussian vorticity distribution, ${ \omega (r) = \frac{\Gamma}{\pi a^2}e^{ -r^2/a^2} }$, the viscous diffusion leads to
    \begin{align} \label{eq:ring:dadt:viscous}
        \frac{\mathrm{d} a^2}{\mathrm{d} t} = 4 \nu
    .\end{align}

    Finally, when viscous effects are ignored through operator splitting, the ring must conserve mass due to Helmholtz' circulation theorem in incompressible flow.
    This is expressed as
    \begin{align*}
        \frac{\mathrm{d} }{\mathrm{d} t} \left( R a^2 \right) = 0
    ,\end{align*}
    which leads to
    \begin{align} \label{eq:ring:dadt:stretching}
        \frac{\mathrm{d} a}{\mathrm{d} t} = - \frac{a}{2R}\frac{\mathrm{d} R}{\mathrm{d} t}
    ,\end{align}
    meaning that the core size must shrink as the ring is stretched by the interaction with neighboring rings.
    We have observed that incorporating this last equation introduces dynamics that are incongruent with the numerical results from DNS, LBM, and VPM discussed in~\cref{sec:res:isolatedring,sec:res:leapfrog}.
    This could be due to the fact that~\cref{eq:ring:dRdt,eq:ring:dZdt} ignore the effects of core size in the mutual interactions between rings, thus~\cref{eq:ring:dadt:stretching} is introducing an unphysical one-way coupling between $\frac{\mathrm{d} a}{\mathrm{d} t}$ and $\frac{\mathrm{d} R}{\mathrm{d} t}$.
    Hence,~\cref{eq:ring:dadt:stretching} was not used in the analytical solutions shown in~\cref{sec:res:isolatedring,sec:res:leapfrog}.

    In summary,~\cref{eq:ring:dRdt,eq:ring:dZdt,eq:ring:V,eq:ring:C,eq:ring:dadt:viscous} form a system of analytical ordinary differential equations that describe the dynamics of an arbitrary number of coaxial, viscous, incompressible vortex rings. In this derivation we have assumed that the core thickness only affects the self-induced velocity of each ring, while its effects are ignored in the interactions between the rings and the core size is not affected by vortex stretching.

    \bibliography{library.bib}

\end{document}